\RequirePackage{ifpdf}
\ifpdf 
\documentclass[pdftex]{sigma}
\else
\documentclass{sigma}
\fi

\usepackage{bm} 

\numberwithin{equation}{section}

\newcommand{\nc}{\newcommand}
\nc{\rnc}{\renewcommand}
\nc{\nn}{\nonumber}
\nc{\db}{\displaybreak[0]\\}
\rnc{\[}{\left[}
\rnc{\]}{\right]}
\rnc{\(}{\left(}
\rnc{\)}{\right)}
\nc{\lt}{\left\{}
\nc{\rt}{\right\}}
\nc{\lam}{\lambda}
\nc{\eps}{\epsilon}
\nc{\veps}{\varepsilon}
\nc{\vp}{\varphi}
\nc{\ch}{\cosh}
\nc{\sh}{\sinh}
\nc{\bra}{\langle}
\nc{\ket}{\rangle}
\rnc{\i}{{\rm i}}
\rnc{\d}{{\rm d}}
\nc{\Res}[2]{{\rm Res}\[\left.#1\right|_{#2}\]}
\nc{\intall}{\int_{-\infty}^\infty}
\newcommand{\be}{\begin{equation}}
\newcommand{\ee}{\end{equation}}
\newcommand{\bea}{\begin{eqnarray}}
\newcommand{\eea}{\end{eqnarray}}
\newcommand{\non}{\nonumber}

\begin{document}

\allowdisplaybreaks

\renewcommand{\thefootnote}{$\star$}

\renewcommand{\PaperNumber}{056}

\FirstPageHeading

\ShortArticleName{One-Point Functions of the Integrable Spin-1 XXZ Chain}

\ArticleName{Quantum Group $\boldsymbol{U_q(sl(2))}$ Symmetry\\ and
Explicit Evaluation of the One-Point Functions \\ of
the Integrable Spin-1 XXZ Chain\footnote{This paper is a
contribution to the Proceedings of the International Workshop ``Recent Advances in Quantum Integrable Systems''. The
full collection is available at
\href{http://www.emis.de/journals/SIGMA/RAQIS2010.html}{http://www.emis.de/journals/SIGMA/RAQIS2010.html}}}

\Author{Tetsuo DEGUCHI and Jun SATO}

\AuthorNameForHeading{T.~Deguchi and J.~Sato}

\Address{Department of Physics, Graduate School of Humanities and Sciences,
Ochanomizu University \\
2-1-1 Ohtsuka, Bunkyo-ku, Tokyo 112-8610, Japan}
\Email{\href{mailto:deguchi@phys.ocha.ac.jp}{deguchi@phys.ocha.ac.jp}, \href{mailto:jsato@sofia.phys.ocha.ac.jp}{jsato@sofia.phys.ocha.ac.jp}}

\ArticleDates{Received October 29, 2010, in f\/inal form May 26, 2011;  Published online June 10, 2011}

\Abstract{We show some symmetry relations among the correlation functions
of the integrable higher-spin XXX and XXZ spin chains, where
we explicitly evaluate the multiple integrals
representing the one-point functions in the spin-1 case.
We review the multiple-integral representations of
correlation functions for the integrable higher-spin XXZ chains derived
in a region of the massless regime including the anti-ferromagnetic point.
Here we make use of the gauge transformations between
the symmetric and  asymmetric $R$-matrices, which correspond to
the principal and homogeneous gradings, respectively, and we
send the inhomogeneous parameters to the set of complete $2s$-strings.
We also give a numerical support for the analytical expression of
the one-point functions in the spin-1 case.}


\Classification{82B23}

\section{Introduction}

The correlation functions of the  spin-1/2 XXZ spin chain
 has attracted much interest during the last decades in mathematical physics,
and several nontrivial results such as their multiple-integral representations
have been obtained explicitly \cite{Korepin,Jimbo-Miwa,Review}.
The Hamiltonian of the XXZ spin chain
under the periodic boundary conditions (P.B.C.) is given by
\begin{gather*}
{\cal H}_{\rm XXZ} =
\sum_{j=1}^{L} \left(\sigma_j^X \sigma_{j+1}^X +
 \sigma_j^Y \sigma_{j+1}^Y + \Delta \sigma_j^Z \sigma_{j+1}^Z  \right) .
\end{gather*}
Here $\sigma_j^{a}$ ($a=X, Y, Z$) are the Pauli matrices def\/ined
on the $j$th site and $\Delta$ denotes the anisotropy
of the exchange coupling. The P.B.C.\ are given by
$\sigma_{L+1}^{a} = \sigma_{1}^{a}$ for $a=X, Y, Z$.

The XXZ Hamiltonian shows the quantum phase transition:
the ground state of the XXZ spin chain depends on $\Delta$.
For $|\Delta| > 1$ the low-lying excitation spectrum at the ground state
has a gap, while for $|\Delta| \leq 1$ it has no gap.
Here we remark that the quantum phase transition
that we have discussed is associated with the behavior of the XXZ spin chain in the thermodynamic limit: $L \rightarrow \infty$.
In terms of the $q$ parameter of the quantum group $U_q(sl_2)$,
we express  $\Delta$ as follows
\begin{gather*}
\Delta = {\frac 1 2}\left(q+q^{-1}\right) .
\end{gather*}
It is often convenient to def\/ine parameters $\eta$
and $\zeta$ by $q = \exp \eta$ with $\eta=i \zeta$.
Here we have $\Delta=\cosh \eta=\cos \zeta$.
In the massive regime $\Delta> 1$, we set $\eta>0$.
In the massless regime $-1 < \Delta \le 1$, we set $\eta= \i \zeta$
where $\zeta$ satisf\/ies $0 \le  \zeta < \pi$.
Here, the XXX limit is given by $\eta \to +0$ or $\zeta \to +0$.
Here we remark that the XXZ Hamiltonian can be derived from
the $R$-matrix of the af\/f\/ine quantum group with $q$ parameter,
$U_q(\widehat{sl_2})$: we derive
the $R$-matrix by solving the intertwining relations,
construct the XXZ transfer matrix from the product of the~$R$ matrices,
and then we derive the XXZ Hamiltonian by
taking the logarithmic derivative of the XXZ transfer matrix.
Thus, the $q$ parameter of the af\/f\/ine quantum group
is related to the ground state of the XXZ spin chain through $\Delta$.

The multiple-integral representations of the XXZ correlation functions
were derived for the f\/irst time by making use of the $q$-vertex operators
through the af\/f\/ine quantum-group symmetry
in the massive regime for the inf\/inite lattice at zero temperature
\cite{Miki,Jimbo-Miwa}.
In the massless regime they were derived
by solving the $q$-KZ equations \cite{Jimbo-Miwa-qKZ,Takeyama}.
Making use of the algebraic Bethe-ansatz techniques
\cite{Slavnov,Korepin,MS2000,KMT1999,MT2000},
the multiple-integral representations were
derived for the spin-1/2 XXZ correlation functions
under a non-zero magnetic f\/ield~\cite{KMT2000}.
Here, they are derived through the thermodynamic limit
after calculating the scalar product for a f\/inite chain.
The multiple-integral representations were extended into those
at f\/inite temperatures~\cite{Goehmann-CF},
and even for a~large f\/inite chain~\cite{Damerau}.
Interestingly, they are factorized
in terms of single integrals~\cite{Jimbo-Miwa-Smirnov}.
We should remark that the multiple-integral representations
of the dynamical correlation functions were also obtained
under f\/inite-temperatures \cite{Sakai}.
Furthermore, the asymptotic expansion of a correlation function
of the XXZ model has been systematically discussed~\cite{KMTK2008}.
Thus, the exact study of the XXZ correlation functions
should play an important role not only
in the mathematical physics of integrable models
but also in many areas of theoretical physics.

Recently,  the form factors of the integrable higher-spin XXX spin chains
and the multiple-integral representations of
correlation functions for the integrable higher-spin XXX and
XXZ chains have been derived by the algebraic Bethe-ansatz method
\cite{Kitanine2001,Castro-Alvaredo,DM1,DM2,DM3}
(see also~\cite{Terras1999}).
The spin-1 XXZ Hamiltonian under the P.B.C.\
is given by the following~\cite{Bougourzi}:
\begin{gather}
{\cal H}_{\text{spin-1 XXZ}}  =  J \sum_{j=1}^{N_s} \bigg\{
{\vec S}_j \cdot {\vec S}_{j+1} - ({\vec S}_j \cdot {\vec S}_{j+1})^2
- {\frac 1 2}(q-q^{-1})^2 [S_j^{z} S_{j+1}^{z} - (S_j^{z} S_{j+1}^{z})^2
+ 2 (S_j^{z})^2] \non \\
\qquad{}   - (q+q^{-1} -2) [(S_j^{x} S_{j+1}^{x} + S_j^{y} S_{j+1}^{y}) S_j^{z} S_{j+1}^{z} + S_j^{z} S_{j+1}^{z} (S_j^{x} S_{j+1}^{x} + S_j^{y} S_{j+1}^{y}) \bigg\}
  .
\label{eq:spin1-XXZ-Hamiltonian}
\end{gather}
Furthermore, the multiple-integral representations have been obtained for
the correlation functions at f\/inite temperature of the integrable spin-1 XXX chain \cite{GSS}.
The solvable higher-spin generalizations of the XXX and XXZ spin chains
have been derived by the fusion method in several references
\cite{Fateev-Zamolodchikov,FusionXXX,BabujianPLA,Babujian,SAA,Babujian-Tsvelick,KR,V-DWA}.
In the region: $0 \le \zeta < \pi/2s$,
the spin-$s$ ground-state should be given
by a set of string solutions \cite{Takhtajan,Sogo}.
Furthermore, the critical behavior should be
given by the SU(2) WZWN model of level $k=2s$ with central charge
$c=3s/(s+1)$
\cite{Johannesson,Johannesson2,Alcaraz-Martins:XXX,Affleck,Dorfel,Avdeev,Alcaraz-Martins:XXZ,Fowler,Frahm,deVega-Woynarovich,KR,KB,KBP,JSuzuki}.
For the integrable higher-spin XXZ spin chain
correlation functions have been discussed in the massive regime
by the method of $q$-vertex operators
\cite{Idzumi0,Idzumi1,Idzumi2,Bougourzi,Konno,Kojima}.

The purpose of this paper is to show some symmetry relations
among the correlation functions of the integrable spin-$s$ XXZ spin chain
by explicitly calculating the multiple-integral representations for
the spin-1 one-point functions.
Associated with the quantum group $U_q(sl(2))$ symmetry,
there are several relations among the expectation values
of products of the matrix elements of the monodromy matrices.
For the spin-1 case, we conf\/irm some of them by evaluating
the multiple integrals analytically and explicitly.
Here we should remark that the derivation
of the multiple-integral representations for
the spin-$s$ XXZ correlation functions
given in the previous papers \cite{DM1,DM2,DM3}
was not completely correct: the application of the formulas
of the quantum inverse-scattering problem  was not valid \cite{DM4,DM5}.
We thus review the revised derivation \cite{DM4,DM5} in the paper.
The spin-$s$ correlation function of an arbitrary entry is now
expressed in terms of {\it a sum} of multiple integrals,
not as a single multiple integral.
Furthermore, we show numerical results which conf\/irm the analytical
expressions of the spin-1 one-point functions.

Let us express by  $\bra E^{00} \ket$,  $\bra E^{11} \ket$ and
$\bra E^{22} \ket$, the expectation values of
$S_1^Z = 1$, $S_1^Z = 0$ and $S_1^Z =-1$, respectively, where
$S_1^Z$ denotes the $Z$-component of the spin operator def\/ined on
the f\/irst site. Then, we have the following:
\begin{gather*}
 \bra E^{22} \ket=\bra E^{00} \ket=\frac{\zeta-\sin\zeta\cos\zeta}{2\zeta\sin^2\zeta},
\qquad
\bra E^{11} \ket
=\frac{\cos\zeta(\sin\zeta-\zeta\cos\zeta)}{\zeta\sin^2\zeta}.
\end{gather*}
We shall show the derivation
of $\bra E^{00} \ket$,  $\bra E^{11} \ket$ and
$\bra E^{22} \ket$, in detail.
Here we remark that the expressions of
$\bra E^{22} \ket$, the emptiness formation probability,
 and $\bra E^{11} \ket$ have been
reported in \cite{DM2} without an explicit derivation.
In fact, although the derivation was not completely correct,
the expressions of the spin-1 one-point functions are correct~\cite{DM4,DM5}.
Here, the quantum group
symmetry as well as the spin inversion symmetry play an important role,
as we shall show explicitly in the present paper.

It is nontrivial to evaluate the multiple integral representations
of the XXX and XXZ models analytically or even numerically.
Let us now return to the spin-1/2 case.
Boos and Korepin have analytically evaluated
the emptiness formation probability $P(n)$ of the XXX spin chain
for up to $n=4$ successive lattice sites~\cite{Boos-Korepin}.
Performing explicit evaluation of the multiple integrals,
they successfully reproduced Takahashi's result that was obtained through
the one-dimensional Hubbard model~\cite{Taka77}.
The method was applied to all the density matrix elements
for up to $n=4$ successive lattice sites
in the XXX chain~\cite{XXXn=4} and also
in the XXZ chain~\cite{Kato2003,Kato2004,Kato3}.
Furthermore, the algebraic method to obtain the correlation functions
of the XXX chain based on the $q$KZ equation has been developed~\cite{BKS}
and the two-point functions up to $n=8$
have been obtained so far~\cite{BST,SS,SST05,SST06}.
At the special anisotropy $\Delta=1/2$, some further results
have been shown for the correlation functions
through explicit evaluation~\cite{RS,KMST02,KMST05b,SS07}.

The paper consists of the following. In Section~\ref{section2} we review the Hermitian elementary matrices~\cite{DM2}, and give the basis
vectors and their conjugate vectors in the spin-1 case
as an illustrative example. We also show a formula for expressing
higher-spin local operators in terms of spin-1/2 local operators
in the spin-1 case, which plays a central role
in the revised  method \cite{DM4,DM5}.
In Section~\ref{section3} we summarize the notation of the fusion transfer matrices and the quantum inverse scattering problem for the spin-$s$ operators.
For an illustration, in Section~\ref{section4},
we show some relations among the expectation values of the
Hermitian elementary matrices in the spin-1 XXX case and then in the spin-1 XXZ case. In particular, we show the spin inversion symmetry.
We also show the transformation which maps
the basis vectors of the spin-1 representation
$V^{(2)}$ constructed in the tensor product
of the spin-1/2 representations $V^{(1)} \otimes V^{(1)}$
to the basis of the three-dimensional vector space ${\bf C}^3$.
The former basis is related to the fusion method,
while the spin-1 XXZ Hamiltonian (\ref{eq:spin1-XXZ-Hamiltonian})
is formulated in terms of the latter basis.
 In Section~\ref{section5},  we review the revised multiple-integral
representations of correlation functions for
the integrable spin-$s$ XXZ spin chain \cite{DM4,DM5}.
Here we remark that necessary corrections to the previous papers
\cite{DM1} and \cite{DM2} are listed in references~[20] and~[21]
of the paper~\cite{DM5}, respectively.
In Section~\ref{section6}, we explicitly calculate the multiple integrals
of the one-point functions for the spin-1 XXZ spin chain for a region in the massless regime. We show some details of the calculation
such as shifting the integral paths.
In Section~\ref{section7} we show that the numerical estimates
of the spin-1 one-point functions obtained
through exact diagonalization of the
spin-1 XXZ Hamiltonian~(\ref{eq:spin1-XXZ-Hamiltonian})
are consistent with the analytical expressions
of the spin-1 one-point functions. Thus, we shall conclude that
the analytical result of the spin-1 one-point functions should be valid.

\section{The quantum group invariance}\label{section2}

We construct the basis vectors of the f\/inite-dimensional spin-$\ell/2$
representation of the quantum group $U_q(sl_2)$
in the tensor product space of the spin-1/2 representations,
and introduce their conjugate vectors. In terms of the basis
and conjugate basis vectors we formulate the spin-$\ell/2$
elementary matrices which have only one nonzero element 1 with respect to
entries of the basis and conjugate basis vectors.
We then illustrate an important formula for reducing the
spin-$\ell/2$ elementary matrices
into a sum of products of
the spin-1/2 elementary operators.

\subsection[Quantum group $U_q(sl_2)$]{Quantum group $\boldsymbol{U_q(sl_2)}$}\label{section2.1}

Let us introduce the quantum group $U_q(sl_2)$
in order to formulate not only the $R$-matrix
of the integrable spin-$s$ XXZ spin chain algebraically
but also the higher-spin elementary matrices,
by which we introduce correlation functions.
Here we remark that the correlation functions
of the spin-$s$ XXZ spin chains are given by
the expectation values of products of
the higher-spin elementary matrices at zero temperature.

The quantum algebra $U_q(sl_2)$
is an associative algebra over ${\bf C}$ generated by
$X^{\pm}$, $K^{\pm}$  with the following relations \cite{Jimbo-QG,Jimbo-Hecke,Drinfeld}:
\begin{gather*}
K K^{-1}   =    K^{-1} K = 1   , \qquad
K X^{\pm} K^{-1}  =  q^{\pm 2} X^{\pm}   , \qquad
{[} X^{+}, X^{-} {]}   =
{\frac   {K - K^{-1}}  {q- q^{-1}} }   .
\end{gather*}
The algebra $U_q(sl_2)$ is also a Hopf algebra over ${\bf C}$
with comultiplication
\begin{gather*}
\Delta (X^{+})   =   X^{+} \otimes 1 + K \otimes X^{+}   ,
 \qquad
\Delta (X^{-})  =  X^{-} \otimes K^{-1} + 1 \otimes X^{-}   ,  \qquad
\Delta(K)   =  K \otimes K    ,
\end{gather*}
and antipode:
$S(K)=K^{-1}$, $S(X^{+})= - K^{-1} X^{+}$, $S(X^{-}) = -  X^{-} K$, and
coproduct: $\epsilon(X^{\pm})=0$ and $\epsilon(K)=1$.

\subsection[Basis vectors of spin-$\ell/2$ representation of $U_q(sl_2)$]{Basis vectors of spin-$\boldsymbol{\ell/2}$ representation of $\boldsymbol{U_q(sl_2)}$}\label{section2.2}

We introduce the $q$-integer for an integer $n$ by
$[n]_q= (q^n-q^{-n})/(q-q^{-1})$.
We def\/ine the $q$-factorial  $[n]_q!$ for integers $n$ by
\begin{gather*}
[n]_q ! = [n]_q   [n-1]_q   \cdots   [1]_q  .
\end{gather*}
For integers $m$ and $n$ satisfying $m \ge n \ge 0$
we def\/ine the $q$-binomial coef\/f\/icients as follows
\begin{gather*}
\left[
\begin{matrix}
m \\
n
\end{matrix}
 \right]_q
= {\frac {[m]_q !} {[m-n]_q !   [n]_q !}}    .
\end{gather*}

Let us denote by $|\alpha \rangle$ for $\alpha=0, 1$,
the basis vectors of the spin-1/2 representation $V^{(1)}$.
Here we remark that
0 and 1 correspond to $\uparrow$ and $\downarrow$, respectively.
In the $\ell$th tensor product space $(V^{(1)})^{\otimes \ell}$
we construct the basis vectors of the $(\ell+1)$-dimensional
irreducible representation of $U_q(sl_2)$,
$|| \ell, n  \rangle$ for $n=0, 1, \ldots, \ell$,  as follows.
We def\/ine the highest weight vector $||\ell, 0 \rangle$ by
\begin{gather*}
||\ell , 0 \rangle = |0 \rangle_1 \otimes |0 \rangle_2 \otimes
\cdots \otimes |0 \rangle_\ell   .
\end{gather*}
Here  $|\alpha \rangle_j$ for $\alpha=0, 1$,
denote the basis vectors of the spin-1/2 representation def\/ined
on the $j$th position in the tensor product $(V^{(1)})^{\otimes \ell}$.
We def\/ine
$|| \ell, n \rangle$ for $n \ge 1$ and evaluate them as follows~\cite{DM1}
\begin{gather*}
|| \ell, n \rangle    =
\big( \Delta^{(\ell-1)} (X^{-}) \big)^n ||\ell, 0 \rangle
{\frac 1 {[n]_q!}}
 =
\sum_{1 \le i_1 < \cdots < i_n \le \ell}
\sigma_{i_1}^{-} \cdots \sigma_{i_n}^{-} | 0 \rangle
q^{i_1+ i_2 + \cdots + i_n  - n \ell + n(n-1)/2}   .
\end{gather*}
Here $\sigma_{j}^{-}$ denotes the Pauli spin operator $\sigma^{-}$ acting
on the $j$th component of the tensor product $(V^{(1)})^{\otimes \ell}$:
we have $\sigma_{j}^{-}=I^{\otimes (j-1)} \otimes \sigma^{-} \otimes
I^{\otimes (\ell-j)}$.
We def\/ine the conjugate vectors explicitly by the following:
\begin{gather*}
\langle \ell, n || =
\left[
\begin{matrix}
\ell \\
n
\end{matrix}
 \right]_q^{-1}   q^{n(\ell-n)}
\sum_{1 \le i_1 < \cdots < i_n \le \ell}
\langle 0 | \sigma_{i_1}^{+} \cdots \sigma_{i_n}^{+}
q^{i_1 + \cdots + i_n - n \ell + n(n-1)/2}    .
\end{gather*}
It is easy to show the normalization conditions \cite{DM1}:
$\langle \ell, n || \, || \ell, n \rangle = 1$.
Let us def\/ine $F(\ell, n)$ by
\begin{gather*}
F(\ell, n)=
\left[
\begin{matrix}
\ell \\
n
\end{matrix}
 \right]_q   q^{-n(\ell-n)}   .
\end{gather*}
We have $\left( || \ell, n \rangle \right)^{t}  || \ell, n \rangle
=F(\ell, n)$, and hence
$\langle \ell, n || =  \left( || \ell, n \rangle \right)^{t} /F(\ell, n)$.
Here the superscript $t$ denotes the matrix transposition.

In the massive regime where $q = \exp \eta$ with real $\eta$,
conjugate vectors $\langle \ell, n || $ are also
Hermitian conjugate to vectors
$|| \ell, n \rangle$.

\subsection[Affine quantum group $U_q(\widehat{sl_2})$]{Af\/f\/ine quantum group $\boldsymbol{U_q(\widehat{sl_2})}$}\label{section2.3}

In order to def\/ine the $R$-matrix in terms of algebraic relations
we now introduce the af\/f\/ine quantum group $U_q(\widehat{sl_2})$.
It is an inf\/inite-dimensional algebra generalizing
the quantum group $U_q(sl_2)$.

The algebra $U_q(\widehat{sl_2})$
is an associative algebra over ${\bf C}$ generated by
$X_i^{\pm}, K_i^{\pm}$ for $i=0,1$ with the following def\/ining relations:
\begin{gather*}
K_i K_i^{-1}   =   K^{-1}_i K_i = 1   , \qquad
K_i X_i^{\pm} K_i^{-1}  =  q^{\pm 2} X_i^{\pm}   ,  \qquad
K_i X_j^{\pm} K_i^{-1}  =  q^{\mp 2} X_j^{\pm}
\qquad i \ne j   ,   \\
 [  X_i^{+}, X_j^{-} {]}   =   \delta_{i,j}
{\frac   {K_i - K_i^{-1}}  {q- q^{-1}} }   ,      \\
(X_i^{\pm})^{3} X_j^{\pm}   -   [3]_q   (X_i^{\pm})^{2} X_j^{\pm} X_i^{\pm}
+ [3]_q   X_i^{\pm} X_j^{\pm} (X_i^{\pm})^2 -
 X_j^{\pm} (X_i^{\pm})^3 = 0, \qquad  i \ne j  .
\end{gather*}
The algebra $U_q(\widehat{sl_2})$ is also a Hopf algebra over ${\bf C}$
with comultiplication:
\begin{gather*}
\Delta (X_i^{+})   =   X_i^{+} \otimes 1 + K_i \otimes X_i^{+}    ,
\!\! \qquad
\Delta (X_i^{-})  =  X_i^{-} \otimes K_i^{-1} + 1 \otimes X_i^{-}   ,
\!\!\qquad
\Delta(K_i)   =   K_i \otimes K_i    ,
\end{gather*}
and antipode:
$S(K_i)=K_i^{-1}$, $S(X_i)= - K_i^{-1} X_i^{+}$,
$S(X_i^{-}) = - X_i^{-} K_i $, and counit:
$\varepsilon(X_i^{\pm})=0$ and $\varepsilon(K_i)=1$ for $i=0, 1$.

The quantum group $U_q(sl_2)$ gives a Hopf subalgebra of
$U_q(\widehat{sl_2})$ generated by
$X_i^{\pm}$, $K_i$ with either $i=0$ or $i=1$. Thus, the
af\/f\/ine quantum group generalizes the quantum group $U_q(sl_2)$.

\subsection{Evaluation representations with principal and homogeneous gradings}\label{section2.4}

We shall introduce two types of representations of $U_q(\widehat{sl_2})$: evaluation representations asso\-cia\-ted with principal grading and
that with homogeneous grading. The former is related to
the symmetric $R$-matrix which leads to  the most concise expression
of the integrable quantum spin
Hamiltonian, while the latter is
related to
the asymmetric $R$-matrix $R^{+}(u)$ which we shall def\/ine
in Section~\ref{section3.2} and suitable for an explicit construction of
representations of the quantum group.
Here and hereafter we denote by~$X^{\pm}$
and~$K$ the generators of~$U_q(sl_2)$.

Let us now introduce a representation of $U_q(\widehat{sl_2})$
associated with homogeneous grading~\cite{Jimbo-Miwa}.
 With a nonzero complex number $\lambda$ we def\/ine
a homomorphism of algebras  $\varphi_{\lambda}^{(p)}$:
$U_q(\widehat{sl_2}) \rightarrow U_q({sl_2})$, as follows.
\begin{gather}
\varphi^{(p)}_{\lambda}(X_0^{\pm})  =
e^{\pm \lambda}   X^{\mp }   ,\!\! \qquad
\varphi^{(p)}_{\lambda}(X_1^{\pm})  = e^{\pm \lambda}  X^{\pm}  ,\!\! \qquad
\varphi^{(p)}_{\lambda}(K_0) = K^{-1}   ,\!\!
\qquad \varphi^{(p)}_{\lambda}(K_1)  = K   .\!\!\!\!
\label{eq:eval-p}
\end{gather}
Thus, from a given f\/inite-dimensional representation
$(\pi^{(\ell)}, V^{(\ell)})$ of the quantum group~$U_q(sl_2)$,
we derive a representation of the af\/f\/ine quantum group
$U_q(\widehat{sl_2})$ by
$\pi^{(\ell)}(\varphi^{(p)}_{\lambda}(a))$ for~$a \in U_q(\widehat{sl_2})$,
where $\varphi^{(p)}_{\lambda}(\cdot)$ is given by~(\ref{eq:eval-p}).
We call it an evaluation representation of the af\/f\/ine quantum group;
more specif\/ically, the spin-$\ell/2$ evaluation representation
with evaluation parameter $\lambda$ associated
with principal grading.
We denote it by $(\pi_{\lambda}^{(\ell   p)}, V^{(\ell)}(\lambda))$ or
$V^{(\ell   p)}(\lambda)$.

Similarly in the case of principal grading,
we now introduce a representation associated with
homogeneous grading \cite{Jimbo-Miwa}.
With a nonzero complex number $\lambda$ we def\/ine
a homomorphism of algebras  $\varphi^{(+)}_{\lambda}$:
$U_q(\widehat{sl_2}) \rightarrow U_q({sl_2})$ by the following:
\begin{gather}
\varphi^{(+)}_{\lambda}(X_0^{\pm})  =
e^{\pm 2 \lambda}   X^{\mp }   , \!\qquad
\varphi^{(+)}_{\lambda}(X_1^{\pm})  =  X^{\pm}   , \!\qquad
\varphi^{(+)}_{\lambda}(K_0) = K^{-1}   ,
\!\qquad \varphi^{(+)}_{\lambda}(K_1)  = K  .\!\!\!\!
\label{eq:eval-h}
\end{gather}
 From a given f\/inite-dimensional representation $(\pi^{(\ell)}, V^{(\ell)})$
of the quantum group $U_q(sl_2)$
we derive a representation of the af\/f\/ine quantum group
$U_q(\widehat{sl_2})$ by
$\pi^{(\ell)}(\varphi^{(+)}_{\lambda}(a))$ for $a \in U_q(\widehat{sl_2})$,
where $\varphi^{(+)}_{\lambda}(\cdot)$ is given by (\ref{eq:eval-h}).
We call it the spin-$\ell/2$ evaluation representation
with evaluation parameter $\lambda$ associated
with homogeneous grading.
We denote it by $(\pi_{\lambda}^{(\ell  +)}, V^{(\ell)}(\lambda))$ or
$V^{(\ell  +)}(\lambda)$.

\subsection[Defining relations of the $R$-matrix]{Def\/ining relations of the $\boldsymbol{R}$-matrix}\label{section2.5}

Let us now def\/ine the $R$-matrix for any given pair of
f\/inite-dimensional representations
of the af\/f\/ine quantum group $U_q(\widehat{ sl_2})$.
Let $(\pi_1, V_1)$ and $(\pi_2, V_2)$ be f\/inite-dimensional
representations of~$U_q(\widehat{ sl_2})$.
We def\/ine the $R$-matrix $R_{12}$ for the tensor product $V_1 \otimes V_{2}$
by the following relations:
\begin{gather}
\pi_{1} \otimes \pi_{2} \left(\tau \circ \Delta(a) \right)
R_{12} =
R_{12}
\pi_{1} \otimes \pi_{2} \left(
\Delta(a) \right),  \qquad a \in U_q(\widehat{sl_2})   .
\label{eq:defR}
\end{gather}
Here $\tau$ denotes the permutation operator:
$\tau( a \otimes b) = b \otimes a$ for $a, b \in U_q(sl_2)$.

For an illustration, let us write down relations (\ref{eq:defR}) of
the $R$-matrices associated with eva\-lua\-tion representations.
We call them intertwining relations.
Associated with principal grading we have
for $a=X_0^{\pm}$, $X_1^{\pm}$ and $K_1$, respectively,
the following relations:
\begin{gather}
R_{12}^{(p)}(\lambda_1- \lambda_2)
\big( e^{\lambda_1} X^{-} \otimes 1 + e^{\lambda_2} K^{-1} \otimes X^{-}
\big)    =
\big( e^{\lambda_2} 1 \otimes X^{-} + e^{\lambda_1}  X^{-} \otimes K^{-1}
\big)R_{12}^{(p)}(\lambda_1- \lambda_2)   , \non \! \\
R_{12}^{(p)}(\lambda_1- \lambda_2)
\big( e^{-\lambda_1} X^{+} \otimes K + e^{-\lambda_2} 1 \otimes X^{+}
\big)   =
\big( e^{-\lambda_2} K \otimes X^{+}\! + e^{-\lambda_1}  X^{+} \otimes 1
\big)R_{12}^{(p)}(\lambda_1- \lambda_2)   ,  \non\! \\
R_{12}^{(p)}(\lambda_1- \lambda_2)
\big( e^{\lambda_1} X^{+} \otimes 1 + e^{\lambda_2} K \otimes X^{+}
\big)    =
\big( e^{\lambda_2} 1 \otimes X^{+} + e^{\lambda_1}  X^{+} \otimes K
\big) R_{12}^{(p)}(\lambda_1- \lambda_2)   ,  \non \\
R_{12}^{(p)}(\lambda_1- \lambda_2)
\big( e^{-\lambda_1} X^{-} \otimes K^{-1} + e^{-\lambda_2} 1 \otimes X^{-}
\big)   \nonumber\\
\qquad {} =
\big( e^{-\lambda_2} K^{-1} \otimes X^{-} + e^{-\lambda_1}  X^{-} \otimes 1
\big)R_{12}^{(p)}(\lambda_1- \lambda_2)   ,  \non \\
R_{12}^{(p)}(\lambda_1- \lambda_2) K \otimes K   =    K \otimes K
R_{12}^{(p)}(\lambda_1- \lambda_2)   .
\label{eq:int-p}
\end{gather}
Associated with homogeneous grading we have
\begin{gather}
R_{12}^{(+)}(\lambda_1- \lambda_2)
\big( e^{2 \lambda_1} X^{-} \otimes 1 + e^{2 \lambda_2} K^{-1} \otimes X^{-}
\big)   \nonumber\\
 \qquad {}=
\big( e^{2 \lambda_2} 1 \otimes X^{-} + e^{2 \lambda_1} X^{-} \otimes K^{-1}
\big)R_{12}^{(+)}(\lambda_1- \lambda_2)   ,  \non \\
R_{12}^{(+)}(\lambda_1- \lambda_2)
\big( e^{- 2 \lambda_1} X^{+} \otimes K + e^{-2 \lambda_2} 1 \otimes X^{+}
\big)  \nonumber\\
\qquad {}=
\big( e^{- 2 \lambda_2} K \otimes X^{+} + e^{- 2 \lambda_1}  X^{+} \otimes 1
\big)R_{12}^{(+)}(\lambda_1- \lambda_2)   ,  \non \\
R_{12}^{(+)}(\lambda_1- \lambda_2)
\big(  X^{+} \otimes 1 + K \otimes X^{+}
\big)    =
\big( 1 \otimes X^{+} + X^{+} \otimes K
\big) R_{12}^{(+)}(\lambda_1- \lambda_2)   ,  \non \\
R_{12}^{(+)}(\lambda_1- \lambda_2)
\big( X^{-} \otimes K^{-1} + 1 \otimes X^{-}
\big)    =
\big( K^{-1} \otimes X^{-} +   X^{-} \otimes 1
\big)R_{12}^{(+)}(\lambda_1- \lambda_2)   ,  \non \\
R_{12}^{(+)}(\lambda_1- \lambda_2) K \otimes K   =   K \otimes K
R_{12}^{(+)}(\lambda_1- \lambda_2)   .
\label{eq:int-h}
\end{gather}
Here $\lambda_1$ and $\lambda_2$ correspond to the ``string centers'' of
the sets of the evaluation parameters associated with the evaluation
representations $\pi_1$ and $\pi_2$.
We have $\lambda_1= \xi_1 - (\ell -1) \eta/2$,
if $\pi_1$ is given by the spin-$\ell/2$ evaluation representation
derived from the tensor product $(V^{(1)})^{\otimes \ell}$ with complete
$\ell$-string $w_j^{(\ell)}$ for $j=1, 2, \ldots, \ell$. Here we shall
def\/ine complete strings in Section~\ref{section3.6}.

We can show that the solution of intertwining relations~(\ref{eq:defR})
is unique. We may therefore def\/ine the $R$-matrix
in terms of relations~(\ref{eq:defR}).

We remark that relations (\ref{eq:int-p}) for the evaluation representation
associated with principal grading
are mapped into those of (\ref{eq:int-h}) associated with homogeneous grading
through a similarity transformation, which we call the gauge transformation.
We shall formulate it in Section~\ref{section3.4}.

\subsection{Conjugate vectors and Hermitian elementary matrices}\label{section2.6}

In order to construct Hermitian elementary matrices
in the massless regime where $q$ is complex and $|q|=1$,
we now introduce another set of dual basis vectors \cite{DM2}.
For a given nonzero integer~$\ell$ we def\/ine
$\widetilde{\langle \ell, n ||}$ for $n=0, 1, \ldots, n$, by
\begin{gather*}
\widetilde{\langle \ell, n ||} =
\left(
\begin{matrix}
\ell \\
n
\end{matrix}
 \right)^{-1}
\sum_{1 \le i_1 < \cdots < i_n \le \ell}
\langle 0 | \sigma_{i_1}^{+} \cdots \sigma_{i_n}^{+}
q^{-(i_1 + \cdots + i_n) + n \ell - n(n-1)/2}     .
\end{gather*}
They are conjugate to $|| \ell, n \rangle$:
$\widetilde{\langle \ell, m ||}   || \ell, n \rangle =
\delta_{m, n} $.
Here we have denoted the binomial coef\/f\/i\-cients
for integers $\ell$ and $n$ with $0 \le n \le \ell$ as follows
\begin{gather*}
\left(
\begin{matrix}
\ell \\
n
\end{matrix}
 \right)
= {\frac {\ell !} {(\ell-n)! n!}}  .
\end{gather*}

We now introduce vectors $\widetilde{|| \ell, n \rangle}$
which are Hermitian conjugate to $\langle \ell, n ||$ when
$|q|=1$
for positive integers $\ell$ with $n=0, 1, \ldots, \ell$.
Setting the norm of $\widetilde{|| \ell, n \rangle}$
such that
$\langle \ell, n ||    \widetilde{|| \ell, n \rangle}=1$,
vectors $\widetilde{|| \ell, n \rangle}$ are given by
\begin{gather*}
\widetilde{|| \ell, n \rangle} =
\sum_{1 \le i_1 < \cdots < i_n \le \ell} \sigma_{i_1}^{-}
\cdots \sigma_{i_n}^{-} | 0 \rangle
q^{-(i_1 + \cdots + i_n) + n \ell - n(n-1)/2}
\left[
\begin{matrix}
\ell \\
n
\end{matrix}
 \right]_q
q^{-n(\ell-n)}
\left(
\begin{matrix}
\ell \\
n
\end{matrix}
 \right)^{-1}   .
\end{gather*}

We have the following normalization conditions:
\begin{gather*}
\widetilde{\langle \ell, n ||}   \widetilde{|| \ell, n \rangle}
= \left[
\begin{matrix}
\ell \\
n
\end{matrix}
 \right]_q^2
\left(
\begin{matrix}
\ell \\
n
\end{matrix}
 \right)^{-2} \qquad \mbox{for} \quad
n=0, 1, \ldots, \ell.
\end{gather*}

In the massless regime where $q$ is complex with $|q|=1$,
we def\/ine elementary matrices
${\widetilde E}^{m,   n   (\ell   +)}$ by
\begin{gather*}
{\widetilde E}^{m,   n   (\ell   + )}
= \widetilde{||\ell, m \rangle}   \langle \ell, n || \qquad
\mbox{for} \quad m, n=0, 1, \ldots, \ell.
\end{gather*}

In the massless regime matrix
$|| \ell, n \rangle \widetilde{\langle \ell, n ||}$ is
Hermitian: $(|| \ell, n \rangle  \widetilde{\langle \ell, n ||})^{\dagger}
= || \ell, n \rangle  \widetilde{\langle \ell, n ||}$.
However, in order to def\/ine projection operators $\tilde{P}$ such that
$P \tilde{P} = P$, we have formulated vectors
 $\widetilde{|| \ell, n \rangle}$.

Associated with principal grading we def\/ine the spin-$\ell/2$ symmetric
elementary matrices ${\widetilde E}^{i,   j   (\ell   p)}$ by \cite{DM4,DM5}
\begin{gather*}
{\widetilde E}^{i,   j   (\ell   p)}
= \widetilde{ || \ell, i \rangle} \langle \ell, j ||
\sqrt{\frac {F(\ell, j)} {F(\ell, i)}}   \qquad
\mbox{for} \quad i, j=0, 1, \ldots, \ell.
\end{gather*}

\subsection{Projection operators}\label{section2.7}

We def\/ine the projection operator acting on from the 1st to
the $\ell$th tensor-product spaces by
\begin{equation}
P^{(\ell)}_{1 2 \cdots \ell}
= \sum_{n=0}^{\ell}  || \ell, n \rangle   \langle \ell, n ||   .
\label{eq:Psum}
\end{equation}
We introduce another projection operator
$\widetilde{P}_{1 2 \cdots \ell}^{(\ell)}$
as follows
\begin{equation}
\widetilde{ P}_{1 2 \cdots \ell}^{(\ell)} = \sum_{n=0}^{\ell}
\widetilde{ || \ell ,    n \rangle} \langle \ell ,   n ||    .
\label{eq:P'sum}
\end{equation}
The projector $\widetilde{ P}_{1 2 \cdots \ell}^{(\ell)}$ is idempotent:
$\big(\widetilde{ P}_{1 2 \cdots \ell}^{(\ell)}\big)^2=
\widetilde{ P}_{1 2 \cdots \ell}^{(\ell)}$.
In the massless regime where $q$ is complex with $|q|=1$,
it is Hermitian:
$\big( \widetilde{ P}_{1 2 \cdots \ell}^{(\ell)} \big)^{\dagger}=
\widetilde{ P}_{1 2 \cdots \ell}^{(\ell)}$.
From (\ref{eq:Psum}) and (\ref{eq:P'sum}),
we show the following properties:
\begin{gather}
P_{1 2 \cdots \ell}^{(\ell)}
\widetilde{ P}_{1 2 \cdots \ell}^{(\ell)}
  =   P_{1 2 \cdots \ell}^{(\ell)}   ,
\label{eq:PP'=P} \\
\widetilde{ P}_{1 \cdots \ell}^{(\ell)}
P_{1 2 \cdots \ell}^{(\ell)}
  =   \widetilde{ P}_{1 2 \cdots \ell}^{(\ell)}   .
\label{eq:P'P=P'}
\end{gather}

\subsection[Spin-$s$ elementary matrices
in terms of the spin-1/2 elementary matrices]{Spin-$\boldsymbol{s}$ elementary matrices
in terms of the spin-1/2 elementary matrices}\label{section2.8}

Let us denote by $e^{a, b}$ such 2-by-2 matrices
that have only one nonzero matrix element 1 at the entry
$(a, b)$ for $a, b= 0, 1$. We call them the spin-1/2 elementary matrices.
We denote by $e^{a, b}_j$ the elementary matrices $e^{a, b}$
acting on the $j$th component of the tensor product
$(V^{(1)})^{\otimes \ell}$.

Let us introduce variables
$\varepsilon_{\alpha}'$ and $\varepsilon_{\beta}$
which take only two values~0 and~1 for $\alpha, \beta=1, 2, \ldots, \ell$.
 We def\/ine diagonal two-by-two matrices $\Phi_j$
 by $\Phi_j={\rm diag}(1, \exp(w_j))$ acting on $V_j^{(1)}$
for $j=0, 1, \ldots, L$. Here $w_j$ ($1 \le j \le L$)
are called the inhomogeneous parameters of the spin-1/2 XXZ spin chain,
and we set $w_0= \lambda_0$ (see also Section~\ref{section3.3}).
We def\/ine the gauge transformation by
a similarity transformation with respect to
the matrix $\chi_{01 \cdots L}= \Phi_0 \Phi_1 \cdots\Phi_L$.
Here, we put inhomogeneous parameters $w_j$ with the complete $\ell$-strings
such as $w_{\ell(k-1)+j}=w_{\ell(k-1)+j}^{(\ell)}= \xi_k - (j-1) \eta$ for
$j=1, 2, \ldots, \ell$ and $k=1, 2, \ldots, N_s$.
Then, we can show the following relation.
\begin{proposition}[\cite{DM4,DM5}]
The spin-$\ell/2$ symmetric elementary matrices associated
with principal grading
are decomposed into a sum of products of the spin-$1/2$ elementary matrices
as follows
\begin{gather}
{\widetilde E}^{i ,   j   (\ell   p)} =
\left(
\left[ \begin{matrix}
\ell \\
i
\end{matrix}
\right]_q
\left[ \begin{matrix}
\ell \\
j
\end{matrix}
\right]_q^{-1}   \right)^{1/2}
e^{-(i-j)(\xi_1-(\ell-1)\eta/2)}
{\widetilde P}_{1 2 \cdots \ell}^{(\ell)}
\sum_{\{ \varepsilon_{\beta} \}} \chi_{1 2 \cdots \ell}
e_1^{\varepsilon_1', \varepsilon_1} \cdots
e_{\ell}^{\varepsilon_{\ell}', \varepsilon_{\ell}}
\chi_{1 2 \cdots \ell}^{-1}  . \label{eq:tildeE(p)-e}
\end{gather}
Here the sum is taken over all sets of $\varepsilon_{\beta}$s
such that the number of integers $\beta$ satisfying $\varepsilon_{\beta}=1$
for $1 \le \beta \le \ell$ is equal to~$j$.
We take a set of $\varepsilon_{\alpha}'$s such that
the number of integers $\alpha$ satisfying $\varepsilon_{\alpha}'=1$
for $1 \le \alpha \le \ell$ is equal to~$i$.
The expression \eqref{eq:tildeE(p)-e} is independent of the order of
$\varepsilon_{\alpha}'$s with respect to $\alpha$.
\label{prop:E-principal}
\end{proposition}

The formula (\ref{eq:tildeE(p)-e}) plays a central role in the revised
derivation of the spin-$\ell/2$ form factors and the
spin-$\ell/2$ XXZ correlation functions \cite{DM4,DM5}.
We shall derive  (\ref{eq:tildeE(p)-e}) in Appendix~\ref{appendixA}.
We recall that the derivation of the multiple-integral representations
of the integrable spin-$s$ XXZ spin chain given in the previous papers
\cite{DM1,DM2,DM3} was not completely correct \cite{DM4,DM5}.
In fact,
the transfer matrix becomes non-regular at $\lambda=\xi_k$ \cite{DM5},
and hence the straightforward application of the QISP formula was not valid.

\subsection{Example: spin-1 case}\label{section2.9}

We shall show reduction formula (\ref{eq:tildeE(p)-e}) for the spin-1 case.

The spin-1 basis vectors $|| 2, n  \rangle$ ($n=0, 1, 2$)
 are given by \cite{DM1}
\begin{alignat*}{3}
& ||2, 0 \rangle   =   |+ + \rangle   , \qquad &&
\langle 2, 0 ||  =   \langle + + |   , &
\non \\
& ||2, 1 \rangle   =   |+ - \rangle + q^{-1} | - + \rangle   ,   \qquad &&
\langle 2, 1 || =  {\frac q {[2]_q} } \left( \langle + - | + q^{-1} \langle - + | \right)   ,&
\non \\
& ||2, 2 \rangle  =   |- - \rangle   , \qquad &&
 \langle 2, 2 || =   \langle - - |   . &
\end{alignat*}
Here $| + - \rangle $ denotes $|0 \rangle_1 \otimes | 1 \rangle_2$,
brief\/ly.
The conjugate vectors $\widetilde{|| 2, n \rangle}$ ($n=0, 1, 2$)
are given by
\begin{alignat*}{3}
& \widetilde{||2, 0 \rangle}   =   |+ + \rangle   ,
\qquad &&
 \widetilde{\langle 2, 0 ||} =  \langle + + |   , &  \\
& \widetilde{||2, 1 \rangle}   =   \left( |+ -  \rangle
+ q | - + \rangle \right) {\frac {[2]_q} {2q}}   ,
 \qquad &&
\widetilde{\langle 2, 1 ||} =  {\frac 1 2 }
\left( \langle + - | + q^{-1} \langle - + | \right)   ,
&  \\
& \widetilde{||2, 2 \rangle}   =   |- - \rangle , \qquad &&
\widetilde{\langle 2, 2 ||} = \langle - - |   . &
\end{alignat*}

Let us derive the projection operator $\widetilde{ P}^{(2)}_{12}$.
Explicitly we have
\begin{gather}
\widetilde{|| 2, 1 \rangle} \langle 2, 1 ||   =
\left( |+ -  \rangle
+ q | - + \rangle \right) {\frac {[2]_q} {2q}}
  \cdot
{\frac q {[2]_q} } \left( \langle + - | + q^{-1} \langle - + | \right)
\non \\
\phantom{\widetilde{|| 2, 1 \rangle} \langle 2, 1 ||}{} =  {\frac {1} {2}}\left(
 |+ -  \rangle  \langle + - | \, +  q^{-1} |+ -  \rangle  \langle - + |
 \,  + q | - + \rangle  \langle + - |
 + |- + \rangle  \langle - + |
 \right) \non \\
\phantom{\widetilde{|| 2, 1 \rangle} \langle 2, 1 ||}{}
  =   {\frac {1} {2}} \big(
 e_1^{0,   0} e_2^{1,   1}  +  q^{-1}    e_1^{0,   1} e_2^{1,   0}
+  q    e_1^{1,   0} e_2^{0,   1} +  e_1^{1,  1} e_2^{0,   0}
\big)    .
\label{eq:tilE11p}
\end{gather}
Here we remark that in the massless regime where $q$ is complex with $|q|=1$,
operator $\widetilde{||2, 1 \rangle} \langle 2, 1 ||$ is Hermitian
while $||2, 1 \rangle \langle 2, 1 ||$ is not.
As a four-by-four matrix we express $\widetilde{P}^{(2)}_{12}$ by
\begin{gather}
\widetilde{P}^{(2)}_{12}   =
||2, 0 \rangle \langle 2, 0 || +
||2, 1 \rangle \langle 2, 1 || +
||2, 2 \rangle \langle 2, 2 ||
 =  \left(
\begin{array}{cccc}
1 & 0 & 0 & 0 \\
0 & 1/2 & q^{-1}/2 & 0 \\
1 & q/2 & 1/2 & 0 \\
0 & 0 & 0 & 1
\end{array}
\right)_{[1, 2]}   . \label{eq:tilde-P}
\end{gather}
Here the symbol $[1, 2]$ at the bottom of the $4 \times 4$ matrix
of (\ref{eq:tilde-P}) denotes that the matrix acts on the
tensor product space $V_1^{(1)} \otimes V_2^{(1)}$.
We note that operator
$|+ -  \rangle  \langle - + |$
corresponds to $e_1^{0 ,  1} e_2^{1 ,  0}$ in (\ref{eq:tilE11p}),
which gives the entry of (1,2) in the four-by-four matrix of (\ref{eq:tilde-P}); i.e., the element in the 2nd row and 3rd column.

For an illustration, let us show reduction formula (\ref{eq:tildeE(p)-e})
for the spin-1 case.
With $\varepsilon_1'=0$ and $\varepsilon_2'=1$,
reduction formula (\ref{eq:tildeE(p)-e})
for $i=j=1$ reads
\begin{gather}
\widetilde{E}^{1 ,  1   (2   p)} = \widetilde{|| 2, 1 \rangle}
\langle 2, 1 ||
  =   \widetilde{P}^{(2)} \chi_{12} \big(
e_1^{0,   0} e_2^{1,   1} + e_1^{0,   1} e_2^{1,  0} \big)
\chi_{12}^{-1}   . \label{eq:reduction-spin-1}
\end{gather}
First, it is straightforward to show
\begin{gather*}
\chi_{12}  e_1^{0,   0} e_2^{1,   1} \chi_{12}^{-1}
= e_1^{0,   0} e_2^{1,   1}   , \qquad
\chi_{12}  e_1^{0,   1} e_2^{1,   0} \chi_{12}^{-1}
= q^{-1} e_1^{0,   1} e_2^{1,  0}.
\end{gather*}
Then, in terms of the four-by-four matrix notation we have{\samepage
\begin{gather}
e_1^{0,   0} e_2^{1,   1} +  q^{-1} e_1^{0,   1} e_2^{1,   0} =
\left(
\begin{array}{cccc}
0 & 0 & 0 & 0 \\
0 & 1 & q^{-1} & 0 \\
0 & 0 & 0 & 0 \\
0 & 0 & 0 & 0 \\
\end{array}
\right)_{[1,2]}   . \label{eq:ee}
\end{gather}
Here $e_1^{0,   0} e_2^{1,   1}$ corresponds to the element in the
2nd row and 2nd column of the $4 \times 4$ matrix~(\ref{eq:ee}).}

Multiplying (\ref{eq:tilde-P}) by (\ref{eq:ee})
and making use of (\ref{eq:tilE11p}),
we have the following relation:
\begin{gather*}
\widetilde{P}^{(2)}_{12} \big(
e_1^{0,   0} e_2^{1,   1} +  q^{-1} e_1^{0,   1} e_2^{1,   0} \big)
= \widetilde{|| 2, 1 \rangle } \langle 2, 1 || .
\end{gather*}
We have thus conf\/irmed
reduction formula~(\ref{eq:tildeE(p)-e}) for
$\ell=2$ and $i_1=j_1=1$, as shown in~(\ref{eq:reduction-spin-1}).

\section{Fusion transfer matrices and higher-spin expectation values}\label{section3}

We construct the monodromy matrices of the
integrable higher-spin XXZ spin chains through the fusion method.
We then evaluate the form factor of a given product of
the higher-spin operators by reducing them into
a sum of products of the spin-1/2 operators and calculate
their scalar products of the spin-1/2 operators through Slavnov's formula.
When we reduce the higher-spin operators, we make use of the
fusion construction where all the elements are constructed from
a sum of products of the spin-1/2 operators multiplied by the projection
operators.

\subsection{Tensor product notation}\label{section3.1}

Let $s$ be an integer or a half-integer.  We shall mainly
consider the tensor product
$V_1^{(2s)} \otimes \cdots \otimes V_{N_s}^{(2s)}$
of $(2s+1)$-dimensional vector spaces $V^{(2s)}_j$
with $L= 2s N_s$.
Here $V_{j}^{(2s)}$ have spectral parameters
$\lambda_j$ for $j=1, 2, \ldots, N_s$.
We denote by $E^{a, \, b}$ a unit matrix that has only one nonzero element
equal to 1 at entry $(a, b)$ where $a, b= 0, 1, \ldots, 2s$.
For a given set of matrix elements
${\cal A}^{a, \, \alpha}_{b, \, \beta}$  for $a,b=0, 1, \ldots, 2s$ and
$\alpha, \beta=0, 1, \ldots, 2s$,
we def\/ine operators $A_{j, k}$ for $1 \le j < k \le N_s$ by
\begin{gather}
{\cal A}_{j, k}   =   \sum_{a,b=1}^{2s} \sum_{\alpha, \beta}
{\cal A}^{a,   \alpha}_{b,   \beta} I_0^{(2s_0)} \otimes
I_1^{(2s)} \otimes \cdots \otimes I_{j-1}^{(2s)}  \non \\
\phantom{{\cal A}_{j, k}   =}{}  \otimes E^{a, b}_j \otimes
I_{j+1}^{(2s)}  \otimes \cdots \otimes I_{k-1}^{(2s)}
\otimes E_k^{\alpha, \beta}  \otimes I_{k+1}^{(2s)}
 \otimes \cdots \otimes I_{r}^{(2s)}   .
\label{defAjk}
\end{gather}

In the tensor product space, $(V^{(2s)})^{\otimes N_s}$,
we def\/ine $\widetilde{E}^{m,   n   (2s   w)}_i$
for $i=1, 2, \ldots, N_s$ and $w=+, p$ by
\begin{gather*}
\widetilde{E}^{m,   n   (2s   w)}_i = \big(I^{(2s)}\big)^{\otimes (i-1)} \otimes
\widetilde{E}^{m,   n   (2s   w)}  \otimes \big(I^{(2s)}\big)^{\otimes (N_s-i)}  .
\end{gather*}
The elementary matrices
$\widetilde{E}^{n, n  (2s  w)}$
for $n=0, 1, \ldots, 2s$ and $w=+, p$,
are Hermitian in the massless regime.

\subsection[Asymmetric and symmetric $R$-matrices]{Asymmetric and symmetric $\boldsymbol{R}$-matrices}\label{section3.2}

Let us introduce the $R$-matrix of the XXZ spin chain~\cite{Korepin,MS2000,KMT1999,KMT2000}.
Let $V_1$ and $V_2$ be two-dimensional vector spaces.
We def\/ine the $R$-matrix $R_{12}^{+}$ acting on $V_1 \otimes V_2$  by
\begin{gather}
{R}_{12}^{+}(\lambda_1-\lambda_2) = \sum_{a,b,c,d=0,1}
R^{+}(u)^{a   b}_{c   d}
  e_1^{a,   c} \otimes e_2^{b,   d} =
\left(
\begin{array}{cccc}
1 & 0 & 0 & 0 \\
0 & b(u) & c^{-}(u) & 0 \\
0 & c^{+}(u) & b(u) & 0 \\
0 & 0 & 0 & 1 \\
\end{array}
\right)_{[1,2]}  , \label{eq:R+}
\end{gather}
where $u=\lambda_1-\lambda_2$,
$b(u) = \sinh u/\sinh(u + \eta)$ and
$c^{\pm}(u) = \exp( \pm u) \sinh \eta/\sinh(u + \eta)$.

We remark that the $R^{+}(\lambda_1-\lambda_2)$
is compatible with the homogeneous grading
of $U_q(\widehat{sl_2})$.
In fact, it is straightforward to see
that the asymmetric $R$-matrix satisf\/ies the intertwining relations
associated with homogeneous grading (\ref{eq:int-h})
for the tensor product of the spin-1/2 representations of $U_q(sl_2)$,
$V_1^{(1)} \otimes V_1^{(1)}$.

We denote by $R^{(p)}(u)$ or simply by $R(u)$ the symmetric $R$-matrix
where $c^{\pm}(u)$ of (\ref{eq:R+}) are replaced by
$c(u)= \sinh \eta/\sinh(u+\eta)$ \cite{DM1}. The symmetric $R$-matrix
is compatible with evaluation representation associated
with principal grading for the af\/f\/ine quantum group
$U_q(\widehat{sl_2})$~\cite{DM1}.
Hereafter we express $R^{+}$
and $R^{(p)}$ by $R^{(1   w)}$ with $w=+$ and $p$, respectively.

\subsection[Monodromy matrix of type $(1,1^{\otimes L})$]{Monodromy matrix of type $\boldsymbol{(1,1^{\otimes L})}$}\label{section3.3}

We now consider the $(L+1)$th tensor product of the spin-1/2 representations,
which consists of the tensor product of auxiliary space $V_0^{(1)}$ and
the $L$th tensor product of quantum spaces $V_j^{(1)}$
for $j=1, 2, \ldots, L$, i.e.\
$V_0^{(1)} \otimes \big( V_1^{(1)} \otimes \cdots \otimes V_L^{(1)} \big)$.
We call it the tensor product of type $(1,1^{\otimes L})$
and denote it by the following symbol:
\begin{gather*}
(1,1^{\otimes L}) = (1, \overbrace{1, 1, \ldots, 1}^{L})   .
\end{gather*}

Applying def\/inition~(\ref{defAjk})
for matrix elements $R(u)^{ab}_{cd}$ of a given $R$-matrix
such as $R^{(1   w)}$ with $w=+$ and $p$,
we def\/ine  $R$-matrices
$R_{j k}(\lambda_j, \lambda_k)=R_{j k}(\lambda_j-\lambda_k)$
 for integers $j$ and $k$ with $0\le j < k \le L$.
For integers $j$, $k$ and $\ell$ with $0 \le j < k < \ell \le L$,
the $R$-matrices satisfy the Yang--Baxter equations
\begin{gather*}
 R_{j k}(\lambda_j-\lambda_k)
R_{j \ell}(\lambda_j-\lambda_{\ell})
R_{k \ell}(\lambda_k-\lambda_{\ell})
=
 R_{k \ell}(\lambda_k-\lambda_{\ell})
R_{j \ell}(\lambda_j-\lambda_{\ell})
R_{j k}(\lambda_j-\lambda_k)   . 
\end{gather*}

We def\/ine the monodromy matrix of type $(1,1^{\otimes L})$
associated with homogeneous grading by
\begin{gather*}
T^{(1,   1   +)}_{0, 1 2 \cdots L}(\lambda_0; w_1, w_2, \ldots, w_L)
= R_{0L}^{+}(\lambda_0-w_L)
\cdots R_{02}^{+}(\lambda_0-w_2) R_{01}^{+}(\lambda_0-w_1)   .
\end{gather*}
Here we have set $\lambda_j=w_j$ for $j=1, 2, \ldots, L$,
where $w_j$ are arbitrary parameters.
We call them inhomogeneous parameters. We have expressed the symbol of
type $(1,1^{\otimes L})$ as $(1,   1)$ in superscript.
The symbol $(1,   1   +)$ denotes that it is consistent with
homogeneous grading.
We express operator-valued matrix elements of
the monodromy matrix as follows
\begin{gather*}
T^{(1, 1  +)}_{0, 1 2 \cdots L}(\lambda; \{ w_j \}_L ) =
\left(
\begin{array}{cc}
A^{(1 +)}_{1 2 \cdots L}(\lambda; \{ w_j \}_L) &
B^{(1 +)}_{1 2 \cdots L}(\lambda; \{ w_j \}_L) \\
C^{(1 +)}_{1 2 \cdots L}(\lambda; \{ w_j \}_L) &
D^{(1 +)}_{1 2 \cdots L}(\lambda; \{ w_j \}_L)
\end{array}
\right)   .
\end{gather*}
Here $\{ w_j \}_L$ denotes the set of $L$ parameters,
$w_1, w_2, \ldots, w_L$.
We also denote the matrix elements of
the monodromy matrix by
$[T^{(1, 1 +)}_{0, 1 2 \cdots L}(\lambda; \{ w_j \}_L )]_{a,b}$
for $a,b=0,1$.

\subsection{Gauge transformations}\label{section3.4}

We derive the monodromy matrix consistent with principal
grading, $T^{(1,   1   p)}_{0, 1 2 \cdots L}(\lambda; \{ w_j \}_L )$,
from that of homogeneous grading via similarity transformation
$\chi_{0 1 \cdots L}$  as follows \cite{DM1}
\begin{gather*}
  T^{(1, 1   +)}_{0, 1 2 \cdots L}(\lambda; \{ w_j \}_L )  =
\chi_{0 1 2 \cdots L}
T^{(1, 1 \, p)}_{0, 1 2 \cdots L}(\lambda; \{ w_j \}_L )
\chi_{0 1 2 \cdots L}^{-1}
\non \\
\qquad{} =
\left(
\begin{array}{cc}
\chi_{1 2 \cdots L}
A^{(1 \,p)}_{1 2 \cdots L}(\lambda; \{ w_j \}_L)
\chi_{1 2 \cdots L}^{-1} &
e^{- \lambda_0} \chi_{1 2 \cdots L}
B^{(1   p)}_{1 2 \cdots L}(\lambda; \{ w_j \}_L)
\chi_{1 2 \cdots L}^{-1}  \\
e^{\lambda_0}
\chi_{1 2 \cdots L}
C^{(1 \, p)}_{1 2 \cdots L}(\lambda; \{ w_j \}_L)
\chi_{1 2 \cdots L}^{-1}  &
\chi_{1 2 \cdots L} D^{(1   p)}_{1 2 \cdots L}(\lambda; \{ w_j \}_L)
\chi_{1 2 \cdots L}^{-1}
\end{array}
\right)  . 
\end{gather*}
Here we recall that
 $\chi_{01 \cdots L}= \Phi_0 \Phi_1 \cdots\Phi_L$ and
$\Phi_j$ are given by  diagonal two-by-two matrices
$\Phi_j={\rm diag}(1, \exp(w_j))$ acting on $V_j^{(1)}$
for $j=0, 1, \ldots, L$, and we set $w_0= \lambda_0$.
In~\cite{DM1} operator
$A^{(1   +)}(\lambda)$ has been written as  $A^{+}(\lambda)$.

We now introduce the gauge transformation for the spin-$s$ representation~\cite{DM5}. We def\/ine diagonal matrix $\Phi^{(2s)}(w)$ on
the basis vectors $|| 2s, n \rangle$ as follows:
\begin{gather*}
\Phi^{(2s)}(w) || 2s, n \rangle =
\exp(n w)   || 2s, n \rangle     \qquad \mbox{for} \quad n= 0, 1, \ldots,
2s.
\end{gather*}
We denote by $\Phi_j^{(2s)}(w)$ the matrix $\Phi^{(2s)}(w)$ def\/ined on the
$j$th component of the tensor product $V_1^{(2s)} \otimes \cdots \otimes
V_{N_s}^{(2s)}$. We def\/ine $\chi^{(2s)}_{1 2 \cdots N_s}$ acting on
the quantum space $V_1^{(2s)} \otimes \cdots \otimes
V_{N_s}^{(2s)}$ by
\begin{gather*}
\chi_{1 2 \cdots N_s}^{(2s)} = \Phi_1^{(2s)}(\Lambda_1) \cdots
\Phi_{N_s}^{(2s)}(\Lambda_{N_s})   .
\end{gather*}
We express $\Lambda_b$ as $\Lambda_b = \xi_b - (2s-1) \eta/2$ for
$b=1, 2, \ldots, N_s$. Here $\xi_b$ denote the inhomogeneous parameters of the
spin-$s$ XXZ spin chains, which will be given in equation~(\ref{eq:ell-strings})
of Section~\ref{section3.6}. We note that $\Lambda_b$ corresponds to the string center of the $2s$-string, $\xi_b - (\beta-1) \eta$ with $\beta=1, 2, \ldots, 2s$, for each $b$
satisfying $1 \le b \le N_s$.

\subsection{Projection operators through fusion}\label{section3.5}

Let  $V_1$ and $V_2$ be the $(2s+1)$-dimensional vector spaces.
We def\/ine permutation operator $\Pi_{1,  2}$ by
\begin{gather*}
\Pi_{1,   2}   v_1 \otimes v_2 =
v_2 \otimes v_1   , \qquad v_1 \in V_1   , \quad v_2 \in V_2   .
\end{gather*}
In the case of spin-1/2 representations,
we def\/ine operator ${\check R}_{12}^{+}(\lambda_1-\lambda_2)$ by
\begin{gather*}
{\check  R}_{12}^{+}(\lambda_1-\lambda_2)
= \Pi_{1,   2}    R_{12}^{+}(\lambda_1 - \lambda_2)   .
\end{gather*}

We now introduce projection operators $P_{12\cdots \ell}^{(\ell)}$ for
$\ell \ge 2$.
We def\/ine  $P_{12}^{(2)}$ by $P_{12}^{(2)} = {\check R}_{1, 2}^{+}(\eta)$.
 For $\ell > 2$ we def\/ine projection operators
 inductively with respect to $\ell$ as follows
\cite{Jimbo-Hecke,V-DWA}
\begin{gather}
P_{1 2 \cdots \ell}^{(\ell)} =
P_{1 2 \cdots \ell-1}^{(\ell-1)} {\check R}^{+}_{\ell-1,  \ell}
((\ell-1)\eta) P_{12\cdots \ell-1}^{(\ell-1)}  .
\label{eq:def-projector}
\end{gather}
The projection operator $P_{12\cdots \ell}^{(\ell)}$
gives a $q$-analogue of the full symmetrizer
of the Young operators for the Hecke algebra~\cite{Jimbo-Hecke}.

Applying projection operator $P_{a_1 a_2 \cdots a_{\ell}}^{(\ell)}$
to the vectors in the tensor product $V_{a_1}^{(1)} \otimes V_{a_2}^{(1)}
\otimes \cdots \otimes V_{a_{\ell}}^{(1)}$, we can construct
the $(\ell+1)$-dimensional vector space
$V_{a_1 a_2 \cdots a_{\ell}}^{(\ell)}$ associated with
the spin-$\ell/2$ representation of $U_q(sl_2)$.
For instance, we have
$P_{a_1 a_2}^{(2)} |+ - \rangle_{a} = (q/[2]_q)||2, 1 \rangle_{a}$.
Here we have introduced
$| + - \rangle_{a} = |0 \rangle_{a_1} \otimes |1 \rangle_{a_2}$.
We denote $V_{a_1 a_2 \cdots a_{\ell}}^{(\ell)}$
also by $V_{a}^{(\ell)}$ or $V_{0}^{(\ell)}$ for short.
Similarly, we denote
$P_{a_1 a_2 \cdots a_{\ell}}^{(\ell)}$ by
$P_{a_1}^{(\ell)}$ for short.

Let us consider the tensor product
$V^{(2s)}_1 \otimes \cdots \otimes V_{N_s}^{(2s)}$,
which gives the quantum space for the higher-spin transfer matrices.
We construct the $b$th component $V_b^{(2s)}$
of the quantum space
from the $2s$th tensor product of the spin-1/2 representations:
$V_{2s(b-1)+1}^{(1)} \otimes \cdots \otimes V_{2s(b-1)+2s}^{(1)}$,
 for $b=1, 2, \ldots, N_s$.
We therefore def\/ine ${P}_{12 \cdots L}^{(2s)}$ and
$\widetilde{P}_{12 \cdots L}^{(2s)}$ by
\begin{gather*}
{P}_{12 \cdots L}^{(2s)}
= \prod_{i=1}^{N_s} {P}^{(2s)}_{2s(i-1)+1}   , \qquad
\widetilde{P}_{12 \cdots L}^{(2s)}
= \prod_{i=1}^{N_s} \widetilde{P}^{(2s)}_{2s(i-1)+1}    .
\end{gather*}
Here we recall $L = 2s N_s $.

\subsection[Higher-spin monodromy matrix of type $(\ell, (2s)^{\otimes N_s})$]{Higher-spin monodromy matrix of type $\boldsymbol{(\ell, (2s)^{\otimes N_s})}$}\label{section3.6}

Let us now introduce complete strings. For a positive integer $\ell$
we call the following set of rapidities $\lambda_j$
 a complete $\ell$-string:
\begin{gather*}
\lambda_j = \Lambda - (2j- \ell-1) \eta/2  \qquad \mbox{for} \quad
j= 1, 2, \ldots, \ell.
\end{gather*}
Here we call parameter $\Lambda$ the string center.

Let us now set inhomogeneous parameters $w_j$ for $j=1, 2, \ldots, L$,
as $N_s$ sets of complete $2s$-strings \cite{DM1}.
We def\/ine $w_{2s (b-1) + \beta}^{(2s)}$ for $\beta = 1, \ldots, 2s$,
as follows
\begin{gather}
w_{2s(b-1)+ \beta}^{(2s)} = \xi_b - (\beta-1) \eta     \qquad
 \mbox{for} \quad b = 1, 2, \ldots, N_s .
\label{eq:ell-strings}
\end{gather}

We now introduce the massless
monodromy matrix of type $(1, (2s)^{\otimes N_s})$
associated with homogeneous grading.
We def\/ine it by
\begin{gather*}
\widetilde{T}^{(1,   2s   +)}_{0,   1 2 \cdots N_s}
 (\lambda_0; \{ \xi_b \}_{N_s}  )
  =   \widetilde{P}_{12 \cdots L}^{(2s)}
R_{0,   1 \ldots L}^{(1,   1   +)}
\big(\lambda_0; \big\{ w_{j}^{(2s)} \big\}_L\big)
\widetilde{P}_{12 \cdots L}^{(2s)}    \\
\phantom{\widetilde{T}^{(1,   2s   +)}_{0,   1 2 \cdots N_s}
 (\lambda_0; \{ \xi_b \}_{N_s}  )}{}   =
\left(
\begin{array}{cc}
\widetilde{A}^{(2s +)}(\lambda; \{ \xi_b \}_{N_s}) &
\widetilde{B}^{(2s +)}(\lambda; \{ \xi_b \}_{N_s}) \\
\widetilde{C}^{(2s +)}(\lambda; \{ \xi_b \}_{N_s}) &
\widetilde{D}^{(2s +)}(\lambda; \{ \xi_b \}_{N_s})
\end{array}
\right)   .
\end{gather*}
Here, the (0,0) element is given by
$ \widetilde{A}^{(2s +)}(\lambda; \{ \xi_b \}_{N_s})=
\widetilde{P}_{12 \cdots L}^{(2s)}
A^{(1 +)}(\lambda; \{ w_j^{(2s)} \}_{L})
\widetilde{P}_{12 \cdots L}^{(2s)}$.

We shall now def\/ine the massless monodromy matrix of type
$(\ell,  (2s)^{\otimes N_s})$ associated with homogeneous grading.
Let us express the tensor product
$V_0^{(\ell)} \otimes \left( V_1^{(2s)} \otimes \cdots
\otimes V_{N_s}^{(2s)} \right)$,  by the following symbol
\begin{gather*}
\big(\ell,   (2s)^{\otimes N_s}\big)
= (\ell,   \overbrace{2s, 2s, \ldots, 2s}^{N_s})  .
\end{gather*}
Here we recall that $V_0^{(\ell)}$ abbreviates
$V_{a_1 a_2 \ldots a_{\ell}}^{(\ell)}$.
For the auxiliary space $V_0^{(\ell)}$
we def\/ine the massless monodromy matrix of type
$(\ell,  (2s)^{\otimes N_s})$ by
\begin{gather*}
\widetilde{T}^{(\ell,  2s  +)}_{0,  1 2 \cdots N_s}
 =  \widetilde{P}^{(\ell)}_{a_1 a_2 \cdots a_{\ell}}
\widetilde{T}_{a_1,  1 2 \cdots N_s}^{(1,  2s  +)}(\lambda_{a_1})
\widetilde{T}_{a_2,  1 2 \cdots N_s}^{(1,  2s  +)}(\lambda_{a_1}-\eta)
\cdots\\
\phantom{\widetilde{T}^{(\ell,  2s  +)}_{0,  1 2 \cdots N_s}=}{}\times
\widetilde{T}_{a_{2s},  1 2 \cdots N_s}^{(1,   2s  +)}
(\lambda_{a_1}-(\ell-1)\eta)
\widetilde{P}^{(\ell)}_{a_1 a_2 \cdots a_{\ell}}  .
\end{gather*}
Here we remark that it is associated with homogeneous grading.

Let us now construct
the higher-spin monodromy matrices associated with principal grading.
From the higher-spin monodromy matrices associated with homogeneous grading
we derive them through the inverse
of the gauge transformation as follows \cite{DM5}
\begin{gather*}
T^{(\ell,   2s   p)}
= \left(\chi_{a_1 \cdots a_{\ell},
1 2 \ldots N_s}^{(\ell,   2s)} \right)^{-1}
T^{(\ell,   2s   +)}(\lambda)
\left(\chi_{a_1 \cdots a_{\ell},   1 2 \ldots N_s}^{(\ell,   2s)} \right)   .
\end{gather*}
Here $\chi^{(\ell,   2s)}_{a_1 \cdots a_{\ell},    1 2 \ldots N_s}$
denote the following:
\begin{gather*}
\chi^{(\ell,   2s)}_{a_1 \cdots a_{\ell},   1 2 \ldots N_s}
= \Phi^{(\ell)}_{a_1 \cdots a_{\ell}}(\Lambda_0)
\Phi^{(2s)}_1(\Lambda_1) \cdots
\Phi^{(2s)}_{N_s}(\Lambda_{N_s})   ,
\end{gather*}
where $\Lambda_0$ denotes the string center,
 $\Lambda_0=\lambda_{a_1} - (\ell-1) \eta/2$.

For an illustration, let us consider the case of $\ell=1$.
For type $(1, (2s)^{\otimes Ns})$
the monodromy matrix associated with homogeneous grading and
that with principal grading are related to each other as follows
\begin{gather*}
T^{(1,   2s   +)}_{0,   1 2 \cdots N_s} (\lambda; \{ \xi_b \}_{N_s})
= \chi_{0,   1 2 \cdots N_s}^{(1,   2s)}
T^{(1,   2s   p)}_{0,   1 2 \cdots N_s}(\lambda; \{ \xi_b \}_{N_s})
 \left( \chi_{0,   1 2 \cdots N_s}^{(1,   2s)} \right)^{-1}   .
\end{gather*}
In terms of the operator-valued matrix elements we have
\begin{gather*}
\left(
\begin{array}{cc}
A^{(2s   +)}_{1 2 \cdots N_s}(\lambda) &
B^{(2s   +)}_{1 2 \cdots N_s}(\lambda) \\
C^{(2s   +)}_{1 2 \cdots N_s}(\lambda) &
D^{(2s   +)}_{1 2 \cdots N_s}(\lambda)
\end{array}
\right)\\
\qquad{}=
\left(
\begin{array}{cc}
\chi_{1 2 \cdots N_s}^{(2s)}
A^{(2s   p)}_{1 2 \cdots N_s}(\lambda)
\left( \chi_{1 2 \cdots N_s}^{(2s)} \right)^{-1}&
e^{-\lambda}
\chi_{1 2 \cdots N_s}^{(2s)}
 B^{(2s   p)}_{1 2 \cdots N_s}(\lambda)
\left( \chi_{1 2 \cdots N_s}^{(2s)} \right)^{-1}
\\
e^{\lambda}
\chi_{1 2 \cdots N_s}^{(2s)}
C^{(2s   p)}_{1 2 \cdots N_s}(\lambda)
\left( \chi_{1 2 \cdots N_s}^{(2s)} \right)^{-1}
&
\chi_{1 2 \cdots N_s}^{(2s)}
D^{(2s   p)}_{1 2 \cdots N_s}(\lambda)
\left( \chi_{1 2 \cdots N_s}^{(2s)} \right)^{-1}
\end{array}
\right)
  .
\end{gather*}

We shall now introduce the spin-1/2 monodromy matrices with
special inhomogeneous parameters.
Let us introduce a set of $2s$-strings with small deviations from
the set of complete $2s$-strings
\begin{gather*}
w_{2s(b-1)+ \beta}^{(2s;   \epsilon)} = \xi_b - (\beta-1) \eta
+ \epsilon r_b^{(\beta)}   \qquad
 \mbox{for} \quad b=1, 2, \dots, N_s \quad \mbox{and} \quad
\beta=1, 2, \ldots, 2s.
\end{gather*}
Here $\epsilon$ is a inf\/initesimally small generic number
and $r_{b}^{(\beta)}$ are generic parameters.
We call the set of rapidities
$w_{2s(b-1)+ \beta}^{(2s;   \epsilon)}$ for $\beta=1, 2, \ldots, 2s$
``almost complete $2s$-strings''.
We denote by $T^{(1,  2s  +;  \epsilon)}(\lambda)$
 the spin-1/2 monodromy matrix $T^{(1,  1  +)}$ with
inhomogeneous parameters $w_j$ being given by the set of almost complete
$2s$-strings: $w_j=w_j^{(2s;  \epsilon)}$ for $j=1, 2, \ldots, L$
\begin{gather*}
T^{(1,  2s  +;  \epsilon)}_{0,  1 2 \cdots L}
(\lambda) = T^{(1,  1  +)}_{0,  1 2 \cdots L}
\big(\lambda; \big\{ w_j^{(2s; \epsilon)} \big\}_{L}\big)  .
\end{gather*}
We express the elements of $T^{(1,  2s  +;  \epsilon)}(\lambda)$
as follows
\begin{gather*}
T^{(1,  2s  +;  \epsilon)}(\lambda) =
\left(
\begin{array}{cc}
A^{(2s +;  \epsilon)}_{1 2 \cdots L}(\lambda) &
B^{(2s +;  \epsilon)}_{1 2 \cdots L}(\lambda) \\
C^{(2s +;  \epsilon)}_{1 2 \cdots L}(\lambda) &
D^{(2s +;  \epsilon)}_{1 2 \cdots L}(\lambda)
\end{array}
\right)  .
\end{gather*}
Here we recall that
$A^{(2s  + ;   \epsilon)}_{1 2 \cdots L}(\lambda)$ denotes
$A^{(1  +)}_{1 2 \cdots L}(\lambda; \{ w_j^{(2s;   \epsilon)} \}_L)$.
We also remark the following:
\begin{gather*}
\widetilde{ A}^{(2s +)}_{1 2 \cdots N_s}(\lambda; \{\xi_p \}_{N_s}) =
\lim_{\epsilon \rightarrow 0}
\widetilde{P}_{1 2 \cdots L}^{(2s)}
A^{(2s +;  \epsilon)}_{1 2 \cdots L}
\big(\lambda; \big\{w_j^{(2s;  \epsilon)} \big\}_{L}\big)
\widetilde{P}_{1 2 \cdots L}^{(2s)}   .
\end{gather*}

\subsection{Series of commuting higher-spin transfer matrices}\label{section3.7}

Suppose that  $|\ell, m \rangle$
for $m=0, 1, \ldots, \ell$, are the orthonormal
basis vectors of $V^{(\ell)}$,
and their dual vectors are given by $\langle \ell, m |$
for $m=0, 1, \ldots, \ell$.
We def\/ine the trace of operator $A$ over the space~$V^{(\ell)}$ by
\begin{gather*}
{\rm tr}_{V^{(\ell)}} A = \sum_{m=0}^{\ell}
\langle \ell, m | A | \ell, m \rangle   .
\end{gather*}

We def\/ine the massless transfer matrix of type
$(\ell, (2s)^{\otimes N_s})$ by
\begin{gather*}
  \widetilde{t}^{(\ell,   2s   +)}_{1 2 \cdots N_s}(\lambda)
= {\rm  tr}_{V^{(\ell)}}
\left( \widetilde{T}^{(\ell,   2s   +)}_{0,   1 2 \cdots N_s}
(\lambda) \right) \non \\
\phantom{\widetilde{t}^{(\ell,   2s   +)}_{1 2 \cdots N_s}(\lambda)}{}  =
\sum_{n=0}^{\ell} {}_a \langle \ell,  n ||
 \widetilde{T}^{(1,   2s   +)}_{a_1,   1 2 \cdots N_s}(\lambda)
\widetilde{T}^{(1,   2s   +)}_{a_2,   1 2 \cdots N_s}(\lambda-\eta)
\cdots \widetilde{T}^{(1,   2s   +)}_{a_{\ell},  1 2 \cdots N_s}
(\lambda-(\ell-1) \eta)
\widetilde{ || \ell, n \rangle}_a    .
\end{gather*}

It follows from the Yang--Baxter equations that
the higher-spin transfer matrices commute in the tensor product space
$V_1^{(2s)} \otimes \cdots \otimes V_{N_s}^{(2s)}$,
which is derived by applying
projection operator $P^{(2s)}_{1 2 \cdots L}$
to $V^{(1)}_1 \otimes \cdots \otimes V_L^{(1)}$.
For instance, for the massless
transfer matrices, making use of (\ref{eq:PP'=P}) and (\ref{eq:P'P=P'})
we show
\begin{gather*}
P^{(2s)}_{1 2 \cdots L}
 \big[ \widetilde{t}^{(\ell,  2s   +)}_{1 2 \cdots N_s}(\lambda),
\widetilde{t}^{(m, \, 2s \, +)}_{1 2 \cdots N_s}(\mu) \big]  = 0
\qquad \mbox{for} \quad
\ell, m \in {\bf Z}_{\ge 0} .
\end{gather*}
Consequently, for the massless transfer matrices,
the eigenvectors
of $\widetilde{t}^{(1,   2s   +)}_{1 2 \cdots N_s}(\lambda)$
constructed by applying  $\widetilde{B}^{(2s   +)}(\lambda)$
 to the vacuum $| 0 \rangle$
also diagonalize the higher-spin transfer matrices, in particular,
the spin-$s$ massless XXZ transfer matrix,
 $\widetilde{t}^{(2s,   2s   +)}_{1 2 \cdots N_s}(\lambda)$.
Thus, we construct the ground state
of the higher-spin XXZ Hamiltonian in terms of operators
$\widetilde{B}^{(2s   +)}(\lambda)$,
which are the (0, 1)-element of the monodromy matrix
$\widetilde{T}^{(1,   2s   +)}$.

\subsection{Algebraic Bethe ansatz for higher-spin massless transfer matrices}\label{section3.8}

In terms of  the vacuum vector $| 0 \rangle$ where all spins are up,
we def\/ine functions $a(\lambda)$ and $d(\lambda)$  by
\begin{gather*}
A^{(1   p)}(\lambda; \{ w_j \}_L) |0 \rangle   =
a(\lambda; \{ w_j \}_L) | 0 \rangle   , \qquad
D^{(1 \, p)}(\lambda; \{ w_j \}_L) |0 \rangle   =
d(\lambda; \{ w_j \}_L) | 0 \rangle     .
\end{gather*}
We have $a(\lambda; \{ w_j \}_L) = 1$ and
\begin{gather*}
d(\lambda; \{ w_j \}_L)  =  \prod^{L}_{j=1} b(\lambda, w_j)   .
\end{gather*}
Here $b(\lambda, \mu)= b(\lambda-\mu)$.
For the homogeneous grading ($w=+$)
 and the principal  grading ($w=p$),
it is easy to show the following relations:
\begin{gather*}
A^{(2s   w)}(\lambda) |0 \rangle   =
\widetilde{A}^{(2s   w)}(\lambda) |0 \rangle =
a^{(2s)}(\lambda; \{ \xi_b \}) | 0 \rangle  ,   \\
D^{(2s   w)}(\lambda) |0 \rangle   =
\widetilde{D}^{(2s   w)}(\lambda) |0 \rangle =
d^{(2s)}(\lambda; \{ \xi_b \}) | 0 \rangle     ,
\end{gather*}
where $a^{(2s)}(\lambda; \{ \xi_b \})$ and
$d^{(2s)}(\lambda; \{ \xi_b \})$ are given by
\begin{gather*}
a^{(2s)}(\lambda; \{ \xi_b \})   =
a\big(\lambda; \big\{ w_j^{(2s)} \big\}\big)=1   ,   \\
d^{(2s)}(\lambda; \{ \xi_b \})   =
d\big(\lambda; \big\{ w_j^{(2s)} \big\}\big)
 =  \prod^{N_s}_{p=1} b_{2s}(\lambda, \xi_p)   .
\end{gather*}
Here we have  def\/ined $b_t(\lambda, \mu)$ by
$b_t(\lambda, \mu) = {\sinh(\lambda-\mu)}/{\sinh(\lambda-\mu+ t \eta)}$ .
Here we recall $b(u)= b_{1}(u)= \sinh u/\sinh(u+\eta)$.

In the massless regime, we def\/ine the Bethe vectors
$| \widetilde{\{\lambda_{\alpha} \}}_{M}^{(2s   w)}    \rangle$
for $w=+$ and $p$, and their dual vectors
$ \langle \widetilde{\{\lambda_{\alpha} \}}_{M}^{(2s   w)} |$
for $w=+$ and $p$,  as follows
\begin{gather}
| \widetilde{ \{\lambda_{\alpha} \}}_{M}^{(2s   w)}    \rangle
  =   \prod_{\alpha=1}^{M}
\widetilde{B}^{(2s   w)}(\lambda_{\alpha}) | 0 \rangle   ,
\label{eq:eigen} \qquad
 \langle \widetilde{ \{\lambda_{\alpha} \}}_{M}^{(2s  w)}  |
 =  \langle  0 |
\prod_{\alpha=1}^{M}
\widetilde{C}^{(2s  w)}(\lambda_{\alpha})   .
\end{gather}
Here we recall
$\widetilde{B}^{(2s  +)}(\lambda_{\alpha})=
\widetilde{P}^{(2s)}_{1 \cdots L}
B^{(1  +)}(\lambda_{\alpha}, \{ w_k^{(2)} \}_L)
\widetilde{P}^{(2s)}_{1 \cdots L}$.
The Bethe vector (\ref{eq:eigen})
gives an eigenvector of the massless transfer matrix
\begin{gather*}
\widetilde{t}^{(1,   2s   w)}(\mu; \{ \xi_b \}_{N_s})
=\widetilde{ A}^{(2s   w)}(\mu; \{\xi_b \}_{N_s}) +
\widetilde{D}^{(2s   w)}(\mu; \{\xi_b \}_{N_s})
\end{gather*}
for $w=+$ and $w=p$  with the following eigenvalue:
\begin{gather*}
\Lambda^{(1, {2s}   w)}(\mu)=
\prod_{j=1}^{M}
{\frac {\sinh(\lambda_j - \mu + \eta)} {\sinh(\lambda_j - \mu)}}
+
\prod_{p=1}^{N_s}
b_{2s}(\mu, \xi_p)   \cdot
\prod_{j=1}^{M}
{\frac {\sinh(\mu - \lambda_j + \eta)} {\sinh(\mu - \lambda_j)}}   ,
\end{gather*}
if rapidities $\{\lambda_j \}_M$ satisfy  the Bethe ansatz equations
\begin{gather*}
\prod_{p=1}^{N_s} b_{2s}^{-1}(\lambda_{j}, \xi_p) =
\prod_{k \ne j} {\frac {b(\lambda_{k}, \lambda_{j})}
                     {b(\lambda_{j}, \lambda_{k})}} , \qquad
j= 1, \ldots, M  . 
\end{gather*}

Let us denote by
$| \{\lambda_{\alpha}(\epsilon) \}_M^{(2s  w;  \epsilon)} \rangle$
the Bethe vector of $M$ Bethe roots $\{ \lambda_j(\epsilon)  \}_M$
for $w=+, p$:
\begin{gather*}
| \{\lambda_{\alpha}(\epsilon) \}_M^{(2s  w;  \epsilon)} \rangle
 =
\prod_{\gamma=1}^{M} B^{(2s   w; \epsilon)}(\lambda_{\gamma}(\epsilon) )
| 0 \rangle
 =
B^{(2s   w; \epsilon)}(\lambda_1(\epsilon) )
\cdots B^{(2s   w; \epsilon)}(\lambda_M(\epsilon) ) | 0 \rangle  ,
\end{gather*}
where rapidities $\{ \lambda_j(\epsilon) \}_M$ satisfy the Bethe ansatz
equations with inhomogeneous parameters $w_j^{(2s; \epsilon)}$ as follows
\begin{gather*}
\frac {a(\lambda_j(\epsilon); \{ w_{k}^{(2s;   \epsilon)} \}_L )}
{d(\lambda_j(\epsilon) ;  \{ w_{k}^{(2s;   \epsilon)} \}_L) } =
\prod_{k=1; k \ne j}^{M}
{\frac {b(\lambda_k(\epsilon), \lambda_j(\epsilon))}
{b(\lambda_j(\epsilon), \lambda_k(\epsilon))}}   .
\end{gather*}
It gives an eigenvector of the transfer matrix
\begin{gather*}
t^{(1, 1   w)}\big(\mu; \big\{ w_j^{(2s;   \epsilon)}  \big\}_{L}\big)
=A^{({2s   w;   \epsilon})}\big(\mu; \big\{w_j^{(2s;   \epsilon)} \big\}_{L}\big) +
D^{({2s   w;   \epsilon})}\big(\mu; \big\{w_j^{(2s;   \epsilon)} \big\}_{L}\big)
\end{gather*}
with the following eigenvalue:
\begin{gather*}
\Lambda^{(1, 1   w)}\big(\mu; \big\{w_j^{(2s;   \epsilon)} \big\}_{L}\big)=
\prod_{j=1}^{M}
{\frac {\sinh(\lambda_j(\epsilon) - \mu + \eta)}
{\sinh(\lambda_j(\epsilon) - \mu)}}\\
\hphantom{\Lambda^{(1, 1   w)}\big(\mu; \big\{w_j^{(2s;   \epsilon)} \big\}_{L}\big)=}{}
+ \prod_{j=1}^{L}
b\big(\mu, w_j^{(2s;   \epsilon)} \big)   \cdot
\prod_{j=1}^{M}
{\frac {\sinh(\mu - \lambda_j(\epsilon) + \eta)}
{\sinh(\mu - \lambda_j(\epsilon))}}   .
\end{gather*}

Let us assume that in the limit of $\epsilon$ going to 0,
the set of Bethe roots $\{ \lambda_j(\epsilon)  \}_M$ approaches~$\{ \lambda_j  \}_M$.
Assuming the continuity of the limiting procedure,
we have
\begin{gather*}
| \widetilde{ \{ \lambda_j \}}_M^{(2s   +)} \rangle
  =   \lim_{\epsilon \rightarrow 0}  \prod_{j=1}^{M} \left(
\widetilde{P}^{(2s)}_{12 \cdots L}
B^{(2s   +;   \epsilon)}( \lambda_j(\epsilon) )
\widetilde{P}^{(2s)}_{12 \cdots L}
\right)
| 0 \rangle
 =   \widetilde{P}^{(2s)}_{12 \cdots L}
\lim_{\epsilon \rightarrow 0}  \prod_{j=1}^{M}
B^{(2s   +;   \epsilon)}(\lambda_j(\epsilon) )
| 0 \rangle
  .
\end{gather*}
Thus, the expectation value with respect to the Bethe state of
$\{ \lambda_j  \}_M$ is given by the limit of that of
$\{ \lambda_j(\epsilon)  \}_M$ sending $\epsilon$ to zero.
For the $B$ operators associated with principal grading, we have
\begin{gather*}
| \widetilde{ \{ \lambda_j \}}_M^{(2s   p)} \rangle
  =   \lim_{\epsilon \rightarrow 0} \prod_{j=1}^{M}
\left(e^{\lambda_j(\epsilon)}
\big(\chi^{(2s)}_{12 \cdots N_s}\big)^{-1}
\widetilde{P}^{(2s)}_{12 \cdots L}
B^{(2s   +;   \epsilon)}(\lambda_j(\epsilon))
\widetilde{P}^{(2s)}_{12 \cdots L}
\chi^{(2s)}_{12 \cdots N_s}
\right)
| 0 \rangle
 \\
\phantom{| \widetilde{ \{ \lambda_j \}}_M^{(2s   p)} \rangle}{}
  =   \big(\chi^{(2s)}_{12 \cdots N_s}\big)^{-1}
\widetilde{P}^{(2s)}_{12 \cdots L}
\chi_{12 \cdots L}   \times
\lim_{\epsilon \rightarrow 0}  \prod_{j=1}^{M}
B^{(2s   p; \epsilon)}( \lambda_j(\epsilon) )
| 0 \rangle   .
\end{gather*}

Let us introduce symbols for the ground state
of the integrable spin-$s$ XXZ spin chain.
We denote it by $| \psi_g^{(2s   p)} \rangle$
associated with principal grading.
It is given by multiplying the projection operator
to such a product of the spin-1/2 $B$
operators with inhomogeneous parameters being given by
the set of complete $2s$-strings that acts on the vacuum:
\begin{gather*}
| \psi_g^{(2s   p)} \rangle =
\big(\chi^{(2s)}_{12 \cdots N_s} \big)^{-1}
\widetilde{P}^{(2s)}_{1 2 \cdots L}
\chi_{12 \cdots L}   \cdot
\prod_{\gamma=1}^{M} B^{(2s   p;   0)}(\lambda_{\gamma}) | 0 \rangle .
\end{gather*}
We denote by $| \psi_g^{(2s   p; 0)} \rangle$
the product of the spin-1/2 $B$ operators with inhomogeneous parameters
given by complete $2s$-strings $w_j^{(2s)}$ which acts on the vacuum state:
\begin{gather*}
| \psi_g^{(2s   p;   0)} \rangle =
\prod_{\gamma=1}^{M} B^{(2s   p;   0)}(\lambda_{\gamma}) | 0 \rangle   .
\end{gather*}

\subsection{Commutation relations with projection operators}\label{section3.9}

Let us discuss an application of the fusion
construction of projection operators (\ref{eq:def-projector}).
Hereafter we assume that rapidity $\lambda$ does not take such discrete values
at which the transfer matrix becomes singular or non-regular,  such as
$w_j^{(2s)}-\eta + n \pi \sqrt{-1}$ ($1 \le j \le L)$ for
$n \in {\bf Z}$~\cite{DM5}.
Here we recall that $w_j^{(2s)}$ are inhomogeneous parameters
forming complete $2s$-strings.

\begin{lemma}
Projection operators $P^{(2s)}_{12 \cdots L}$ and
$\widetilde{P}^{(2s)}_{12 \cdots L}$
commute with the matrix elements of the monodromy matrix
$T^{(1,   1   +)}_{0, 12 \cdots L}(\lambda; \{ w_j^{(2s)} \}_L)$
such as $A^{(2s   +;   0)}(\lambda)$
\begin{gather}
 P_{12 \cdots L}^{(2s)}
T^{(1 ,   1   +)}_{0, 12 \cdots L}\big(\lambda; \big\{ w_j^{(2s)} \big\}_L\big)
  P_{12 \cdots L}^{(2s)}
  =
P_{12 \cdots L}^{(2s)}
  T^{(1,   1   +)}_{0, 12 \cdots L}
\big(\lambda; \big\{ w_j^{(2s)} \big\}_L\big)
  , \label{eq:commute} \\
  P_{12 \cdots L}^{(2s)}
T^{(1,   1   +)}_{0, 12 \cdots L}\big(\lambda;
\big\{ w_j^{(2s)} \big\}_L\big)
 \widetilde{P}_{12 \cdots L}^{(2s)}
  =
P_{1 2 \cdots L}^{(2s)}   T^{(1,1)}_{0, 1 2 \cdots L}
\big(\lambda; \big\{ w_j^{(2s)} \big\}_L\big)   . \label{eq:commute2}
\end{gather}
For instance we have
$P_{12 \cdots L}^{(2s)} B^{(2s   +;   0)}(\lambda)
P_{12 \cdots L}^{(2s)}
= P_{12 \cdots L}^{(2s)}  B^{(2s   +;   0)}(\lambda).
$
\end{lemma}

We show (\ref{eq:commute}) and (\ref{eq:commute2}) by the Yang--Baxter equation.
In fact,  $T^{(1,  1  +)}_{0, 12 \cdots L}(\lambda; \{ w_j^{(2s)} \}_L)$
commutes with the projection operator $P_{12 \cdots L}^{(2s)}$ thanks to
the fusion construction of projection ope\-ra\-tors~(\ref{eq:def-projector})~\cite{DM1}.
We derive (\ref{eq:commute2}) making use of (\ref{eq:PP'=P}).

\subsection[Quantum inverse scattering problem (QISP) for the spin-$s$ operators]{Quantum inverse scattering problem (QISP) for the spin-$\boldsymbol{s}$ operators}\label{section3.10}

We can express any given spin-$s$ local operator
in terms of the spin-1/2 global operators
such as $A, B, C$ and $D$; i.e.\
we have the QISP formulas for the spin-$s$ local
operators \cite{DM5}.
For an illustration, we show the case of $b=1$, i.e.,
we express one of the spin-$s$ elementary matrices
in terms of the spin-1/2 global operators.
\begin{lemma}[\cite{DM4,DM5}]
For a pair of integers $i$ and $j$ satisfying
$1 \le i, j \le \ell$, the spin-$\ell/2$ elementary matrix
associated with principal grading is decomposed into a sum of
products of the matrix elements of the spin-$1/2$ monodromy matrix
as follows
\begin{gather*}
 \widetilde{E}^{i ,   j   (\ell   p)}_1 =
\left(
\left[ \begin{matrix}
\ell \\
i
\end{matrix}
\right]_q
\left[ \begin{matrix}
\ell \\
j
\end{matrix}
\right]_q^{-1} \right)^{1/2}
e^{-(i-j)\xi_1} \, q^{(i-j)(\ell-1)/2}  \cdot
 P_{1 \cdots \ell}^{(\ell)}   \cdot   \chi_{1 2 \cdots \ell}
 \\
\qquad{} \times
\sum_{ \{ \varepsilon_{\beta} \}}
T^{(1,   \ell   p;   \epsilon)}_{\varepsilon_1, \varepsilon'_1}
\big(w_1^{(\ell;  \epsilon)}\big) \cdots
T^{(1,  \ell  p;  \epsilon)}_{\varepsilon_{\ell}, \varepsilon'_{\ell}}
\big(w_{\ell}^{(\ell;  \epsilon)}\big)
\prod_{k=1}^{\ell}
\big(A^{(\ell  p;  \epsilon)}\big(w_{k}^{(\ell;  \epsilon)}\big)
+ D^{(\ell  p;  \epsilon)}\big(w_{k}^{(\ell;  \epsilon)}\big)
\big)^{-1}
\chi_{1 2 \cdots \ell}^{-1}
 .
\end{gather*}
Here the sum is taken over all sets of $\varepsilon_{\beta}$ such that
the number of integers $\beta$ satisfying $\varepsilon_{\beta}=1$ and
$1 \le \beta \le \ell$ is given by $j$. We take a set of
$\varepsilon'_{\alpha}$ such that
 the number of integers $\alpha$ satisfying $\varepsilon'_{\alpha}=1$ and
$1 \le \alpha \le \ell$ is given by $i$. We have expressed the
element of $(\alpha, \beta)$ in the monodromy matrix
$T^{(1,  \ell  p;  \epsilon)}(\lambda)$ by
$T^{(1,  \ell  p;  \epsilon)}_{\alpha, \beta}(\lambda)$
for $\alpha, \beta=0, 1$.
\end{lemma}

For an illustration, let us consider the spin-1/2 formula
\cite{KMT1999,KMT2000} (see also \cite{MT2000,GK2000}):
\begin{gather}
e_n^{i_n,   j_n} = \prod_{j=1}^{n-1} t^{(1   p)}_{1 2 \cdots L}(w_j) \cdot
{\rm tr}_0 \big( e_0^{i_n,   j_n}
 R^{(1   p)}_{0,   1 2 \cdots L}(w_n)  \big)
\prod_{j=1}^{n} \big( t^{(1   p)}_{1 2 \cdots L}(w_j) \big)^{-1}   .
\label{eq:formula-1/2}
\end{gather}
Here we recall that the spin-1/2 transfer matrix $t^{(1   p)}_{1 2 \cdots L}(\lambda)$ is given by
the trace of the monodromy matrix of type $(1, 1^{\otimes L})$:
$t^{(1   p)}_{1 2 \cdots L}(\lambda)
=A^{(1  p)}(\lambda) + D^{(1   p)}(\lambda)$.
We remark that the expres\-sion~(\ref{eq:formula-1/2})
holds if inhomogeneous parameters $w_j$ $(1 \le j \le L$)
take generic values.
Multiplying the expressions of formula (\ref{eq:formula-1/2})
for $n=1, 2, \ldots, m$, we have
\begin{gather}
e_1^{i_1, j_1} e_2^{i_2, j_2} \cdots e_m^{i_m, j_m}   =
{\rm tr}_0 \big( e_0^{i_1, j_1}
R^{(1   p)}_{0,   1 2 \cdots L}(w_1)  \big)
{\rm tr}_0  \big( e_0^{i_2, j_2}
R^{(1   p)}_{0,   1 2 \cdots L}(w_2)  \big)  \cdots \non \\
\phantom{e_1^{i_1, j_1} e_2^{i_2, j_2} \cdots e_m^{i_m, j_m}   =}{}  \times
{\rm tr}_0 \big( e_0^{i_m, j_m}
R^{(1   p)}_{0,   1 2 \cdots L}(w_m) \big)
\prod_{j=1}^{m} \big( t^{(1   p)}_{1 2 \cdots L}(w_j) \big)^{-1}   .
\label{eq:QISP}
\end{gather}
Here we note that we have $R_{0 n}^{(1   p)}(0)=\Pi_{0,    n}$
from the normalization condition of the $R$-matrices,
where $\Pi_{0,    n}$ denotes
the permutation operator acting on the 0th and $n$th sites (see also Section~\ref{section3.5}).
Thus, we have
\begin{gather}
\prod_{j=1}^{L} t^{(1   p)}_{1 2 \cdots L}(w_j)= I^{\otimes L}.
\label{eq:identity}
\end{gather}

We note that the QISP formulas~(\ref{eq:formula-1/2}) hold
if the inhomogeneous  parameters are generic.
If we send them to a set of complete $2s$-strings such as
$w_j^{(2s)}$, then the transfer matrix becomes non-regular or singular,
and relations such as~(\ref{eq:identity}) do not hold.
Instead of complete $2s$-strings,
we therefore put ``almost complete $2s$-strings'', $w_j^{(2s;  \epsilon)}$,
into inhomogeneous parameters $w_j$. Here parameters $w_j^{(2s;   \epsilon)}$
are generic, and hence the QISP formula~(\ref{eq:formula-1/2}) holds.

\subsection[Expectation value of a local operator
through the limit: $\epsilon \rightarrow 0$]{Expectation value of a local operator
through the limit: $\boldsymbol{\epsilon \rightarrow 0}$}\label{section3.11}

In the massless regime,
we def\/ine the expectation value of product of operators
$\prod\limits_{k=1}^{m} \widetilde{E}_k^{i_k,  j_k  (2s  p)}$
with respect to an eigenstate
$| \widetilde{ \{\lambda_{\alpha} \}}_M^{(2s  p)} \rangle$ by
\begin{gather}
\langle
\prod_{k=1}^{m}
\widetilde{E}_k^{i_k,   j_k   (2s   p)} \rangle \left(
\{ \lambda_{\alpha} \}_M^{(2s \, p)} \right)
=
{\frac {\langle \widetilde{ \{\lambda_{\alpha} \}}_M^{(2s \, p)} |
\prod\limits_{k=1}^{m} \widetilde{E}_k^{i_k,   j_k   (2s   p)}
 | \widetilde{ \{\lambda_{\alpha} \}}_M^{(2s   p)} \rangle}
 {\langle \widetilde{ \{\lambda_{\alpha} \}}_M^{(2s   p)} |
 \widetilde{ \{\lambda_{\alpha} \}}_M^{(2s  p)} \rangle}}   .
\label{eq:defEXP}
\end{gather}

In order to evaluate (\ref{eq:defEXP})
we make use of the following formulas.

\begin{proposition}[\cite{DM4,DM5}]
Let us take a pair of integers $i_1$ and $j_1$ satisfying
$1 \le i_1, j_1 \le \ell$.
For arbitrary parameters $\{\mu_{\alpha} \}_N$ and
$\{\lambda_{\beta} \}_M$ with $i_1-j_1=N-M$
we have
\begin{gather}
 \langle 0 | \prod_{\alpha=1}^{N} C^{(\ell \, p)}(\mu_a) \cdot
\widetilde{E}^{i_1,   j_1   (\ell   p)}_1 \cdot \prod_{\beta=1}^{M}
B^{(\ell   p)}(\lambda_{\beta}) | 0 \rangle
\non \\
 \qquad{} =
\sqrt{
\left[ \begin{matrix}
\ell \\
i_1
\end{matrix}
\right]_q
\left[ \begin{matrix}
\ell \\
j_1
\end{matrix}
\right]^{-1}_{q}}
\sum_{ \{ \varepsilon_{\beta} \} }
\langle 0 | \prod_{\alpha=1}^{N} C^{(\ell   p;   0)}(\mu_a)
  \cdot
e_1^{\varepsilon_1', \varepsilon_1} \cdots
e_{\ell}^{\varepsilon_{\ell}', \varepsilon_{\ell}}
  \cdot   \prod_{\beta=1}^{M}
B^{(\ell   p;   0)}(\lambda_{\beta}) | 0 \rangle   .
\label{eq:<E(p)>-<e>}
\end{gather}
Here we take the sum over all sets of $\varepsilon_{\beta}$
such that the number of integers $\beta$ with $\varepsilon_{\beta}=1$
for $1 \le \beta \le \ell$ is given by $j_1$.
We take a set of $\varepsilon_{\alpha}'$ such that
 the number of integers $\alpha$ satisfying $\varepsilon_{\alpha}'=1$
for $1 \le \alpha \le \ell$ is given by $i_1$.
Each summand is symmetric with respect to
exchange of~$\varepsilon_{\alpha}'$; i.e.,
the following expression is independent of any
permutation $\pi \in {\cal S}_{\ell}$:
\begin{gather}
\langle 0 | \prod_{\alpha=1}^{N} C^{(\ell   p;   0)}(\mu_a)
  \cdot
e_1^{\varepsilon_{\pi 1}',   \varepsilon_1} \cdots
e_{\ell}^{\varepsilon_{\pi \ell}',   \varepsilon_{\ell}}
  \cdot   \prod_{\beta=1}^{M}
B^{(\ell   p;   0)}(\lambda_{\beta}) | 0 \rangle   .
\label{eq:sym-epsilon_a'}
\end{gather}
\label{prop:redEexp}
\end{proposition}
Here we remark that ${\cal S}_n$ denotes the symmetric group of $n$ elements.

We evaluate the expectation value of a given spin-$s$ local operator
for a Bethe-ansatz eigenstate
$\widetilde{| \{\lambda_{\alpha} \}}_M^{(2s)} \rangle$,
as follows. We f\/irst express the spin-$s$ local operators
in terms of the spin-1/2 local operators via
formula~(\ref{eq:tildeE(p)-e}). Through Proposition~\ref{prop:redEexp}  the expectation value of  the spin-$s$ local operators
is reduced into those of  the spin-$1/2$ local operators.
We now assume that the Bethe roots $\{ \lambda_{\alpha}(\epsilon)\}_M$
are continuous with respect to small parameter $\epsilon$.
It follows from the assumption that each entry of the Bethe eigenstate
$|\{ \lambda_k(\epsilon) \}_M^{(2s;   \epsilon)} \rangle$
is continuous with respect to~$\epsilon$.
Then, we apply the spin-1/2 QISP formula
with generic inhomogeneous parameters $w_j^{(2s; \epsilon)}$
such as formula~(\ref{eq:QISP}).
We next calculate the scalar product for the Bethe state
$|\{ \lambda_k(\epsilon) \}_M^{(2s;   \epsilon)} \rangle$.
It has the same inhomogeneous parameters $w_j^{(2s; \epsilon)}$
as the global operators appearing in the QISP formula,
so that we can make use of Slavnov's formula of scalar products
for the spin-1/2 case.
Calculating explicitly the determinant of the scalar product
with Slavnov's formula,  we can show that
the expression of the scalar product is continuous
with respect to~$\epsilon$ at~$\epsilon=0$.
Thus, sending~$\epsilon$ to~0, we obtain
the expectation value of
the spin-$s$ local operator~(\ref{eq:defEXP}).

\begin{corollary}
Suppose that $i_1$ and $j_1$ are integers satisfying
$1 \le i_1, j_1 \le \ell$, and $\{ \mu_k \}_N$ are arbitrary parameters.
Let us assume that Bethe roots
$\{ \lambda_{\gamma}(\epsilon) \}_M$ are continuous at $\epsilon=0$
and $\lim\limits_{\epsilon \rightarrow 0} \lambda_{\gamma}(\epsilon)
=\lambda_{\gamma}$ for $\gamma=1, 2, \ldots, M$.
We have the following:
\begin{gather}
  \langle 0 | \prod_{k=1}^{N} C^{(\ell   p)}(\mu_k) \cdot
\widetilde{E}^{i_1,   j_1   (\ell   p)}_1 \cdot \prod_{\gamma=1}^{M}
B^{(\ell   p)}(\lambda_{\gamma}) | 0 \rangle
=
\sqrt{
\left[ \begin{matrix}
\ell \\
i_1
\end{matrix}
\right]_{q}
\left[ \begin{matrix}
\ell \\
j_1
\end{matrix}
\right]^{-1}_{q}}
\phi_{\ell}(\{ \lambda_{\gamma} \}_M; \{w_j^{(\ell)} \}_L)
\non \\
{} \times   \sum_{ \{ \varepsilon_{\beta} \} }
\lim_{\epsilon \rightarrow 0}
\langle 0 | \prod_{k=1}^{N} C^{(\ell   p;   \epsilon)}(\mu_k)
 \cdot
T^{(\ell   p;   \epsilon)}_{\varepsilon_1,   \varepsilon_1'}\big(w_1^{(\ell;   \epsilon)}\big) \cdots
T^{(\ell    p;   \epsilon)}_{\varepsilon_{\ell},   \varepsilon_{\ell}'}
\big(w_{\ell}^{(\ell;   \epsilon)}\big)
 \cdot   \prod_{\gamma=1}^{M}
B^{(\ell   p;   \epsilon)}(\lambda_{\gamma}(\epsilon)) | 0 \rangle   .
\label{eq:main-formula}
\end{gather}
Here we take the sum over all sets of $\varepsilon_{\beta}$s
such that the number of integers $\beta$ satisfying $\varepsilon_{\beta}=1$
for $1 \le \beta \le \ell$ is given by $j_1$.
We take a set of  $\varepsilon_{\alpha}'$ such that
 the number of integers $\alpha$ satisfying $\varepsilon_{\alpha}'=1$
for $1 \le \alpha \le \ell$ is given by $i_1$.
We have defined  $\phi_{m}(\{ \lambda_{\gamma} \})$ by
$\phi_{m} (\{ \lambda_{\gamma} \}_M; \{ w_j \}_L)
= \prod\limits_{j=1}^{m} \prod\limits_{\alpha=1}^M
b(\lambda_{\alpha}-w_j)$ with $b(u)=\sinh(u)/\sinh(u+\eta)$.
\label{cor:main}
\end{corollary}

We can evaluate the form factors and the expectation values of a spin-$\ell/2$
operator through Corollary~\ref{cor:main}~\cite{DM5}.
The corrections of the form factors given in the paper \cite{DM1}
are listed in refe\-ren\-ce~[20] of the paper \cite{DM5} (see also~\cite{DM4}).
Corrections for the paper~\cite{DM2}
are listed in reference~[21] of the paper~\cite{DM5}.

For an illustration, let us consider the spin-1 case.
We calculate the one-point function
$\langle \widetilde{E}_1^{1,  1  (2  p)} \rangle$.
Here we have $i_1=j_1=1$. Setting $\varepsilon_1'=0$ and
$\varepsilon_2'=1$,  we have
\begin{gather*}
\langle \psi_g^{(2   p)}|   \widetilde{E}_1^{1,   1   (2   p)}
| \psi_g^{(2 \, p)} \rangle =
\langle \psi_g^{(2   p;   0)}|
e_1^{0,  0} e_2^{1,   1} | \psi_g^{(2   p;   0)} \rangle +
\langle \psi_g^{(2   p;   0)} | e_1^{0,   1} e_2^{1,   0}
| \psi_g^{(2   p;   0)} \rangle   .
\end{gather*}
Here we have taken the sum over sets $\{ \varepsilon_{\beta} \}$ such as
$\{\varepsilon_{1}=0, \varepsilon_{2}=1 \}$
and $\{\varepsilon_{1}=1, \varepsilon_{2}=0 \}$.
Making use of the spin-1/2 QISP formula we have
\begin{gather*}
e_1^{0,  0} e_2^{1,  1}  =
A^{(2  p ;  \epsilon)}\big(w_1^{(2;  \epsilon)}\big)
D^{(2  p ;  \epsilon)}\big(w_2^{(2;  \epsilon)}\big)
\prod_{j=1}^{2} \left(
t^{(2 p;  \epsilon)}_{1 2 \cdots L}\big(w_j^{(2; \epsilon)}\big) \right)^{-1},
\non \\
e_1^{0,  1} e_2^{1,  0}  =
C^{(2  p ;  \epsilon)}\big(w_1^{(2;  \epsilon)}\big)
B^{(2  p ;  \epsilon)}\big(w_2^{(2;  \epsilon)}\big)
\prod_{j=1}^{2} \left( t^{(2  p;  \epsilon)}_{1 2 \cdots L}
\big(w_j^{(2;  \epsilon)}\big) \right)^{-1}  .
\end{gather*}
Therefore  we have
\begin{gather*}
\langle \widetilde{E}_1^{1,  1  (2  p)} \rangle
 =
\phi_2\big(\{ \lambda_{\gamma} \}; \big\{w_j^{(2)} \big\}_L\big)
\Bigg(
\lim_{\epsilon \rightarrow 0}
{\frac {\langle {\psi}_g^{(2  p;  \epsilon)} |
A^{(2  p ;  \epsilon)}(w_1^{(2:  \epsilon)})
D^{(2  p ;  \epsilon)}
(w_2^{(2:   \epsilon)})
 | {\psi}_g^{(2  p;  \epsilon)} \rangle}
{\langle {\psi}_g^{(2 p;  \epsilon)}
| {\psi}_g^{(2  p;  \epsilon)} \rangle}   } \non \\
\phantom{\langle \widetilde{E}_1^{1,  1  (2  p)} \rangle = }{}  +
\lim_{\epsilon \rightarrow 0}
{\frac {\langle {\psi}_g^{(2  p ;  \epsilon)} |
C^{(2  p ;  \epsilon)}(w_1^{(2:  \epsilon)})
B^{(2  p ;  \epsilon)}
(w_2^{(2:  \epsilon)})
 | {\psi}_g^{(2  p;  \epsilon)} \rangle}
{\langle {\psi}_g^{(2 p; \, \epsilon)} \,
| {\psi}_g^{(2 \, p; \, \epsilon)} \rangle}   } \Bigg)
\, .
\end{gather*}

\section{Quantum group symmetry relations in the spin 1 case}\label{section4}

We show some important topics.
We derive symmetry relations among the expectation values of
products of the spin-1/2 operators from the spin inversion symmetry.
In particular, we show how to transform the basis vectors constructed
in the $2s$th tensor product space of the spin-1/2 representations to
the $2s+1$-dimensional vectors in ${\bf C}^{2s+1}$.

\subsection{Rotation symmetry of the XXX spin chain and
irreducible components of operators}\label{section4.1}

Let us consider the XXX case where the SU(2) symmetry holds for the total spin operators.
The tensor product of two spin-1/2 representations of
$sl(2)$ decomposes into the direct sum of spin-1 and spin-0 representations;
i.e.,  $V(1)\otimes V(1)= V(2) \oplus V(0)$.
Here we recall that  $V(1)\otimes V(1)$ is four-dimensional, and
 the components  $V(2)$ and  $V(0)$ are three-dimensional and one-dimensional,
respectively. In the spin-1 representation $V(2)$ we have the basis vectors and basis covectors as follows
\begin{alignat*}{3}
& || 2, 0 \rangle   =   || +  + \rangle   , \qquad &&
\langle 2, 0 || = \langle +  + ||    , &  \nonumber \\
& || 2, 1 \rangle   =   || +  - \rangle + || -  + \rangle
 , \qquad && \langle 2, 1 || = {\frac 1 2}
\left( \langle +  - || + \langle -  + || \right)   , &
 \nonumber \\
& || 2, 2 \rangle   =  || -  - \rangle   ,
\qquad && \langle 2, 2 || = \langle -  - ||   .&
\end{alignat*}
Here we recall that
$|| - + \rangle$ denotes $| 1 \rangle_1 \otimes | 0 \rangle_2$.

In the spin-0 representation $V(0)$ we have the basis vectors and basis covectors as follows
\begin{gather*}
|| 0, 0 \rangle  =  | +  - \rangle - | -  + \rangle
, \qquad \langle 0, 0 || = {\frac 1 2}
\left( \langle +  - || - \langle -  + || \right)  .
\end{gather*}
In terms of the basis of the spin-1 irreducible representation
we express the symmetric projection operator as follows
\begin{gather*}
P^{(2)} = || 2, 0 \rangle \langle 2, 0 || +
|| 2, 1 \rangle \langle 2, 1 || +
|| 2, 2 \rangle \langle 2, 2 ||   .
\end{gather*}

In the spin-$s$ XXX case
we def\/ine  elementary matrices by
\begin{gather*}
E^{m,   n   (2s)} = || 2s, m \rangle \langle 2s, n ||   .
\end{gather*}

In the tensor product $V(1)\otimes V(1)$ there are
sixteen elementary matrices $e_1^{j_1, k_1} e_2^{j_2, k_2}$ for
$j_1, j_2, k_1, k_2 = \pm$.
For an illustration we express the operator $e_1^{+ -} e_2^{- +}$ in terms of
the basis vectors and their covectors  as follows
\begin{alignat*}{3}
& \langle 2, 1 || e_1^{+ -} e_2^{- +} || 2, 1 \rangle   =   \frac 1 2   ,
\qquad &&
\langle 0, 0 || e_1^{+ -} e_2^{- +} || 2, 1 \rangle  =  \frac 1 2  ,  & \\
& \langle 2, 1 || e_1^{+ -} e_2^{- +} || 0, 0 \rangle   =   - \frac {1} 2   ,
\qquad &&
\langle 0, 0 || e_1^{+ -} e_2^{- +} || 0, 0 \rangle =  - \frac {1} 2   .
\end{alignat*}
In terms of the bases of vectors and covectors, we have
\begin{gather}
e_1^{+ -} e_2^{- +} =  \frac 1 2 \big(|| 2, 1 \rangle \langle 2, 1 || +
|| 0 , 0 \rangle \langle 2, 1 ||
- || 2, 1 \rangle \langle 0, 0 || -  || 0 , 0 \rangle \langle 0, 0 || \big)
  .  \label{irrep}
\end{gather}
Applying the projection operators to the right-hand-side of (\ref{irrep}),
we have
\begin{gather*}
P^{(2)} e_1^{+ -} e_2^{- +} P^{(2)}
 = {\frac 1 2} ||2, 1 \rangle \langle 2, 1 ||
= {\frac 1 2 } E^{1,   1  (2)}  .
\end{gather*}
Similarly, we have
\begin{gather*}
e_1^{+ +} e_2^{- -} =  \frac 1 2 \big(|| 2, 1 \rangle \langle 2, 1 || +
|| 0 , 0 \rangle \langle 2, 1 ||
+ || 2, 1 \rangle \langle 0, 0 || +  || 0 , 0 \rangle \langle 0, 0 || \big)
  .
\end{gather*}
We thus have
\begin{gather*}
{P}^{(2)} e_1^{+ +} e_2^{- -} {P}^{(2)}
= {\frac 1 2} ||2, 1 \rangle \langle 2, 1 ||
= {\frac 1 2 } E^{1,   1   (2+)}   .
\end{gather*}
In terms of irreducible components, we have
\begin{gather*}
P^{(2)} e_1^{+ -} e_2^{- +} P^{(2)}
= P^{(2)} e_1^{- -} e_2^{+ +} P^{(2)}   .
\end{gather*}
We thus have
\begin{gather*}
P^{(2)} e_1^{- +} e_2^{+ -} P^{(2)} =
P^{(2)} e_1^{+ -} e_2^{- +} P^{(2)} =
P^{(2)} e_1^{+ +} e_2^{- -} P^{(2)}
= P^{(2)} e_1^{- -} e_2^{+ +} P^{(2)}  .
\end{gather*}

We shall evaluate the expectation values of spin-$s$ local operators
by reducing them into those of the spin-1/2 local operators.
Applying formula (\ref{eq:tildeE(p)-e}) to the case of $\varepsilon_1'=0$
and $\varepsilon_2'=1$, which correspond to $+$ and $-$,
respectively, we have
\begin{gather*}
\langle \psi_g^{(2)} |   E^{1,  1  (2)} | \psi_g^{(2)} \rangle =
\langle \psi_g^{(2;  0)}|   e_1^{+ -} e_2^{- +}    | \psi_g^{(2;  0)} \rangle + \langle \psi_g^{(2;  0)} |  e_1^{+ +} e_2^{- -}   | \psi_g^{(2;  0)} \rangle   .
\end{gather*}
Here we remark that
the vector $| \psi_g^{(2s;  0)} \rangle$ is given by
$| \psi_g^{(2s;  0)} \rangle =
\prod\limits_{\gamma=1}^{M} B^{(2s;  0)}(\lambda_{\gamma}) | 0 \rangle$,
while  the vector $| \psi_g^{(2s)} \rangle$ is given by
multiplying the projection operator:
$| \psi_g^{(2s)} \rangle = P^{(2s)}_{1 2 \cdots L}
\prod\limits_{\gamma=1}^{M} B^{(2s;  0)}(\lambda_{\gamma}) | 0 \rangle$.

\subsection{Spin inversion symmetry}\label{section4.2}

For even $L$ we may assume the spin inversion symmetry:
$U | \psi_g^{(2s  p;  0)} \rangle = \pm | \psi_g^{(2s  p;  0)} \rangle$ for $U=\prod\limits_{j=1}^{L} \sigma_j^{x}$.
Here we recall that associated with the ground state of the integrable spin-$s$ XXZ spin chain the vector $| \psi_g^{(2s  p;  0)} \rangle$ is given by
$| \psi_g^{(2s  p;  0)} \rangle =
\prod\limits_{\gamma=1}^{M} B^{(2s  p;  0)}(\lambda_{\gamma}) | 0 \rangle$.

We derive symmetry relations as follows \cite{DM4,DM5}
\begin{gather}
\langle \psi_g^{(2s   p;  0)} |
e_1^{\varepsilon_1', \varepsilon_1} \cdots
e_{2s}^{\varepsilon_{2s}', \varepsilon_{2s}}
| \psi_g^{(2s   p;   0)} \rangle
= \langle \psi_g^{(2s   p;   0)} |
e_1^{1- \varepsilon_1',   1 - \varepsilon_1} \cdots
e_{2s}^{ 1 - \varepsilon_{2s}',   1 - \varepsilon_{2s}}
| \psi_g^{(2s   p;   0)} \rangle   . \label{eq:spin-inv}
\end{gather}

Applying the spin-inversion symmetry (\ref{eq:spin-inv}) we derive symmetry
relations among the expectation values of local or global operators~\cite{DM4,DM5}.

For an illustration, let us evaluate the one-point function
in the spin-1 case with $i_1=j_1=1$,
$\langle E_1^{1,  1  (2  p)} \rangle$.
Setting $\varepsilon_1'=0$ and $\varepsilon_2'=1$ we decompose the
spin-1 elementary matrix in terms of a~sum of products of the spin-1/2 ones
\begin{gather*}
\langle \psi_g^{(2  p)} | E_1^{1,  1  (2  p)}
| \psi_g^{(2  p)} \rangle
=
\langle \psi_g^{(2  p ;  0)} | e_1^{0,  0} e_2^{1,  1}
| \psi_g^{(2  p ;  0)} \rangle
+ \langle \psi_g^{(2  p ;  0)} | e_1^{0,  1} e_2^{1,  0}
| \psi_g^{(2  p ;  0)} \rangle
 .
\end{gather*}
Through the symmetry relations (\ref{eq:sym-epsilon_a'})
with respect to $\varepsilon_{\alpha}'$
we have the following equalities:
\begin{gather*}
  \langle \psi_g^{(2  p ;  0)} | e^{0,  0}_{1} e_2^{1,  1}
| \psi_g^{(2  p ;  0)} \rangle
= \langle \psi_g^{(2 p ;  0)} | e^{1,  0}_{1}e_2^{0,  1}
| \psi_g^{(2  p ;  0)} \rangle
 , \non \\
 \langle \psi_g^{(2  p ;  0)} | e^{1,  1}_{1}e_2^{0,  0}
| \psi_g^{(2  p ;  0)} \rangle
= \langle \psi_g^{(2  p ;  0)} | e^{0,  1}_{1} e_2^{1,  0}
| \psi_g^{(2  p ;  0)} \rangle .
\end{gather*}
From spin-inversion symmetry (\ref{eq:spin-inv}) we have
\begin{gather*}
  \langle \psi_g^{(2  p ; 0))} | e^{0,  0}_{1}e_2^{1,  1}
| \psi_g^{(2  p ;  0))} \rangle
= \langle \psi_g^{(2  p ;  0)} | e^{1,  1}_{1}e_2^{0,  0} | \psi_g^{(2  p ;  0)} \rangle
 , \non \\
 \langle \psi_g^{(2,  p ;  0) )} | e^{0,  1}_{1}e_2^{1,  0}
| \psi_g^{(2  p ;  0)} \rangle
= \langle \psi_g^{(2  p ;  0)} | e^{1,  0}_{1}e_2^{0,  1} | \psi_g^{(2  p ;  0)} \rangle
\end{gather*}
and hence we have the equalities of the four terms.
We therefore obtain the following:
\begin{gather*}
\langle \psi_g^{(2)} | E_1^{1,  1  (2  p)} | \psi_g^{(2)} \rangle
= 2  \langle \psi_g^{(2  p ;  0)} | e_1^{0,  0} e_2^{1,  1} | \psi_g^{(2  p ;  0)} \rangle .
\end{gather*}
We thus derive the double-integral representation
of the one-point function $\langle E_1^{1,  1  (2  p)} \rangle$
of~\cite{DM2}, as we shall show in Section~\ref{section6}.

\subsection[Transformation from $V^{(2s)}$
to the $(2s+1)$-dimensional vector space ${\bf C}^{2s+1}$]{Transformation from $\boldsymbol{V^{(2s)}}$
to the $\boldsymbol{(2s+1)}$-dimensional vector space $\boldsymbol{{\bf C}^{2s+1}}$}\label{section4.3}

We shall express the spin-$s$ massless XXZ transfer matrix
as a $(2s+1)^{N_s} \times (2s+1)^{N_s}$ matrix acting on the
tensor product of the $(2s+1)$-dimensional
vector spaces ${\bf C}^{2s+1}$; i.e.,
acting on $({\bf C}^{2s+1})^{\otimes N_s}$.

In Section~\ref{section3.6} we have def\/ined the spin-$s$ XXZ transfer matrix through
the fusion method. It is expressed in terms of operators
def\/ined on the $L$th tensor product space of the
spin-1/2 representations, $(V^{(1)})^{\otimes L}$, and given by a
$2^{L} \times 2^{L}$ matrix.
We have constructed them by applying the projection operators to
the spin-1/2 XXZ transfer matrix with inhomogeneous parameters
given by complete strings $w_j^{(2s)}$.

We now formulate the spin-$s$ XXZ transfer matrix
in terms of the basis of the $(2s+1)$-dimensional
vector space ${\bf C}^{2s+1}$ such as
$|2s,   m))$ for $m=0, 1, \ldots, 2s$.
As the basis vectors of the $(2s+1)$-dimensional representation of $U_q(sl_2)$
we introduce vectors $| 2s, m \rangle$ with the following normalization:
\begin{gather*}
| 2s, m \rangle = || 2s, m \rangle / \sqrt{
\left(
\begin{matrix}
2s \\
m
\end{matrix}
\right)} \qquad \mbox{for} \quad m=0, 1, \ldots, 2s .
\end{gather*}
We denote by $\langle 2s, m |$ the transposition of $| 2s, m \rangle $
\begin{gather*}
\langle 2s, m | = \left( | 2s, m \rangle \right)^{t} \qquad \mbox{for} \quad
m=0, 1, \ldots, 2s.
\end{gather*}
Let us denote the complex conjugate of a complex number $z$ by ${\bar z}$.
We express the Hermitian conjugate of a vector $| 2s, m \rangle$ by
\begin{gather*}
\overline{\langle 2s, m |} = \left( | 2s, m \rangle \right)^{\dagger}
\qquad \mbox{for} \quad m=0, 1, \ldots, 2s.
\end{gather*}

Let us introduce the transformation $S$:
$V^{(2s)} \rightarrow {\bf C}^{2s+1}$. We def\/ine it by
\begin{gather*}
S = \sum_{m=0}^{2s} | 2s, m))   \overline{\langle 2s, m |}   .
\end{gather*}
It maps the basis of the spin-$s$ representation
$V^{(2s)}$ constructed in the tensor product space
$V_1^{(1)} \otimes \cdots \otimes V_{2s}^{(1)}$; i.e., $|2s, m \rangle$ for
$m=0, 1, \ldots, 2s$,
to that of the ($2s+1$)-dimensional representation~${\bf C}^{2s+1}$; i.e., $|2s, m))$ for $m=0, 1, \ldots, 2s$.
We can show the following relations:
\begin{gather}
\overline{S}   \widetilde{E}^{i,   i   (2s   p)}
\overline{ S}^{\dagger} = E^{i,   j} \qquad \mbox{for} \quad
i, j = 0, 1, \ldots, 2s.   \label{eq:tildeE(p)-E}
\end{gather}
Here we recall that $E^{i,   j}$ denote the $(2s+1)$-by-$(2s+1)$
unit matrices which have only one nonzero element 1 at the entry of $(i, j)$
for $i, j = 0, 1, \ldots, 2s$.

For an illustration, let us consider the spin-1 case.
The basis vectors of ${\bf C}^3$ are given by
\begin{gather*}
|2,   0))=(1, 0, 0)^{t}, \qquad |2,   1 ))=(0, 1, 0)^{t}, \qquad
|2,   2))=(0, 0, 1)^{t}.
\end{gather*}
In the spin-1 case the transformation $S$:
$V^{(2)} \rightarrow {\bf C}^{3}$ is given by
\begin{gather*}
S = \sum_{m=0}^{2} | 2, m))   \overline{\langle 2, m |}   .
\end{gather*}
In the massless regime where $q$ is complex with $|q|=1$,
explicitly we have
\begin{gather*}
S= \left(
\begin{array}{cccc}
1 & 0 & 0 & 0 \\
0 & {\frac 1 {\sqrt{2}}} & {\frac q {\sqrt{2}}}  & 0 \\
0 & 0 & 0 & 1
\end{array}
\right)   .
\end{gather*}
Taking the Hermitian conjugate of $S$ we have
\begin{gather*}
S^{\dagger}
= \sum_{m=0}^{2} | 2, m \rangle ((2, m|
= \left(
\begin{array}{ccc}
1 & 0 & 0  \\
0 & {\frac 1 {\sqrt{2}}} & 0 \vspace{1mm}\\
0 & {\frac {1} {\sqrt{2} q} }  & 0 \\
0 & 0 & 1
\end{array}
\right)   .
\end{gather*}
 It is straightforward to show the following:
\begin{gather*}
S S^{\dagger} =
\left(
\begin{array}{ccc}
1 & 0 & 0 \\
0 & 1 & 0 \\
0 & 0 & 1
\end{array}
\right)   ,
\qquad
S^{\dagger} S =
\left(
\begin{array}{cccc}
1 & 0 & 0 & 0 \\
0 & {\frac 1 2} & {\frac q 2} & 0 \vspace{1mm} \\
0 & {\frac 1 {2q} } & {\frac 1 2} & 0 \\
0 & 0 & 0 & 1
\end{array}
\right)   .
\end{gather*}
In terms of the bras and kets we have
\begin{gather*}
S S^{\dagger}   =  \sum_{m=0}^{2} | 2, m))   \overline{\langle 2, m |}
\sum_{n=0}^{2} | 2, n \rangle ((2, n|  \\
\hphantom{S S^{\dagger}}{}
  =   \sum_{m=0}^{2}  \sum_{n=0}^{2} \delta(m, n)   | 2, m))  ((2, n|
  =   \sum_{m=0}^{2}  | 2, m))   ((2, m|   .
\end{gather*}
Similarly, we have
\begin{gather*}
S^{\dagger} S = \sum_{m=0}^{2} | 2, m \rangle   \overline{ \langle 2, m |}   .
\end{gather*}

In order to transform the conjugate vectors $\widetilde{||2,  m \rangle }$
it is also useful to introduce the complex conjugates
of transformations $S$ and
$S^{\dagger}$:
\begin{gather*}
\overline{ S} = \sum_{m=0}^{2} | 2, m))   \langle 2, m |
=
\left(
\begin{array}{cccc}
1 & 0 & 0 & 0 \\
0 & {\frac 1 {\sqrt{2}}} & {\frac 1 {\sqrt{2} q}}  & 0 \\
0 & 0 & 0 & 1
\end{array}
\right)   ,
\\
\overline{ S}^{\dagger}
= \sum_{m=0}^{2} \overline{ | 2, m \rangle}    ((2, m|
= \left(
\begin{array}{ccc}
1 & 0 & 0  \\
0 & {\frac 1 {\sqrt{2}}} & 0 \vspace{1mm}\\
0 & {\frac {q} {\sqrt{2}} }  & 0 \\
0 & 0 & 1
\end{array}
\right)  .
\end{gather*}
They are related to the projection operator $\widetilde{P}^{(2)}$.
We have
\begin{gather*}
\overline{ S}^{\dagger} \overline{S }
  =
\left(
\begin{array}{cccc}
1 & 0 & 0 & 0 \\
0 & {\frac 1 2} & {\frac 1 {2 q}} & 0  \vspace{1mm}\\
0 & {\frac q {2} } & {\frac 1 2} & 0 \\
0 & 0 & 0 & 1
\end{array}
\right)
  =   \sum_{m=0}^{2} \overline{ | 2, m \rangle}    \langle 2, m |
 = \widetilde{ P}^{(2)}   .
\end{gather*}

The spin-1 elementary matrices $\widetilde{E}^{i,  j  (2   p)}$
are transformed into the $3 \times 3$ unit matrices $E^{i, j}$ as
\begin{gather*}
\overline{S}   \widetilde{E}^{i,  i  ( 2   p)}
\overline{ S}^{\dagger} = E^{i, j} \qquad \mbox{for} \quad
i, j = 0, 1, 2.
\end{gather*}
For instance we have
\begin{gather*}
\overline{S}    \widetilde{E}^{1, 1   ( 2    p)}
\overline{ S}^{\dagger} = E^{1, 1} =
\left(
\begin{array}{ccc}
0 & 0 & 0 \\
0 & 1 & 0 \\
0 & 0 & 0
\end{array}
\right)   .
\end{gather*}
We have thus conf\/irmed relations (\ref{eq:tildeE(p)-E}).

Let us introduce the transformation which
maps the tensor product of
the spin-$s$ representations:
$V_1^{(2s)} \otimes \cdots \otimes V_{N_s}^{(2s)}$
to the tensor product of the $(2s+1)$-dimensional representations:
${\bf C}_1^{2s+1} \otimes \cdots \otimes {\bf C}_{N_s}^{2s+1}$.
We def\/ine it by the tensor product of transformation $S$ as follows
\begin{gather*}
S_1 \otimes \cdots \otimes S_{N_s}: \
V_1^{(2s)} \otimes \cdots \otimes V_{N_s}^{(2s)} \rightarrow
 {\bf C}_1^{2s+1} \otimes \cdots \otimes {\bf C}_{N_s}^{2s+1}   .
\end{gather*}
We also def\/ine its complex conjugate
\begin{gather*}
\overline{S}_1 \otimes \cdots \otimes \overline{S}_{N_s}: \
V_1^{(2s)} \otimes \cdots \otimes V_{N_s}^{(2s)} \rightarrow
 {\bf C}_1^{2s+1} \otimes \cdots \otimes {\bf C}_{N_s}^{2s+1}  .
\end{gather*}
Let us consider the spin-$s$ ground state with $(2s+1)$-dimensional entries,
${| \Psi_G^{(2s)} \rangle}$. For the spin-1 case, it gives
the ground state of the spin-1 XXZ Hamiltonian~(\ref{eq:spin1-XXZ-Hamiltonian}). In terms of the ground state constructed by
the fusion method,  ${| \psi_g^{(2s   p)} \rangle}$,
it is given by
\begin{gather*}
{| \Psi_G^{(2s)} \rangle} = \overline{S}_1  \otimes \cdots \otimes \overline{S}_{N_s} {| \psi_g^{(2s   p)} \rangle}   .
\end{gather*}
Here we recall that ${| \psi_g^{(2s  p)} \rangle}$ denotes the
ground state of the integrable spin-$s$ XXZ spin chain constructed
through the fusion method,
where the evaluation representations are associated with principal grading.
In terms of the eigenvector with $(2s+1)$-dimensional entries,
the expectation value of a given local
operator $E$ with $(2s+1)$-dimensional entries is given by
\begin{gather*}
\langle \Psi_G^{(2s)} |  E  | \Psi_G^{(2s)} \rangle
= \langle \psi_g^{(2s  p)} |
\overline{S}_1^{\dagger}
 \otimes \cdots \otimes \overline{S}_{N_s}^{\dagger}
 E
\overline{S}_1 \otimes \cdots \otimes \overline{S}_{N_s}
 | \psi_g^{(2s  p)} \rangle .
\end{gather*}
Therefore, the operator $E$ corresponds to the operator $E^{(2s  p)}$
in the fusion construction as follows
\begin{gather*}
E^{(2s  p)} = \overline{S}_1^{\dagger}
 \otimes \cdots \otimes \overline{S}_{N_s}^{\dagger}   E
\overline{S}_1 \otimes \cdots \otimes \overline{S}_{N_s}   .
\end{gather*}
For instance, from (\ref{eq:tildeE(p)-E}) we have the following:
\begin{gather*}
 \widetilde{E}^{i,  i  (2s  p)}
 = \overline{ S}^{\dagger} E^{i,  j} \overline{S}
\qquad \mbox{for} \quad
i, j = 0, 1, \ldots, 2s.
\end{gather*}
Similarly, we have the following relations for
the spin-$s$ XXZ transfer matrices def\/ined
as $(2s+1)^{N_s} \times (2s+1)^{N_s}$ matrices
${t}^{(\ell,  2s)}_{1 2 \cdots N_s}$,
to those of the fusion construction:
\begin{gather*}
\widetilde{t}^{(\ell,  2s  p)}_{1 2 \cdots N_s}
(\lambda) = \overline{S}_1^{\dagger}
 \otimes \cdots \otimes \overline{S}_{N_s}^{\dagger}
{t}^{(\ell,   2s)}_{1 2 \cdots N_s}
(\lambda)
\overline{S}_1 \otimes \cdots \otimes \overline{S}_{N_s}  \qquad
\mbox{for} \quad \ell=1, 2, \ldots, 2s.
\end{gather*}

\section[Multiple-integral representations for spin-$s$ case]{Multiple-integral representations for spin-$\boldsymbol{s}$ case}\label{section5}

We introduce some useful symbols for expressing the correlation
functions of the integrable spin-$s$ XXZ spin chain. We derive
the multiple-integral representation of the spin-$s$ correlation
functions by mainly following the procedures
of~\cite{DM2} except for the formula of reducing the higher-spin
form factors into the spin-1/2 scalar products such as in Corollary~\ref{cor:main}.

Let us sketch the main procedures for deriving the multiple-integral representation of the spin-$s$ XXZ correlation functions.
First, we introduce the spin-$s$ elementary operators
as the basic blocks for constructing the local operators
of the integrable spin-$s$ XXZ spin chain.
Secondly, we reduce them into a sum of products of the spin-1/2
elementary operators, which we express
through the spin-1/2 QISP formula
in terms of the matrix elements of the spin-1/2 monodromy matrix,
and evaluate their scalar products
through Slavnov's formula of the Bethe-ansatz scalar products.
Here, the expectation value of a physical quantity is expressed
as a sum of the ratios of the Bethe-ansatz scalar products
to the norm of the Bethe-ansatz eigenvector. Furthermore,
the ratios are expressed in terms of the determinants of some matrices.
Thirdly, solving the integral equations for the matrices
in the thermodynamic limit, we derive the multiple-integral
representation of the correlation functions.

Let us summarize the multiple-integral representations of
correlation functions for the integrable spin-$s$ XXZ spin chain
in a region of the massless regime with $0 \le \zeta < \pi/2s$~\cite{DM2}.
We show the revised expression~\cite{DM4,DM5}.
Here we recall that
in the massless regime we set $\eta= i \zeta$ with $0 \le \zeta < \pi$.

We express any correlation function of the integrable spin-$s$ XXZ chain
in terms of the linear combination of products of the spin-$s$
elementary matrices. They are  def\/ined by
\begin{gather*}
F_m^{(2s  p)}(\{i_k, j_k \}) =  \bra \psi_g^{(2s  p)} |
\prod_{k=1}^{m} \widetilde{E}_k^{i_k ,  j_k  (2s  p)}
|\psi_g^{(2s  p)}  \ket / \bra \psi_g^{(2s  p)}
| \psi_g^{(2s  p)} \ket  ,
\end{gather*}
where $\widetilde{E}_k^{i_k,  j_k  (2s  p)}$
denotes the $(2s+1) \times (2s+1)$ elementary matrix
whose entries are all zero except for the
$(i_k, j_k)$ element which is given by $1$,
for each $k$ with $1 \le k \le m$. Here integers
$i_k$ and $j_k$ satisfy $1 \le i_k , j_k \le 2s$.
We recall that $| {\psi}_g^{(2s  p)} \ket$
denotes the spin-$s$ ground state.

Let us consider a product of the spin-$\ell/2$ elementary matrices associated
with principal grading,
 ${E}_1^{i_1 ,  j_1  (\ell  p)}
\cdots {E}_m^{i_m,  j_m  (\ell  p)}$, for which we shall
evaluate the zero-temperature spin-$s$ XXZ correlation functions.
We introduce variables $\varepsilon_{\alpha}^{[k]'}$
and $\varepsilon_{\beta}^{[k]}$ which take only two values 0 or 1
for $k=1, 2, \ldots, m$ and $\alpha, \beta=0, 1, \ldots, \ell$.
For the $m$th product of elementary matrices,
we introduce sets of $\varepsilon_{\alpha}^{[k] '}$s
and $\varepsilon_{\beta}^{[k]}$s ($1 \le k \le m$) such that
the number of $\alpha$s satisfying $\varepsilon_{\alpha}^{[k]'}=1$ and
$1 \le \alpha \le 2s$ is given by $i_k$
and the number of $\beta$s satisfying $\varepsilon_{\beta}^{[k]}=1$ and
$1 \le \beta \le 2s$ by $j_k$, respectively.
We then express them by integers $\varepsilon_j'$s and $\varepsilon_j$s for
$j=1, 2, \ldots, 2sm$ as follows:
\begin{gather*}
\varepsilon_{2s (k-1)+ \alpha}'    =   \varepsilon_{\alpha}^{[k]'} \qquad \mbox{for} \quad \alpha=1, 2, \ldots, 2s;  \quad k=1, 2, \ldots, m,  \nonumber \\
\varepsilon_{2s (k-1)+ \beta}   =   \varepsilon_{\beta}^{[k]} \quad
\mbox{for} \quad
\beta=1, 2, \ldots, 2s;  \quad k=1, 2, \ldots, m.
\end{gather*}
We express the $m$th product of $(2s+1) \times (2s+1)$
elementary matrices
in terms of a sum of $2sm$th products of the $2 \times 2$ elementary matrices
with entries $\{ \epsilon_j, \epsilon_j' \}$; i.e.,
 $e_{1}^{\varepsilon_1', \varepsilon_1} \cdots
e_{2sm}^{\varepsilon_{2sm}', \varepsilon_{2sm}}$ \cite{DM2,DM5}.

For given sets of $\varepsilon_j$ and $\varepsilon_j'$ for
$j=1, 2, \ldots, 2sm$ we def\/ine
$\mbox{\boldmath$\alpha$}^{-}$ by the set of integers $j$
satisfying $\varepsilon_j'=1$ and
$\mbox{\boldmath$\alpha$}^{+}$ by
the set of integers $j$ satisfying $\varepsilon_j=0$:
\begin{gather*}
\mbox{\boldmath$\alpha$}^{-}(\{ \varepsilon_j' \})
= \{ j ;   \varepsilon_j'=1 \}
  , \qquad
\mbox{\boldmath$\alpha$}^{+}(\{ \varepsilon_j \})
= \{ j ;   \varepsilon_j=0 \}    .
\end{gather*}
We denote by $\alpha_{-}$ and $\alpha_{+}$
the number of elements of the set $\mbox{\boldmath$\alpha$}^{-}$
and $\mbox{\boldmath$\alpha$}^{+}$, respectively.
Due to the ``charge conservation'', we have
\begin{gather}
\alpha_{-} + \alpha_{+} = 2sm. \label{eq:charge}
\end{gather}
Precisely, we have $\alpha_{-}= \sum\limits_{k=1}^{m} i_k$ and
$\alpha_{+}= 2sm - \sum\limits_{k=1}^{m} j_k$.
Here we recall that for the $R$-matrix of the XXZ spin chain
matrix elements $R(u)^{a  b}_{c  d}$ vanish if $a+b \ne c+d$,
which we call the charge conservation.
It follows from the charge conservation that the correlation function
$F_m^{(2s  p)}(\{ \varepsilon_j, \varepsilon_j' \})$ vanishes unless
the two sums are equal: $\sum\limits_{k=1}^{m} i_k = \sum\limits_{k=1}^{m} j_k$.
We therefore obtain relation~(\ref{eq:charge}).
We remark that the charge conservation of the $R$-matrix
corresponds to the ``ice rule'' of the six-vertex model, which is
def\/ined as a two-dimensional ferro-electric lattice model.

For sets  ${\bm \alpha}^{-}$ and ${\bm \alpha}^{+}$
we def\/ine ${\tilde \lambda}_j$ for $j \in {\bm \alpha}^{-}$
and ${\tilde \lambda}'_{j'}$ for $j' \in {\bm \alpha}^{+}$,
by the following sequence:
\begin{gather*}
({\tilde \lambda}'_{j'_{\max}}, \dots,
{\tilde \lambda}'_{j'_{\min}},  {\tilde \lambda}_{j_{\min}}, \dots,
{\tilde \lambda}_{j_{\max}})
=(\lambda_1, \ldots, \lambda_{2sm})   .
\end{gather*}

Let us recall the  assumption
that in the region $0 \le \zeta < \pi/2s$
the spin-$s$ ground state $| \psi_g^{(2s  p)} \rangle$ is
given by $N_s/2$ sets of the $2s$-strings:
\begin{gather*}
\lambda_{a}^{(\alpha)}
= \mu_a - (\alpha- 1/2) \eta + \delta_a^{(\alpha)}   \qquad
\mbox{for} \quad a=1, 2, \ldots, N_s/2 \quad
\mbox{and} \quad  \alpha = 1, 2, \ldots, 2s .
\end{gather*}
Here we also assume that string deviations $\delta_a^{(\alpha)}$ are
very small for large $N_s$.
In terms of rapidities forming strings, $\lambda_{a}^{(\alpha)}$,
the spin-$s$ ground state associated with the principal grading
is given by
\begin{gather*}
 | \psi_g^{(2s   p)} \rangle =
\prod_{a=1}^{N_s/2} \prod_{\alpha=1}^{2s}
\widetilde{B}^{(2s   p)}
(\lambda_a^{(\alpha)}; \{\xi_b \}_{N_s}) | 0 \rangle .
\end{gather*}
Here we have $M$ Bethe roots with $M= 2s  N_s/2 = s N_s$.
The density of string centers, $\rho(\lam)$, is given by
\begin{gather*}
\rho(\lam)=
{\frac 1 {2 \zeta \cosh(\pi \lam/\zeta)}},
\end{gather*}
which has simple poles at $\lam=\i\zeta(n+1/2)$, for $n\in\mathbb{Z}$
with the residues $(-1)^n/(2\pi \i)$.

We def\/ine the $(j,k)$ element of a matrix
$S=S\big( (\lambda_j)_{2sm}; (w_j^{(2s)})_{2sm} \big)$ by
\begin{gather*}
S_{j,k} = \rho\big(\lambda_j - w_k^{(2s)} + \eta/2\big)
\delta(\alpha(\lambda_j), \beta(k))  \qquad {\rm for} \quad
j, k= 1, 2, \ldots, 2sm   .
\end{gather*}
Here $\delta(\alpha, \beta)$ denotes the Kronecker delta.
We def\/ine $\beta(j)$ by
\begin{gather*}
\beta(j)=j-2s[[(j-1)/2s]],
\end{gather*}
where the Gauss symbol $[[x]]$ is def\/ined by the greatest integer less than
or equal to a real num\-ber~$x$.
We def\/ine
 $\alpha(\lambda_j)$ by $\alpha(\lambda_j)= \gamma$ ($1\leq\gamma\leq 2s$)
if $\lambda_j$ is related to the integral variable $\mu_j$ by
$\lambda_j=\mu_j - (\gamma - 1/2) \eta$, or $\lambda_j \approx w_{k}^{(2s)}$
where $\beta(k)=\gamma$ ($1 \le \gamma \le 2s$)~\cite{DM2}.
We remark that $\mu_j$ correspond to the centers
of complete $2s$-strings $\lambda_j$.
When we evaluate $\alpha(\lam_j)$,
we assume that the integral paths of
$\int_{-\infty-\i(\gamma-1)\zeta \pm \i \eps}^{\infty-\i(\gamma-1)\zeta \pm \i \eps}$ are replaced by those of
$\int_{-\infty-\i(\gamma-1/2)\zeta}^{\infty-\i(\gamma-1/2)\zeta}$ for
$\gamma=1,2,\dots,2s$, respectively. Here we remark that when
we deform the integral path of
$\int_{-\infty-\i(\gamma-1)\zeta + \i \eps}^{\infty-\i(\gamma-1)\zeta + \i \eps}$ to that of
$\int_{-\infty-\i(\gamma-1/2)\zeta}^{\infty-\i(\gamma-1/2)\zeta}$
(for $\gamma = 1, 2, \dots, 2s$),
we may have the contribution of a simple pole at $\lambda= w_k^{(2s)}$ with
integer $k$ satisfying $\beta(k)=\gamma$.

With the above notations,
we express correlation functions for the massless spin-$s$ XXZ chain
in the form of multiple integrals as follows
\begin{gather}
F_m^{(2s  p)}(\{ \varepsilon_j, \varepsilon_j' \}) =
C(\{i_k, j_k \})
\left( \int_{-\infty+ \i \epsilon}^{\infty+ \i \epsilon}
+ \int_{-\infty - \i \zeta + \i \epsilon}^{\infty - \i \zeta  + \i \epsilon}
+ \cdots
+ \int_{-\infty - i(2s-1) \zeta + \i \epsilon}
^{\infty - i(2s-1) \zeta  + \i \epsilon} \right)  \d \lambda_1
\cdots\nonumber\\
\phantom{F_m^{(2s  p)}(\{ \varepsilon_j, \varepsilon_j' \}) =}{}
\times \left( \int_{-\infty+ \i \epsilon}^{\infty+ \i \epsilon}
+ \cdots
+ \int_{-\infty - i(2s-1) \zeta + \i \epsilon}^{\infty - i(2s-1) \zeta  + \i \epsilon}
\right)  \d \lambda_{\alpha_+}
\nn \\
\phantom{F_m^{(2s  p)}(\{ \varepsilon_j, \varepsilon_j' \}) =}{}
\times
\left( \int_{-\infty - \i \epsilon}^{\infty - \i \epsilon}
+ \int_{-\infty - \i \zeta - \i \epsilon}^{\infty - \i \zeta  - \i \epsilon}
+ \cdots
+ \int_{-\infty - i(2s-1) \zeta - \i \epsilon}
^{\infty - i(2s-1) \zeta  - \i \epsilon} \right)  \d \lambda_{\alpha_+ + 1}
\cdots
\nn \\
\phantom{F_m^{(2s  p)}(\{ \varepsilon_j, \varepsilon_j' \}) =}{}
\times
\left( \int_{-\infty - \i \epsilon}^{\infty - \i \epsilon}
+ \cdots
+ \int_{-\infty - i(2s-1) \zeta - \i \epsilon}
^{\infty - i(2s-1) \zeta  - \i \epsilon}
\right)  \d \lambda_{2sm}
\nn \\
\phantom{F_m^{(2s  p)}(\{ \varepsilon_j, \varepsilon_j' \}) =}{}
\times
  \sum_{{\bm \alpha}^{+}(\{ \epsilon_j \})}
 Q(\{ \varepsilon_j, \varepsilon_j' \};
\lambda_1, \ldots, \lambda_{2sm})
  {\rm det} S(\lambda_1, \ldots, \lambda_{2sm})  .
\label{eq:Multiple-integral_represenation}
\end{gather}
Here we have def\/ined
$Q(\{ \varepsilon_j, \varepsilon_j' \};
\lambda_1, \ldots, \lambda_{2sm})$ by
\begin{gather}
Q(\{ \varepsilon_j, \varepsilon_j' \};
\lambda_1, \ldots, \lambda_{2sm})) \nonumber\\
\qquad{}  =
(-1)^{\alpha_{+}}
{\frac { \prod\limits_{j \in {\bm \alpha}^{-}} \left( \prod\limits_{k=1}^{j-1}
\sh\big({\tilde \lambda}_{j} - w_k^{(2s)} + \eta\big)
\prod\limits_{k=j+1}^{2sm} \sh\big({\tilde \lambda}_{j} - w_k^{(2s)} \big) \right)}
{\prod\limits_{1 \le k < \ell \le 2sm}
\sh(\lambda_{\ell} - \lambda_{k} + \eta + \epsilon_{\ell, k})} } \nn \\
  \qquad\quad{} \times
{\frac { \prod\limits_{j' \in {\bm \alpha}^{+}} \left( \prod\limits_{k=1}^{j'-1}
\sh\big({\tilde \lambda}'_{j'} - w_k^{(2s)} - \eta\big)
\prod\limits_{k=j'+1}^{2sm} \sh\big({\tilde \lambda}'_{j'} - w_k^{(2s)} \big) \right)}
{\prod\limits_{1 \le k < \ell \le 2sm}
\sh\big(w_{k}^{(2s)} - w_{\ell}^{(2s)}\big)} }  . \label{eq:Q}
\end{gather}
Here we have set $\epsilon_{k, \ell}=\i\eps$
for ${\rm Im}(\lam_k-\lam_\ell)>0$
and $\epsilon_{k, \ell}=-\i\eps$ for ${\rm Im}(\lam_k-\lam_\ell)<0$,
where $\eps$ is an inf\/initesimally small positive real number:
$0 < \eps \ll 1$. The normalization factor $C$ is given by
\begin{gather*}
C(\{i_k, j_k \}) =
\prod_{k=1}^{m} \left( \sqrt{ F(\ell, i_k)/F(\ell, j_k)}
q^{i_k(\ell-i_k)/2-j_k(\ell-j_k)/2} \right)
=
\prod_{b=1}^m \sqrt{
\begin{bmatrix}2s\\i_b\end{bmatrix}_q
\begin{bmatrix}2s\\j_b\end{bmatrix}_q^{-1}} . 
\end{gather*}
where $q=e^\eta=e^{\i\zeta}$.

Here we should remark that in~(\ref{eq:Multiple-integral_represenation})
the sum of ${\bm \alpha}^{+}(\{ \varepsilon_j \})$
is taken over all sets $\{ \varepsilon_j \}$ corresponding to
$\{ \varepsilon_{\beta}^{[k]} \}$ $(1 \le k \le m)$ such that the number
of integers $\beta$ satisfying $\varepsilon_{\beta}^{[k]}=1$
with $1 \le \beta \le 2s$ is given by $j_k$ for each $k$ satisfying
$1 \le k \le m$.
In (\ref{eq:Q}) we take a set ${\bm \alpha}^{-}(\{ \varepsilon_j' \})$
corresponding to $\varepsilon_{\alpha}^{[k]  '}$
for $k=1, 2, \ldots, m$, where the number of integers $\alpha$ satisfying
$\varepsilon_{\alpha}^{[k]  '}=1$ and $1 \le {\alpha} \le 2s$
is given by $i_k$ for each $k$ ($1 \le k \le m$).

\section[Multiple integrals of the spin-1 one-point functions ($s=1$)]{Multiple integrals of the spin-1 one-point functions ($\boldsymbol{s=1}$)}\label{section6}

We calculate analytically
the integrals for the spin-1 one-point functions.
Considering the residues which are derived
when we shift the integral paths,
we explicitly evaluate the double integrals expressing the spin-1
one-point functions.
Hereafter, we shall often denote the spin-$s$
elementary matrices $\widetilde{ E}^{i,  j  (2s  p)}$
 by $E^{i,  j}$ for simplicity.

\subsection[$\bra E^{ 2 2} \ket$: The emptiness formation probability]{$\boldsymbol{\bra E^{ 2 2} \ket}$: The emptiness formation probability}\label{section6.1}

Let us evaluate the emptiness formation probability (EFP)
$\bra \widetilde{E}^{2,   2  (2  p)} \ket$.
In this case we have
\begin{gather*}
i_1=j_1=2;  \qquad (\veps_1, \veps_2)=(1, 1), \qquad
(\veps'_1, \veps'_2)=(1, 1); \qquad C=1;
\nn\\
{\bm \alpha}^{+}=\varnothing, \qquad {\bm \alpha}^{-}=\{1,2\}; \qquad
(\tilde{\lam}_1,\tilde{\lam}_2)=(\lam_1,\lam_2).
\end{gather*}
Here the symbol $\varnothing$ denotes the empty set.  We evaluate EFP as follows
\begin{gather*}
\langle \widetilde{E}_1^{2,   2   (2   p)} \rangle
 =
\phi_2\big(\{ \lambda_{\gamma} \}; \big\{w_j^{(2)} \big\}_L\big)
\Bigg(
\lim_{\epsilon \rightarrow 0}
{\frac {\langle {\psi}_g^{(2  p;  \epsilon)} |
D^{(2  p ;  \epsilon)}\big(w_1^{(2;  \epsilon)}\big)
D^{(2  p ;  \epsilon)}\big(w_2^{(2;  \epsilon)}\big)
 | {\psi}_g^{(2  p;  \epsilon)} \rangle}
{\langle {\psi}_g^{(2  p;  \epsilon)}
| {\psi}_g^{(2  p;  \epsilon)} \rangle}   }
 \Bigg)
 .
\end{gather*}

Let us denote  the integral path
$\int_{-\infty+\i\alpha}^{\infty+\i\alpha}$ by $\int_{C_{\i\alpha}}$.
The multiple-integral formula reads
\begin{gather*}
\bra E^{22} \ket=
\left( \int_{C_{-\i\eps}}+\int_{C_{-\eta-\i\eps}}
\right)\d\lam_1
\left( \int_{C_{-\i\eps}}+\int_{C_{-\eta-\i\eps}}
\right)\d\lam_2
Q(\lam_1,\lam_2)\det S(\lam_1,\lam_2),
\end{gather*}
where $Q(\lam_1,\lam_2)$ and $S(\lam_1,\lam_2)$ are
expressed in terms of $\vp(x)=\sh(x)$ as
\begin{gather*}
Q(\lam_1,\lam_2)
=\frac{\vp(\lam_1-w^{(2)}_2)\vp(\lam_2-w^{(2)}_1+\eta)}
{\vp(\lam_2-\lam_1+\eta+\eps_{21})\vp(w^{(2)}_1-w^{(2)}_2)}, \\
S(\lam_1,\lam_2)=
\begin{pmatrix}
\rho(\lam_1-w^{(2)}_1+\eta/2)\,\delta_{\alpha(\lam_1),1}
& \rho(\lam_1-w^{(2)}_2+\eta/2) \,\delta_{\alpha(\lam_1),2} \\
\rho(\lam_2-w^{(2)}_1+\eta/2) \,\delta_{\alpha(\lam_2),1}
& \rho(\lam_2-w^{(2)}_2+\eta/2) \,\delta_{\alpha(\lam_2),2}
\end{pmatrix}   .
\end{gather*}
We now shift the integral paths $C_{-\i\eps}$ and $C_{-\eta-\i\eps}$
into $C_{-\eta/2}$ and $C_{-3\eta/2}$, respectively.
During the contour deformation
each of the integral paths does not cross any pole of the integrand,
and hence we have
\begin{gather*}
\bra E^{22} \ket=
\left( \int_{C_{-\eta/2}}+\int_{C_{-3\eta/2}}
\right)\d\lam_1
\left( \int_{C_{-\eta/2}}+\int_{C_{-3\eta/2}}
\right)\d\lam_2
Q(\lam_1,\lam_2)\det S(\lam_1,\lam_2) .
\end{gather*}
We now denote
$C_{-\eta/2}$ and $C_{-3\eta/2}$ by $C_1$ and $C_2$, respectively.
After expanding the above expression with respect to
the types of integral paths, we have four terms. However,
only two of them survive due to the Kronecker deltas in the matrix $S$
\begin{gather*}
\bra E^{22} \ket =
\int_{C_1}\d\lam_1\int_{C_1}\d\lam_2 Q(\lam_1,\lam_2)
\left|\begin{matrix}
\rho(\lam_1-w^{(2)}_1+\eta/2) & \hspace{11.8mm} 0  \\
\rho(\lam_2-w^{(2)}_1+\eta/2) & \hspace{11.8mm} 0
\end{matrix}\hspace{13mm}\right|
\nn\\
\phantom{\bra E^{22} \ket =}{}+
\int_{C_1}\d\lam_1\int_{C_2}\d\lam_2 Q(\lam_1,\lam_2)
\left|\begin{matrix}
\rho(\lam_1-w^{(2)}_1+\eta/2) & 0  \\
0 & \rho(\lam_2-w^{(2)}_2+\eta/2)
\end{matrix}\right|
\nn\\
\phantom{\bra E^{22} \ket =}{}+
\int_{C_2}\d\lam_1\int_{C_1}\d\lam_2 Q(\lam_1,\lam_2)
\left|\begin{matrix}
0 & \rho(\lam_1-w^{(2)}_2+\eta/2)  \\
\rho(\lam_2-w^{(2)}_1+\eta/2) & 0
\end{matrix}\right|
\nn\\
\phantom{\bra E^{22} \ket =}{}+
\int_{C_2}\d\lam_1\int_{C_2}\d\lam_2 Q(\lam_1,\lam_2)
\left|\hspace{13mm}\begin{matrix}
0 & \hspace{12mm}\rho(\lam_1-w^{(2)}_2+\eta/2) \\
0 & \hspace{12mm}\rho(\lam_2-w^{(2)}_2+\eta/2)
\end{matrix}\right|
\nn\\
\phantom{\bra E^{22} \ket}{}=
\int_{C_1}\d\lam_1\int_{C_2}\d\lam_2 Q(\lam_1,\lam_2)
\rho(\lam_1-w^{(2)}_1+\eta/2)\rho(\lam_2-w^{(2)}_2+\eta/2)
\nn\\
\phantom{\bra E^{22} \ket =}{}-
\int_{C_2}\d\lam_1\int_{C_1}\d\lam_2 Q(\lam_1,\lam_2)
\rho(\lam_1-w^{(2)}_2+\eta/2)\rho(\lam_2-w^{(2)}_1+\eta/2),
\end{gather*}
Substituting $w^{(2)}_1=\xi_1$, $w^{(2)}_2=\xi_1-\eta$, we have
\begin{gather*}
\bra E^{22} \ket=\intall\d\mu_1\intall\d\mu_2
(Q_{12}-Q_{21})\rho(\mu_1-\xi_1)\rho(\mu_2-\xi_1)\nn\\
\phantom{\bra E^{22} \ket}{} =\intall\d x_1\intall\d x_2
(Q_{12}-Q_{21})\rho(x_1)\rho(x_2)
\end{gather*}
where $x_1=\mu_1- \xi_1$, $x_2=\mu_2- \xi_1$, and $Q_{12}$ and $Q_{21}$
are given by
\begin{gather}
Q_{12}=Q(\mu_1-\eta/2, \mu_2-3\eta/2) =\frac1{\vp(\eta)}\frac{\vp(\mu_1-\xi_1+\eta/2)\vp(\mu_2-\xi_1-\eta/2)}
{\vp(\mu_2-\mu_1-\i\eps)}\nn\\
\phantom{Q_{12}}{}
=\frac1{\vp(\eta)}\frac{\vp(x_1+\eta/2)\vp(x_2-\eta/2)}
{\vp(x_2-x_1-\i\eps)}, \label{eq:Q12} \\
Q_{21} =Q(\mu_1-3\eta/2, \mu_2-\eta/2) =\frac1{\vp(\eta)}\frac{\vp(\mu_1-\xi_1-\eta/2)\vp(\mu_2-\xi_1+\eta/2)}
{\vp(\mu_2-\mu_1+2\eta+\i\eps)}\nn\\
\phantom{Q_{21}}{}
=\frac1{\vp(\eta)}\frac{\vp(x_1-\eta/2)\vp(x_2+\eta/2)}
{\vp(x_2-x_1+2\eta+\i\eps)}   .\nonumber
\end{gather}
Thus, we have
\begin{gather*}
\bra E^{22} \ket = I_{12} - I_{21}
\end{gather*}
where $I_{12}$ and $I_{21}$ are given by
\begin{gather*}
I_{12}=\intall\d x_1\intall\d x_2
Q_{12}(x_1,x_2)\rho(x_1)\rho(x_2), \\
 I_{21}=\intall\d x_1\intall\d x_2
Q_{21}(x_1,x_2)\rho(x_1)\rho(x_2).
\end{gather*}

The integrand $Q_{12}$ is transformed into $Q_{21}$
when we shift the integral path as
$x_1\to x_1-\eta$ and $x_2\to x_2+\eta$.
First we shift the integral path in $I_{12}$ as $x_1\to x_1-\eta$.

\smallskip

\centerline{\includegraphics[width=40mm]{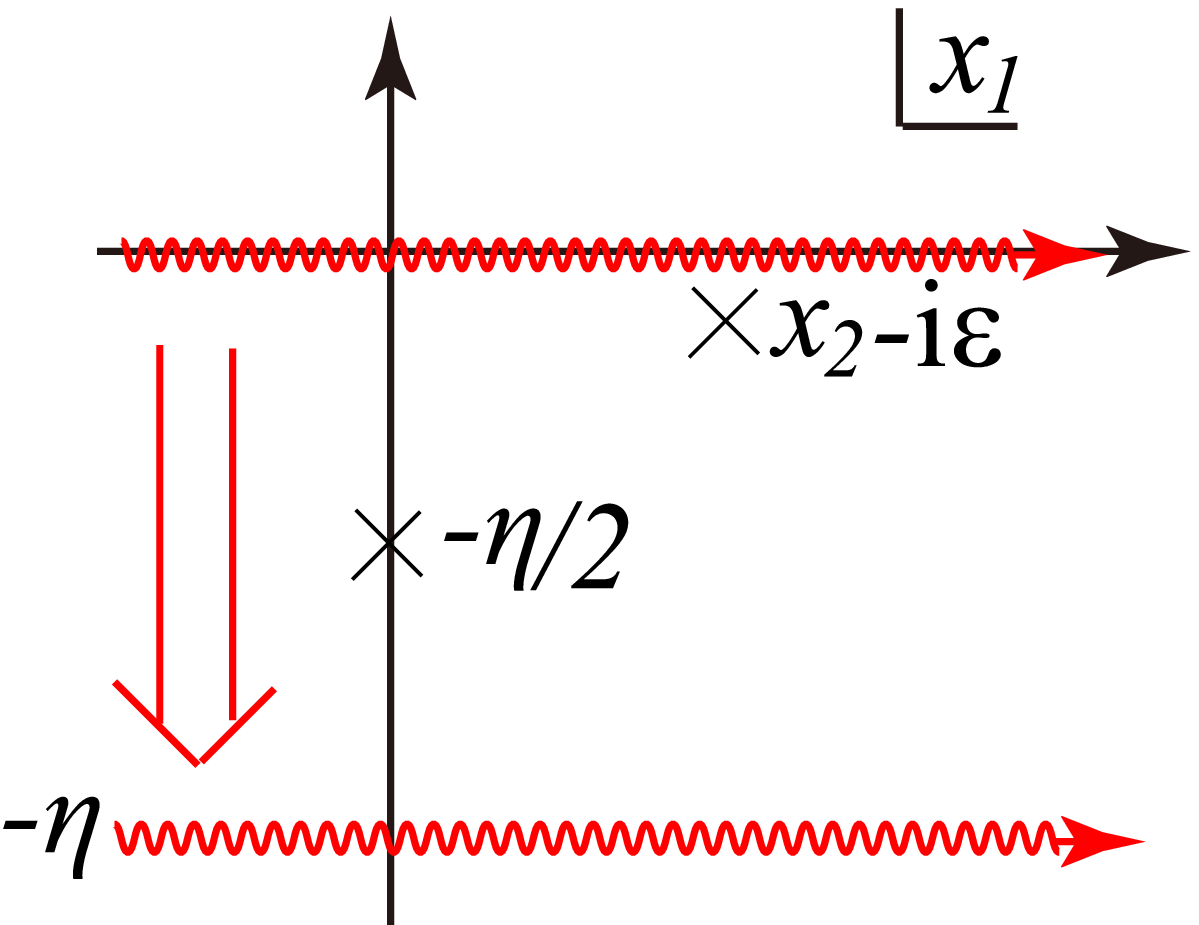}}

Here we note that due to the sign in front of $\eps$
in the denominator of (\ref{eq:Q12}), the integrand $Q_{12}$
has a pole at $x_1=x_2- i \eps$ as a function of $x_1$.
Here we recall that
$\eps$ is an inf\/initesimally small positive real number.
We therefore express the integral $I_{12}$ in terms of a sum of
two integrals, $J_1+J_2$, as follows
\begin{gather*}
I_{12} =\intall\d x_2\rho(x_2)\intall\d x_1Q_{12}(x_1,x_2)\rho(x_1) \nn\\
\phantom{I_{12}}{} =\intall\d x_2\rho(x_2)\left(
\int_{-\infty-\eta}^{\infty-\eta}\d x_1Q_{12}(x_1,x_2)\rho(x_1)
-2\pi\i   \Res{Q_{12}(x_1,x_2)\rho(x_1)}{x_1=x_2-\i\eps}
\right) \nn\\
\phantom{I_{12}}{}
=\intall\d x_2\rho(x_2)
\intall\d x_1Q_{12}(x_1-\eta,x_2)(-1)\rho(x_1) \nn\\
\phantom{I_{12}=}{}
-2\pi\i   \intall\d x_2\rho(x_2)
\Res{Q_{12}(x_1,x_2)\rho(x_1)}{x_1=x_2-\i\eps} \equiv J_1+J_2.
\end{gather*}
Here we have made use of the anti-periodicity: $\rho(x+n\eta)=(-1)^n\rho(x)$.
We also remark that the simple pole at $x_1=-\eta/2$ due to $\rho(x_1)$
is canceled by the factor $\vp(x_1+\eta/2)$ in $Q_{12}(x_1,x_2)$.

Let us f\/irst consider the single integral $J_2$ derived from the pole
at $x_1=x_2-\eps$.
Explicitly evaluating the integral $J_2$ we have
\begin{gather*}
J_2 =-2\pi\i   \intall\d x_2\rho(x_2)^2 \frac 1 {\vp(\eta)}
\frac{\vp(x_2+\eta/2)\vp(x_2-\eta/2)}{-1} \nn\\
\phantom{J_2}{}
=\frac{\pi}{4\zeta^2\sin\zeta}\intall\d x\frac{\ch2x-\cos\zeta}{\ch^2(\pi x/\zeta)}
=\frac{\zeta-\sin\zeta\cos\zeta}{2\zeta\sin^2\zeta}.
\end{gather*}
Here we have made use of formula \eqref{int2}.

\centerline{\includegraphics[width=40mm]{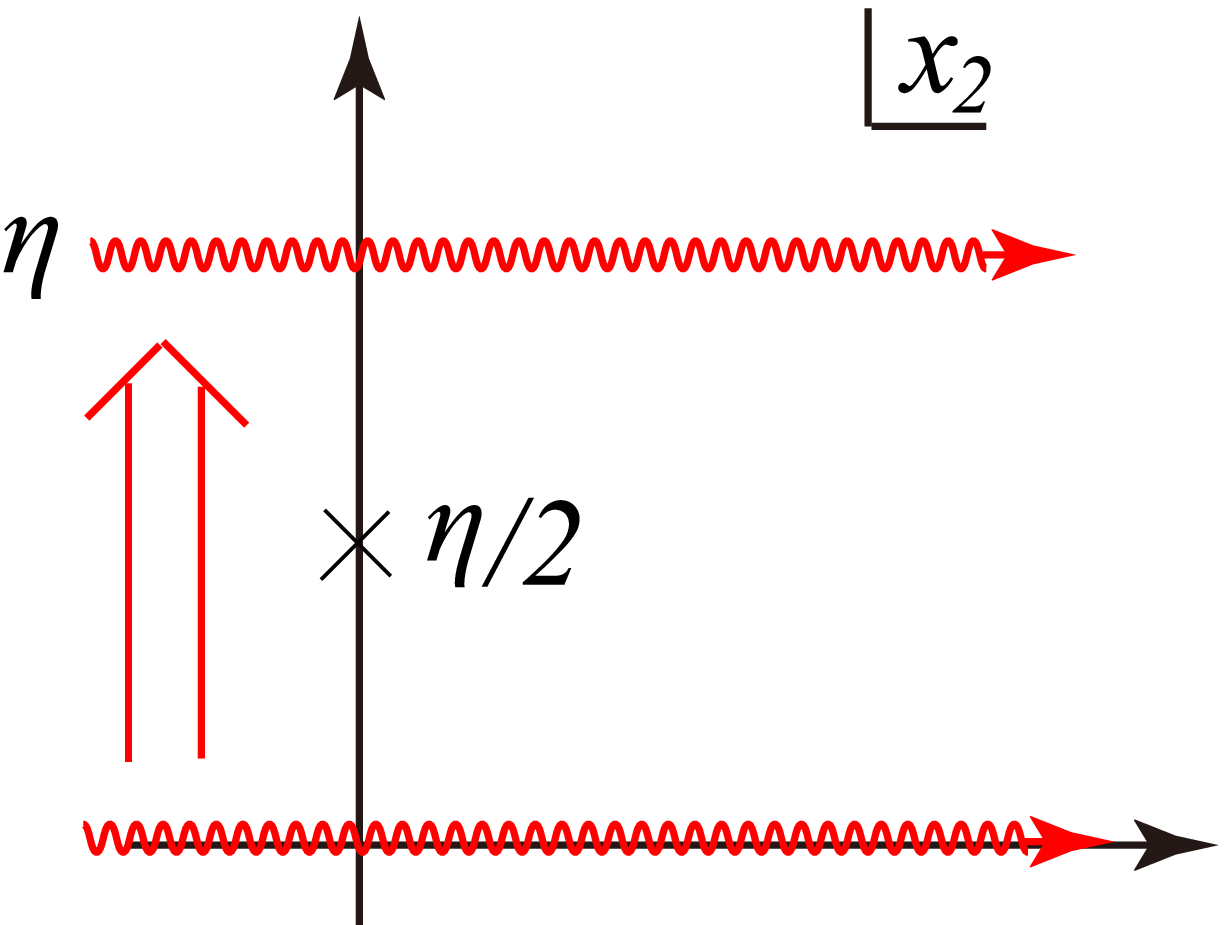}}

Let us next consider the double integral $J_1$.
We shift the integral path in $J_1$ as $x_2\to x_2+\eta$.
We derive the wanted integral $I_{21}$ as follows
\begin{gather*}
J_1 =\intall\d x_1\intall\d x_2
Q_{12}(x_1-\eta,x_2)(-1)\rho(x_1)\rho(x_2) \\
\phantom{J_1}{} =\intall\d x_1\int_{-\infty+\eta}^{\infty+\eta}\d x_2
Q_{12}(x_1-\eta,x_2)(-1)\rho(x_1)\rho(x_2) \\
\phantom{J_1}{} =\intall\d x_1\intall \d x_2
Q_{12}(x_1-\eta,x_2+\eta)(-1)^2\rho(x_1)\rho(x_2) \\
\phantom{J_1}{}
=\intall\d x_1\intall \d x_2
Q_{21}(x_1,x_2)\rho(x_1)\rho(x_2) =I_{21}.
\end{gather*}
Here we note that the simple pole at $x_2=\eta/2$ due to $\rho(x_2)$
is canceled by the factor $\vp(x_2-\eta/2)$ in $Q_{12}$.

Finally we have the analytical expression of the one-point
function $\bra E^{22} \ket$ as follows
\begin{gather*}
\bra E^{22} \ket =I_{12}-I_{21}
=
(J_1+J_2)-I_{21}
=
J_2
=\frac{\zeta-\sin\zeta\cos\zeta}{2\zeta\sin^2\zeta}.
\end{gather*}

\subsection[$\bra E^{11} \ket = 2 \bra e_1^{1,  1} e^{0,  0}_2  \ket$]{$\boldsymbol{\bra E^{11} \ket = 2 \bra e_1^{1,  1} e^{0,  0}_2  \ket}$}\label{section6.2}

Let us calculate a spin-1 one-point function, $\bra E^{11} \ket$.
Setting $\varepsilon_1'=1$ and $\varepsilon_2'=0$
in formula (\ref{eq:tildeE(p)-e}) we have
\begin{gather*}
 i_1=j_1=1; \qquad (\varepsilon_1, \varepsilon_2)= (0,1)  , (1,0)  ;
\qquad C=1;
\\
{\bm \alpha}^{+}=\{1 \}, \, \{ 2 \}   ;
\qquad {\bm \alpha}^{-}=\{1\}   ;
\qquad
(\tilde{\lam}'_1,\tilde{\lam}_1)=(\lam_1,\lam_2),  \qquad
(\tilde{\lam}'_2,\tilde{\lam}_1)=(\lam_1,\lam_2) .
\end{gather*}
Here $(\varepsilon_1, \varepsilon_2)= (1,0)$ corresponds to
 ${\bm \alpha}^{+}=\{ 2 \}$, and hence we have
$(\tilde{\lam}'_2,\tilde{\lam}_1)=(\lam_1,\lam_2)$.

Applying formula (\ref{eq:main-formula})
we express $\bra E^{11} \ket$ as follows
\begin{gather}
  \bra \widetilde{E_1}^{1 ,  1  (2   p)} \ket
= \langle {\psi_g^{(2  p)}} |
\widetilde{E_1}^{1 ,  1 \, (2  p)}
| {\psi_g^{(2 \, p)} } \rangle /
\langle {\psi_g^{(2  p)}} | {\psi_g^{(2  p)} } \rangle
\non \\
\qquad{}   =
\phi_2\big(\{ \lambda_{\gamma} \}; \big\{w_j^{(2)} \big\}_L\big)
\Bigg( \lim_{\epsilon \rightarrow 0}
{\frac
{\bra  \{ \lambda_{\alpha}(\epsilon) \}_M^{(2  p;  \epsilon)}  |
  B^{(2  p ;   \epsilon)}(w_1^{(2;   \epsilon)})
  C^{(2  p ;   \epsilon)}(w_2^{(2;   \epsilon)})
 |  \{ \lambda_{\alpha}(\epsilon) \}_M^{(2  p;  \epsilon)}   \ket}
  {\bra \{ \lambda_{\alpha}(\epsilon) \}_M^{(2  p;  \epsilon)}
 | \{ \lambda_{\alpha}(\epsilon) \}_M^{(2  p;  \epsilon)}  \ket} }
\non \\
\qquad\quad{} +
 \lim_{\epsilon \rightarrow 0}
{\frac
{\bra \{ \lambda_{\alpha}(\epsilon) \}_M^{(2  p ;  \epsilon)}|
  D^{(2  p ;  \epsilon)}(w_1^{(2 ;  \epsilon)})
  A^{(2  p ;  \epsilon)}(w_2^{(2 ;  \epsilon)})
| \{ \lambda_{\alpha}(\epsilon) \}_M^{(2  p ;  \epsilon)}  \ket}
{\bra \{ \lambda_{\alpha}(\epsilon) \}_M^{(2  p ;  \epsilon)}  |
      \{ \lambda_{\alpha}(\epsilon) \}_M^{(2  p ;  \epsilon)} \ket} }
\Bigg)
 . \label{eq:BC+DA}
\end{gather}
Considering the spin inversion symmetry
and the quantum group invariance
we evaluate $\bra E^{11} \ket$ by
\begin{gather}
\bra \widetilde{E_1}^{1 ,  1  (2   p)} \ket
=
2   \phi_2\big(\{ \lambda_{\gamma} \}; \big\{w_j^{(2)} \big\}_L\big)\nonumber\\
\phantom{\bra \widetilde{E_1}^{1 ,  1  (2   p)} \ket=}{} \times
\lim_{\epsilon \rightarrow 0}
{\frac
{\bra \{ \lambda_{\alpha}(\epsilon) \}_M^{(2  p ;  \epsilon)}|
  D^{(2  p ;  \epsilon)}(w_1^{(2 ;  \epsilon)})
  A^{(2  p ;  \epsilon)}(w_2^{(2 ;  \epsilon)})
| \{ \lambda_{\alpha}(\epsilon) \}_M^{(2  p ;  \epsilon)}  \ket}
{\bra \{ \lambda_{\alpha}(\epsilon) \}_M^{(2  p ;  \epsilon)}  |
      \{ \lambda_{\alpha}(\epsilon) \}_M^{(2  p ;  \epsilon)} \ket} }  .
\label{eq:DA}
\end{gather}

Let us brief\/ly review how we reduce (\ref{eq:BC+DA}) to (\ref{eq:DA}).
It follows from formula (\ref{eq:<E(p)>-<e>}) that we have
\begin{gather*}
\langle {\psi_g^{(2  p)}} | \widetilde{E}^{1,  1  (2  p)}_1
| {\psi_g^{(2  p)}} \rangle
= \langle \psi_g^{(2  p;  0} | e^{1,  0}_1e^{0,  1}_2
| \psi_g^{(2  p;  0} \rangle
+ \langle \psi_g^{(2  p;  0} | e^{1,  1}_1e^{0,  0}_2
| \psi_g^{(2  p;  0} \rangle   .
\end{gather*}
Due to the spin inversion symmetry (\ref{eq:spin-inv}) we have
\begin{gather*}
\langle \psi_g^{(2  p;  0} | e^{1,  0}_1e^{0,  1}_2
| \psi_g^{(2  p;  0} \rangle
=
\langle \psi_g^{(2  p;  0} | e^{0,  1}_1e^{1,  0}_2
| \psi_g^{(2  p;  0} \rangle  .
\end{gather*}
We have the following symmetry relation
(\ref{eq:sym-epsilon_a'}) due to the quantum group invariance $U_q(sl_2)$:
\begin{gather*}
\langle \psi_g^{(2  p;  0)} | e^{0,  1}_1e^{1,  0}_2
| \psi_g^{(2  p;  0)} \rangle
=
\langle \psi_g^{(2  p;  0)} | e^{1,  1}_1e^{0,  0}_2
| \psi_g^{(2  p;  0)} \rangle   .
\end{gather*}
Therefore, we have
\begin{gather*}
\langle \psi_g^{(2  p;  0)} | e^{1,  0}_1e^{0,  1}_2
| \psi_g^{(2  p;  0} \rangle
=
\langle \psi_g^{(2  p;  0)} | e^{1,  1}_1e^{0,  0}_2
| \psi_g^{(2  p;  0} \rangle.
\end{gather*}
Here we also recall that $|\psi_g^{(2  p ;  \epsilon)} \ket
=| \{ \lambda_{\alpha}(\epsilon) \}_M^{(2  p;  \epsilon)} \ket$.

Let us consider the case of
$(\varepsilon_1', \varepsilon_2')= (1,0)$
and $(\varepsilon_1, \varepsilon_2)= (1,0)$, which corresponds to
 ${\bm \alpha}^{+}=\{ 2 \}$ and ${\bm \alpha}^{-}=\{ 1 \}$,
and hence we have $(\tilde{\lam}'_2,\tilde{\lam}_1)=(\lam_1,\lam_2)$.
The multiple-integral formula reads
\begin{gather*}
\bra E^{11} \ket=2
\left( \int_{C_{+\i\eps}}+\int_{C_{-\eta+\i\eps}}
\right)\d\lam_1
\left( \int_{C_{-\i\eps}}+\int_{C_{-\eta-\i\eps}}
\right)\d\lam_2
Q(\lam_1,\lam_2)\det S(\lam_1,\lam_2)  ,
\end{gather*}
where $Q(\lam_1,\lam_2)$ and $\det S(\lam_1,\lam_2)$ are given by
\begin{gather}
Q(\lam_1,\lam_2)
 =-\frac{\vp(\lam_2-w^{(2)}_2)\vp(\lam_1-w^{(2)}_1-\eta)}
{\vp(\lam_2-\lam_1+\eta+\eps_{21})\vp(w^{(2)}_1-w^{(2)}_2)}
 =\frac{\vp(\lam_1-\xi_1-\eta) \vp(\lam_2-\xi_1+ \eta )}
{\vp(\lam_1-\lam_2-\eta+\eps_{12})\vp(\eta)}, \label{eq:Q-DA} \\
S(\lam_1,\lam_2)=
\begin{pmatrix}
\rho(\lam_1-w^{(2)}_1+\eta/2)\,\delta_{\alpha(\lam_1),1}
& \rho(\lam_1-w^{(2)}_2+\eta/2) \,\delta_{\alpha(\lam_1),2} \\
\rho(\lam_2-w^{(2)}_1+\eta/2) \,\delta_{\alpha(\lam_2),1}
& \rho(\lam_2-w^{(2)}_2+\eta/2) \,\delta_{\alpha(\lam_2),2}
\end{pmatrix}  .\nonumber
\end{gather}
Here we recall that  $\int_{C_{\i\alpha}}$ denotes the integral path
$\int_{-\infty+\i\alpha}^{\infty+\i\alpha}$ and also that
$\vp(x)=\sh(x)$.

Let $\Gamma_j$ denote a small contour rotating counterclockwise
around $\lam=w^{(2)}_j$ for each $j$.
We shift the integral paths $C_{-\i\eps}\to C_1$,
$C_{-\eta-\i\eps}\to C_2$
and $C_{+\i\eps}\to C_1-\Gamma_1$, $C_{-\eta+\i\eps}\to C_2-\Gamma_2$,
where $C_1=C_{-\eta/2}$ and $C_2=C_{-3\eta/2}$. For instance, we have
\begin{gather*}
\int_{C_{+\i\eps}} \d\lam_1
= \int_{C_1} \d\lam_1  - \int_{\Gamma_1} \d\lam_1   .
\end{gather*}
Expanding the determinant of matrix $S$, we thus obtain
\begin{gather*}
\bra E^{11} \ket/2 =
\( \int_{C_1}-\int_{\Gamma_1}\)\d\lam_1\int_{C_2}\d\lam_2
Q(\lam_1,\lam_2) \rho(\lam_1-w^{(2)}_1+\eta/2)\rho(\lam_2-w^{(2)}_2+\eta/2)
\\
\phantom{\bra E^{11} \ket/2 =}{}
-\( \int_{C_2}-\int_{\Gamma_2}\)\d\lam_1\int_{C_1}\d\lam_2 Q(\lam_1,\lam_2)
\rho(\lam_1-w^{(2)}_2+\eta/2)\rho(\lam_2-w^{(2)}_1+\eta/2) .
\end{gather*}
The one-point function $\bra E^{11} \ket$ is now expressed in terms
of $J_1$, $J_2$, $K_1$ and $K_2$, as follows.
\begin{gather*}
\bra E^{11} \ket = 2 (- K_1+ K_2 + J_1 - J_2) .
\end{gather*}
Here we shall give def\/initions of integrals $J_1$, $J_2$, $K_1$
and $K_2$ and calculate them shortly in the following.
For $K_1$ and $K_2$,
making use of the formula: $2\pi\i\Res{\rho(\lam-w+\eta/2)}{\lam=w}=1$,
we have
\begin{gather*}
K_1 \equiv\int_{\Gamma_1}\d\lam_1\int_{C_2}\d\lam_2
Q(\lam_1,\lam_2) \rho(\lam_1-w^{(2)}_1+\eta/2)\rho(\lam_2-w^{(2)}_2+\eta/2)
\\
\phantom{K_1}{} =\int_{C_2}\d\lam_2 Q(w^{(2)}_1,\lam_2) \rho(\lam_2-w^{(2)}_2+\eta/2)
=\intall\d\mu_2 Q(\xi_1,\mu_2-3\eta/2) \rho(\mu_2-\xi_1)
\\
\phantom{K_1}{}
=\intall\d x \rho(x),
\end{gather*}
and
\begin{gather*}
K_2 \equiv\int_{\Gamma_2}\d\lam_1\int_{C_1}\d\lam_2
Q(\lam_1,\lam_2) \rho(\lam_1-w^{(2)}_2+\eta/2)\rho(\lam_2-w^{(2)}_1+\eta/2)
\\
\phantom{K_2}{}
=\int_{C_1}\d\lam_2 Q(w^{(2)}_2,\lam_2) \rho(\lam_2-w^{(2)}_1+\eta/2)
=\intall\d\mu_2 Q(\xi_1-\eta,\mu_2-\eta/2) \rho(\mu_2-\xi_1)\\
\phantom{K_2}{}
=2\ch\eta\intall\d x \rho(x)\frac{\vp(x+\eta/2)}{\vp(x+3\eta/2)}
=-2\ch\eta\intall\d x \rho(x)\frac{\vp(x-\eta/2)}{\vp(x+\eta/2)}.
\end{gather*}
We have def\/ined the integrals $J_1$ and $J_2$ by
\begin{gather*}
J_1 \equiv\int_{C_1}\d\lam_1\int_{C_2}\d\lam_2
Q(\lam_1,\lam_2) \rho(\lam_1-w^{(2)}_1+\eta/2)\rho(\lam_2-w^{(2)}_2+\eta/2)
\\
\phantom{J_1}{}
=\intall\d x_1\intall\d x_2\rho(x_1)\rho(x_2)\left(-\frac1{\vp(\eta)}\right)
\frac{\vp(x_1-3\eta/2)\vp(x_2-\eta/2)}{\vp(x_2-x_1-\i\eps)}, \\
J_2 \equiv\int_{C_2}\d\lam_1\int_{C_1}\d\lam_2
Q(\lam_1,\lam_2) \rho(\lam_1-w^{(2)}_2+\eta/2)\rho(\lam_2-w^{(2)}_1+\eta/2)
\\
\phantom{J_2}{} =\intall\d x_1\intall\d x_2\rho(x_1)\rho(x_2)\left(-\frac1{\vp(\eta)}\right)
\frac{\vp(x_1-5\eta/2)\vp(x_2+\eta/2)}{\vp(x_2-x_1+2\eta+\i\eps)}.
\end{gather*}

As in the case of $\bra E^{22} \ket$, we transform
the integral $J_1$ into $J_2$
by shifting the integral path as
$x_1 \to x_1-\eta$ and $x_2 \to x_2+\eta$.
First we shift the integral path in $J_1$ as $x_1\to x_1-\eta$.
There are two simple poles at $x_1=x_2-\i\eps$ and $x_1=-\eta/2$.
Using $2\pi\i \Res{\rho(x)}{x=-\eta/2}=-1$,
we can calculate the residues as
\begin{gather*}
 2\pi\i\Res{\rho(x_1)\rho(x_2)\left(-\frac1{\vp(\eta)}\right)
\frac{\vp(x_1-3\eta/2)\vp(x_2-\eta/2)}{\vp(x_2-x_1-\i\eps)}}{x_1=-\eta/2}
\\
\qquad{} =-2(\ch\eta)\rho(x_2)\frac{\vp(x_2-\eta/2)}{\vp(x_2+\eta/2)},
\end{gather*}
and
\begin{gather*}
 2\pi\i\Res{\rho(x_1)\rho(x_2)\left(-\frac1{\vp(\eta)}\right)
\frac{\vp(x_1-3\eta/2)\vp(x_2-\eta/2)}{\vp(x_2-x_1-\i\eps)}}{x_1=x_2-\i\eps}
\\
\qquad {} =\frac{2\pi\i}{\vp(\eta)}\rho(x_2)^2\vp(x_2-3\eta/2)\vp(x_2-\eta/2).
\end{gather*}
Thus we have
\begin{gather*}
J_1 =-I_1-I_2+\int_{-\infty-\eta}^{\infty-\eta}
\d x_1\intall\d x_2\rho(x_1)\rho(x_2)\left(-\frac1{\vp(\eta)}\right)
\frac{\vp(x_1-3\eta/2)\vp(x_2-\eta/2)}{\vp(x_2-x_1-\i\eps)} \\
\phantom{J_1}{} =-I_1-I_2+\intall\d x_1\intall\d x_2(-1)\rho(x_1)\rho(x_2)\left(-\frac1{\vp(\eta)}\right)
\frac{\vp(x_1-5\eta/2)\vp(x_2-\eta/2)}{\vp(x_2-x_1+\eta-\i\eps)},
\end{gather*}
where
\begin{gather*}
I_1 =\intall\d x\left[
-2(\ch\eta)\rho(x)\frac{\vp(x-\eta/2)}{\vp(x+\eta/2)}
\right], \\
I_2 =\intall\d x\left[
\frac{2\pi\i}{\vp(\eta)}\rho(x)^2\vp(x-3\eta/2)\vp(x-\eta/2)
\right].
\end{gather*}
Next we shift the integral path as $x_2\to x_2+\eta$.
Here we remark that the simple pole at $x_2=\eta/2$ of $\rho(x_2)$
has zero residue due to the factor $\vp(x_2-\eta/2)$ of the integrand.
Thus we have
\begin{gather*}
J_1
=-I_1-I_2+\intall\d x_1\intall\d x_2(-1)^2\rho(x_1)\rho(x_2)\(-\frac1{\vp(\eta)}\)
\frac{\vp(x_1-5\eta/2)\vp(x_2+\eta/2)}{\vp(x_2-x_1+2\eta-\i\eps)} \\
\phantom{J_1}{} =-I_1-I_2+J_2,
\end{gather*}
where we have omitted the inf\/initesimal $\eps$ since we can shift the integral path without crossing the poles.
Thus,  we have
\begin{gather*}
\bra E^{11} \ket =2(-K_1+K_2+J_1-J_2)
=2(-K_1+K_2-I_1-I_2)
=2(-K_1-I_2),
\end{gather*}
where we have used the fact that $K_2=I_1$.
Using the formula (\ref{int1}), we have $K_1=1/2$.
Next we consider the integral $I_2$.
Shifting the integral path of $x$ as $x \to x+ \i\pi$, we have
\begin{gather*}
I_2/(2\pi \i)  =\int_{-\infty+\i\pi}^{\infty+\i\pi}
    {\frac {\vp(x-\eta/2) \vp(x- 3 \eta/2)} {\vp(\eta)}} \rho(x)^2 \d x\\
\phantom{I_2/(2\pi \i)  =}{}
+ 2 \pi\i \Res{ \frac{\vp(x-\eta/2) \vp(x-3\eta/2)}{\vp(\eta)} \rho(x)^2 }
{x=\i\pi/2} \\
\phantom{I_2/(2\pi \i) }{}
 = \int_{-\infty}^{\infty}
    {\frac {\vp(x+\eta/2) \vp(x- \eta/2)} {\vp(\eta)}} (-1)^2 \rho(x)^2 \d x
+ 2 \pi\i  \frac{\vp(-\eta)}{\vp(\eta) (2 \pi \i)^2}
\\
\phantom{I_2/(2\pi \i)}{}
  ={\frac 1 {4 \zeta^2 \vp(\eta)}}
   \int_{-\infty}^{\infty}
 {\frac {\sinh(x+\eta/2) \sinh(x- \eta/2)}
{\cosh^2(\pi x/\zeta )}} \d x - {\frac 1 {2 \pi \i}}    .
\end{gather*}
Making use of the formula:
$\sinh(x+\eta/2) \sinh(x- \eta/2) = (\cosh 2x - \cosh \eta)/2$ we have
\begin{gather*}
I_2   = 2 \pi \i   \left( \int_{-\infty}^{\infty} \frac {\cosh 2x}
{\cosh^2(\pi x/\zeta)} \d x - \cosh \eta
\int_{-\infty}^{\infty} \frac {1} {\cosh^2(\pi x/\zeta)} \d x \right) - 1
 =\frac{\zeta-\sin\zeta\cos\zeta}{ 2 \zeta\sin^2\zeta}- 1 ,
\end{gather*}
where we have used the formula (\ref{int2}).
Finally, we obtain
\begin{gather*}
\bra E^{11} \ket =2(-K_1-I_2)
=\frac{\cos\zeta(\sin\zeta-\zeta\cos\zeta)}{\zeta\sin^2\zeta}.
\end{gather*}

\subsection[$\bra E^{00} \ket$]{$\boldsymbol{\bra E^{00} \ket}$}\label{section6.3}
In this case we have
\begin{gather*}
 i_1=j_1=0; \qquad (\veps_1, \veps_2) = (0, 0), \qquad
(\veps_1', \veps'_2) = (0, 0);
\qquad C=1;
\\
{\bm \alpha}^{+}=\{1, 2\}; \qquad {\bm \alpha}^{-}=\varnothing; \qquad
(\tilde{\lam}'_2,\tilde{\lam}_1)=(\lam_1,\lam_2).
\end{gather*}
The multiple-integral formula reads
\begin{gather*}
\bra E^{00} \ket=2
\left( \int_{C_{+\i\eps}}+\int_{C_{-\eta+\i\eps}}
\right)\d\lam_1
\left( \int_{C_{+\i\eps}}+\int_{C_{-\eta+\i\eps}}
\right)\d\lam_2
Q(\lam_1,\lam_2)\det S(\lam_1,\lam_2),
\end{gather*}
where $Q(\lam_1,\lam_2)$ and $S(\lam_1,\lam_2)$ are given by
\begin{gather*}
Q(\lam_1,\lam_2)
=\frac{\vp(\lam_2-w^{(2)}_2)\vp(\lam_1-w^{(2)}_1-\eta)}
{\vp(\lam_2-\lam_1+\eta+\eps_{21})\vp(w^{(2)}_1-w^{(2)}_2)}, \\
S(\lam_1,\lam_2)=
\begin{pmatrix}
\rho(\lam_1-w^{(2)}_1+\eta/2)\,\delta_{\alpha(\lam_1),1}
& \rho(\lam_1-w^{(2)}_2+\eta/2) \,\delta_{\alpha(\lam_1),2} \\
\rho(\lam_2-w^{(2)}_1+\eta/2) \,\delta_{\alpha(\lam_2),1}
& \rho(\lam_2-w^{(2)}_2+\eta/2) \,\delta_{\alpha(\lam_2),2}
\end{pmatrix}.
\end{gather*}
Here we recall that  $\int_{C_{\i\alpha}}$ denotes
the integral path $\int_{-\infty+\i\alpha}^{\infty+\i\alpha}$
and $\vp(x)=\sh(x)$.

We now shift the integral paths
$C_{+\i\eps} \to C_1-\Gamma_1$, $C_{-\eta+\i\eps} \to C_2-\Gamma_2$,
where $C_1=C_{-\eta/2}$, $C_2=C_{-3\eta/2}$ and
$\Gamma_j$ is a small contour rotating counterclockwise
around $\lam=w^{(2)}_j$.
Expanding the determinant of matrix $S$, we obtain
\begin{gather*}
\bra E^{00} \ket =
\( \int_{C_1}-\int_{\Gamma_1}\)\d\lam_1\left( \int_{C_2}-\int_{\Gamma_2}\right)\d\lam_2
Q(\lam_1,\lam_2) \rho(\lam_1-w^{(2)}_1+\eta/2)\rho(\lam_2-w^{(2)}_2+\eta/2)
\\
\phantom{\bra E^{00} \ket=}{}{}-
\left( \int_{C_2}\!-\int_{\Gamma_2}\!\right)\d\lam_1\left( \int_{C_1}\!-\int_{\Gamma_1}\!\right)\d\lam_2 Q(\lam_1,\lam_2)
\rho(\lam_1-w^{(2)}_2+\eta/2)\rho(\lam_2-w^{(2)}_1+\eta/2) \\
\phantom{\bra E^{00} \ket}{}
=I_1-I_2-I_3-I_4+I_5+I_6,
\end{gather*}
where
\begin{gather*}
 I_1=\int_{C_1}\d\lam_1\int_{C_2}\d\lam_2 Q(\lam_1,\lam_2) \rho^1_1 \rho^2_2, \qquad
I_2=\int_{C_2}\d\lam_1\int_{C_1}\d\lam_2 Q(\lam_1,\lam_2) \rho^1_2 \rho^2_1, \\
I_3=Q\big(w^{(2)}_2,w^{(2)}_1\big), \qquad
I_4=\int_{C_2}\d\lam_2 Q\big(w^{(2)}_1,\lam_2\big) \rho^2_2, \qquad
I_5=\int_{C_1}\d\lam_2 Q\big(w^{(2)}_2,\lam_2\big) \rho^2_1, \\
 I_6=\int_{C_2}\d\lam_1 Q\big(\lam_1,w^{(2)}_1\big) \rho^1_2,
\end{gather*}
and $\rho^j_k=\rho(\lam_j-w^{(2)}_k+\eta/2)$.
Shifting the integral path as for the former cases, we have $I_1-I_2=-K_1-K_2$ where
\begin{gather*}
K_1 =\intall \d x_2 2\pi\i
\Res{\frac{1}{\vp(\eta)}\frac{\vp(x_1-3\eta/2)\vp(x_2-\eta/2)}{\vp(x_2-x_1-\i\eps)}\rho(x_1)\rho(x_2)}
{x_1=x_2-\i\eps}\nn\\
\phantom{K_1}{} =\frac{\sin\zeta\cos\zeta-\zeta}{2\zeta\sin^2\zeta}+1, \nn\\
K_2 =\intall \d x_2 2\pi\i
\Res{\frac{1}{\vp(\eta)}\frac{\vp(x_1-3\eta/2)\vp(x_2-\eta/2)}{\vp(x_2-x_1-\i\eps)}\rho(x_1)\rho(x_2)}
{x_1=-\eta/2}\nn\\
 \phantom{K_2}{}
 =2\ch\eta\intall\d x\frac{\vp(x-\eta/2)}{\vp(x+\eta/2)}\rho(x).
\end{gather*}
The other terms are calculated as
\begin{gather*}
I_3 =-1, \qquad
I_4 =-\intall\rho(x)\d x, \qquad
I_5 =K_2, \qquad
I_6 =-\intall\rho(x)\d x.
\end{gather*}
Summing up all the contributions, we have
\begin{align}
\bra E^{00} \ket=\frac{\zeta-\sin\zeta\cos\zeta}{2\zeta\sin^2\zeta}.
\end{align}
Here we can conf\/irm that the relation $\bra E^{22} \ket=\bra E^{00} \ket$
by directly evaluating the integral.

\subsection[$\bra E^{11} \ket$ through $2\bra e_1^{0, 1} e_2^{1, 0}  \ket$]{$\boldsymbol{\bra E^{11} \ket}$ through $\boldsymbol{2\bra e_1^{0, 1} e_2^{1, 0}  \ket}$}\label{section6.4}

We evaluate $\bra E^{11} \ket$ by calculating the multiple integral
representing $\bra e_1^{0, 1} e_2^{1, 0} \ket$.
Here we recall that due to the spin inversion symmetry we have
$\bra E^{11} \ket = 2 \bra e_1^{0, 1} e_2^{1, 0}  \ket$.
In this case we have
\begin{gather*}
 i_1=j_1=1; \qquad (\veps_1, \veps_2) = (0, 1),  \qquad
(\veps_1', \veps'_2) = (1, 0); \qquad C=1 ;
\\
{\bm \alpha}^{+}=\{2\}; \qquad {\bm \alpha}^{-}=\{ 2 \}; \qquad
(\tilde{\lam}'_2, \tilde{\lam}_2)=(\lam_1,\lam_2).
\end{gather*}
The multiple-integral formula reads
\begin{gather*}
\bra E^{11} \ket=2
\left( \int_{C_{+\i\eps}}+\int_{C_{-\eta+\i\eps}}
\right)\d\lam_1
\left( \int_{C_{-\i\eps}}+\int_{C_{-\eta-\i\eps}}
\right)\d\lam_2
Q(\lam_1,\lam_2)\det S(\lam_1,\lam_2)  ,
\end{gather*}
where $Q(\lam_1,\lam_2)$ and $\det S(\lam_1,\lam_2)$ are given by
\begin{gather*}
Q(\lam_1,\lam_2)
 =-\frac{\vp(\lam_2-w_1^{(2)} + \eta)\vp(\lam_1-w_1^{(2)}-\eta)}
{\vp(\lam_2-\lam_1+\eta+\eps_{21})\vp\big(w_1^{(2)}-w_2^{(2)}\big)}
 = \frac{\vp(\lam_1-\xi_1-\eta) \vp(\lam_2-\xi_1 + \eta)}
{\vp(\lam_1-\lam_2 - \eta+\eps_{12}) \vp(\eta)},  \\
S(\lam_1,\lam_2) =
\begin{pmatrix}
\rho(\lam_1-w^{(2)}_1+\eta/2) \delta_{\alpha(\lam_1),1}
& \rho(\lam_1-w^{(2)}_2+\eta/2) \,\delta_{\alpha(\lam_1),2} \\
\rho(\lam_2-w^{(2)}_1+\eta/2)  \delta_{\alpha(\lam_2),1}
& \rho(\lam_2-w^{(2)}_2+\eta/2)  \delta_{\alpha(\lam_2),2}
\end{pmatrix}  .
\end{gather*}
Here we remark that we have the same  $Q(\lam_1,\lam_2)$ as in equation~(\ref{eq:Q-DA})
for the case of $\bra e_1^{1,  1} e_2^{0,  0} \ket$.
We therefore obtain
\begin{gather*}
\bra  E^{1,  1} \ket
=\frac{\cos\zeta(\sin\zeta-\zeta\cos\zeta)}{\zeta\sin^2\zeta}.
\end{gather*}
We have thus conf\/irmed the quantum group invariance
$\bra e_1^{0, 1} e_2^{1, 0}  \ket=\bra e_1^{1, 1} e_2^{0, 0} \ket$
through the multiple-integral representation.

Finally in Section~\ref{section6} we give an important remark:
through an explicit evaluation of the multiple integrals
of $\bra \widetilde{E}^{1 ,   1  (2  p)} \ket$
we  have shown the following relations:
\begin{gather*}
\bra e_1^{0, 0} e_2^{1, 1}  \ket = \bra e_1^{1, 1} e_2^{0, 0}  \ket
  , \qquad
\bra e_1^{1, 0} e_2^{0, 1}  \ket = \bra e_1^{0, 0} e_2^{1, 1}  \ket   .
\end{gather*}
It follows that in the spin-1 case,
every one-point function is expressed in terms of a single multiple integral,
which corresponds to the expectation value of
a single product of the local spin-1/2 operators.
In general, however, the spin-$s$ correlation function of an arbitrary entry
is expressed in terms of
the expectation values of a sum of products of the local spin-1/2 operators
such as shown in~(\ref{eq:Multiple-integral_represenation}).
Here we recall that  the sum over sets
${\bm \alpha}^{+}(\{ \varepsilon_{\beta} \} )$
in~(\ref{eq:Multiple-integral_represenation}) corresponds to
the sum over sequences $\{ \varepsilon_{\beta} \}$
 in the reduction formula of Corollary~\ref{cor:main}.

\section{Consistency with numerical estimates\\ of the spin-1 one-point functions}\label{section7}

We now show that the analytical expressions of the spin-1 one-point functions
 are consistent with their numerical estimates,
which are obtained by the method of numerical exact diagonalization
of the integrable spin-1 XXZ Hamiltonian.

\begin{figure}[t]
\centering
\includegraphics[width=50mm]{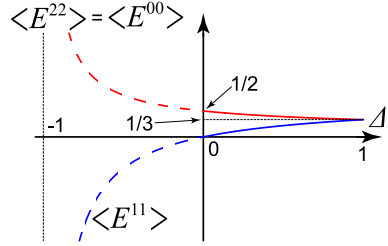}
\caption{One-point functions of spin-1 XXZ chain in the massless regime
obtained by the explicit evaluations of multiple integrals.
The red and blue lines represent those
for $\bra E^{22} \ket =\bra E^{00} \ket$ and
$\bra E^{11} \ket$, respectively.}\label{result}
\end{figure}

Let us f\/ist summarize the analytical results derived in Section~\ref{section6}.
Evaluating the multiple integrals explicitly,
we have obtained all the one-point function for the integrable spin-1 XXZ
chain as
\begin{gather*}
\bra \widetilde{E}^{2,   2   (2   p)} \ket =
\bra \widetilde{E}^{0,   0   (2   p)} \ket=
\frac{\zeta-\sin\zeta\cos\zeta}{2\zeta\sin^2\zeta},
\qquad
\bra \widetilde{ E}^{1,    1   (2  p)} \ket
=\frac{\cos\zeta(\sin\zeta-\zeta\cos\zeta)}{\zeta\sin^2\zeta},
\end{gather*}
which are shown in Fig.~\ref{result}.
In particular,  via evaluation of the multiple integrals,
 we have conf\/irmed the uniaxial symmetry relation:
\begin{gather}
\bra E^{22} \ket=\bra E^{00} \ket   . \label{eq:uniaxial}
\end{gather}
Through the direct evaluation of the multiple integrals
we conf\/irm the identity:
$\bra E^{22} \ket + \bra E^{11} \ket + \bra E^{00} \ket=1$.
Here we recall that assuming the uniaxial symmetry~(\ref{eq:uniaxial})
the analytical expression of $\bra E^{00} \ket$ has
been given in~\cite{DM2}.

Furthermore, we have conf\/irmed the relations among the correlation functions
due to the quantum group $U_q(sl_2)$ symmetry
and the spin inversion symmetry as follows
\begin{gather*}
\bra \widetilde{E}^{1 ,    1   (2   p)} \ket
= 2   \bra e_1^{0, 0} e_2^{1, 1}  \ket
= 2   \bra e_1^{1, 1} e_2^{0, 0}  \ket
=  2   \bra e_1^{0, 1} e_2^{1, 0}  \ket
=   2   \bra e_1^{1, 0} e_2^{0, 1} \ket   .
\end{gather*}

In the XXX limit $\Delta\to 1$ we have $\bra E^{22} \ket=\bra E^{11} \ket = \bra E^{00} \ket =1/3$, which has been shown by Kitanine in the XXX case~\cite{Kitanine2001}.
In the free Fermion limit $\Delta\to 0$ we have
$\bra E^{22} \ket=\bra E^{00} \ket =1/2$, and $\bra E^{11} \ket=0$.
Here we should remark
that we consider the region $0\leq\zeta<\pi/(2s)$ with $s=1$,
namely, $0 < \Delta \leq 1$.

\begin{figure}[t]
\centering
\includegraphics[width=80mm]{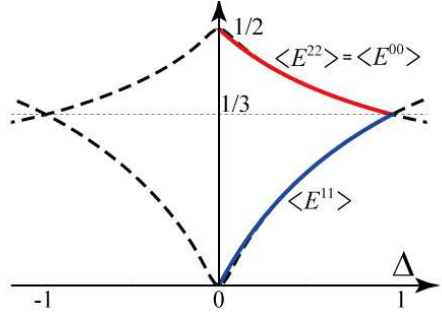}
\caption{Comparison with the exact numerical diagonalization.
The red and blue lines represent analytical results obtained by the multiple integrals
for $\bra E^{22} \ket=\bra E^{00} \ket$ and $\bra E^{11} \ket$, respectively.
The black dotted lines represent those obtained by
exact diagonalization
with the system size $N_s=8$.}\label{ed}
\end{figure}

Finally, we conf\/irm the analytical results
by comparing them with the numerical results of
exact diagonalization, which are shown in Fig.~\ref{ed}.
In Fig.~\ref{ed}, the red and blue lines represent
the analytical results obtained by evaluating the multiple integrals
of the one-point functions,
$\bra E^{22} \ket=\bra E^{00} \ket$ and $\bra E^{11} \ket$, respectively.
The black dotted lines represent the numerical estimates
of the one-point functions which are obtained by the method of
exact diagonalization of the integrable spin-1 XXZ Hamiltonian
with the system size of $N_s=8$. We numerically obtain the ground-state
eigenvector of the integrable spin-1 XXZ Hamiltonian, and
calculate the numerical estimates of the one-point functions.

We have found that
the numerical and analytical results of the spin-1 one-point functions
agree quite well in the region $0< \Delta \leq 1$, as shown in Fig.~\ref{ed}.
We thus conclude that the numerical results should support the
validity of the multiple-integral representations
for the spin-1 one-point functions.

\appendix

\section{Derivation of reduction formula (\ref{eq:tildeE(p)-e})}\label{appendixA}

 For the spin-$\ell/2$ Hermitian elementary matrices
associated with homogeneous grading,
$\widetilde{E}^{i,  j  (\ell,  +)}$,
we introduce coef\/f\/icients ${\widetilde g}_{i,j}$ by
\begin{gather*}
\widetilde{|| \ell, i \rangle}
\langle \ell, j ||
=
\sum_{\{ \varepsilon_{\alpha}' \}_{\ell}}
\sum_{ \{ \varepsilon_{\beta} \}_{\ell} }
{\widetilde g}_{i,j}( \{ \varepsilon_{\alpha}' \},
\{ \varepsilon_{\beta} \} )
e_1^{\varepsilon_1', \, \varepsilon_1} \cdots
e_{\ell}^{\varepsilon_{\ell}', \, \varepsilon_{\ell}} \, .
\end{gather*}
Then, we have
\begin{gather*}
{\widetilde g}_{i,j}( \{ \varepsilon_{\alpha}' \},
\{ \varepsilon_{\beta} \})
=
\left[
\begin{matrix}
\ell \\
i
\end{matrix}
\right]_q
\left[
\begin{matrix}
\ell \\
j
\end{matrix}
\right]_q^{-1}
\left(
\begin{matrix}
\ell \\
i
\end{matrix}
\right)_q   q^{i(i-1)/2-j(j-1)/2}
  q^{-(a(1) + \cdots +a(i) - i) + (b(1) + \cdots + b(j) -j)}.
\end{gather*}
We derive the reduction formula
for the Hermitian elementary operators
$\widetilde{E}^{i,  j  (\ell,  +)}$ as follows
\begin{gather}
  \widetilde{E}^{i,  j  (\ell,  +)}
 = {\widetilde P}^{(\ell)} \widetilde{E}^{i,  j  (\ell, \, +)}
 =
\left[
\begin{matrix}
\ell \\
i
\end{matrix}
\right]_q
\left[
\begin{matrix}
\ell \\
j
\end{matrix}
\right]_q^{-1}
\left(
\begin{matrix}
\ell \\
i
\end{matrix}
\right)   q^{i(i-1)/2-j(j-1)/2} \widetilde{ || \ell, i \rangle}
\label{g:sum} \\
 {} \times
\sum_{ \{ \varepsilon_{\beta} \}_{\ell} }
\sum_{ \{ \varepsilon_{\alpha}' \}_{\ell} }
\left(
\langle \ell, i || \sigma_{a(1)}^{-}  \cdots \sigma_{a(i)}^{-}
|| \ell, 0 \rangle q^{-(a(1) + \cdots + a(i) -i)}
\right) \langle \ell, 0 || \sigma_{b(1)}^{+}  \cdots \sigma_{b(j)}^{+}
q^{b(1) + \cdots + b(j) -i}  .
\non
\end{gather}
Here $\{ \varepsilon_{\beta} \}_{\ell}$
is given by a sequence of 0 or 1 such that
the number of integers $\beta$ for $1 \le \beta \le \ell$
satisfying $\varepsilon_{\beta}=1$ is given by
$j$, and $\{ \varepsilon_{\alpha}' \}_{\ell}$
is given by a sequence of 0 or 1 such that
the number of integers $\alpha$ satisfying
$\varepsilon_{\alpha}'=1$ is given by $i$.

We can show the following property:
\begin{lemma}[\cite{DM5}]
Let ${\bm \alpha}^{-}$ be a set of distinct integers
$\{a(1), \ldots, a(i) \}$ satisfying
$1 \le a(1) <  \cdots < a(i) \le \ell$,
we have the following:
\begin{gather*}
\langle \ell, i || \sigma_{a(1)}^{-} \cdots   \sigma_{a(i)}^{-}
||\ell, 0 \rangle
q^{-(a(1) + \cdots + a(i)) + i }
= \left[
\begin{matrix}
\ell \\
i
\end{matrix} \right]^{-1}
q^{-i(i-1)/2}   , 
\end{gather*}
which is independent of the set ${\bm \alpha}^{-} =
\{a(1), a(2), \ldots, a(i)\}$.
\label{lem:independance}
\end{lemma}

Applying Lemma~\ref{lem:independance} we show that
the inside of the parentheses (or the round brackets)
of equation~(\ref{g:sum}) is independent of~$a(k)$s.
Making use of the following:
\begin{gather*}
\sum_{\{ \varepsilon_{\alpha}' \}_{\ell} } 1 =
\left(
\begin{matrix}
\ell \\
i
\end{matrix}
\right)   ,
\end{gather*}
where $\varepsilon_{\alpha}'$ are such a sequence of 0 or 1 that
the number of $\varepsilon_{\alpha}'=1$ is given by~$i$.
We thus have
\begin{gather*}
\widetilde{E}^{i,   j   (\ell,   +)}
  =
\left[
\begin{matrix}
\ell \\
i
\end{matrix}
\right]_q
\left[
\begin{matrix}
\ell \\
j
\end{matrix}
\right]_q^{-1}
\left(
\begin{matrix}
\ell \\
i
\end{matrix}
\right)^{-1}   q^{i(i-1)/2-j(j-1)/2} \widetilde{ || \ell, i \rangle}
\langle \ell, i ||
 \\
\phantom{\widetilde{E}^{i,   j   (\ell,   +)}=}{} \times
\left(
\begin{matrix}
\ell \\
i
\end{matrix}
\right)
\sum_{ \{ \varepsilon_{\beta} \}_{\ell} }
 \sigma_{a(1)}^{-}  \cdots \sigma_{a(i)}^{-}
|| \ell, 0 \rangle  \langle \ell, 0 || \sigma_{b(1)}^{+}
 \cdots \sigma_{b(j)}^{+}    q^{-(a(1) + \cdots + a(i) -i)}
q^{b(1) + \cdots + b(j) -i}  \\
 \phantom{\widetilde{E}^{i,   j   (\ell,   +)}}{} =
\left[
\begin{matrix}
\ell \\
i
\end{matrix}
\right]_q
\left[
\begin{matrix}
\ell \\
j
\end{matrix}
\right]_q^{-1}
 q^{i(i-1)/2-j(j-1)/2}   \widetilde{ || \ell, i \rangle}
\langle \ell, i ||   e^{-(i-j) \xi_1}
\sum_{ \{ \varepsilon_{\beta} \}_{\ell} }
\chi_{1 \cdots \ell}    e_1^{\varepsilon_1',   \varepsilon_{1}}
\cdots e_{\ell}^{\varepsilon_{\ell}', \, \varepsilon_{\ell}}
\, \chi_{1 \cdots \ell}^{-1}  . \non \\
\end{gather*}
Here we recall that $\{ \varepsilon_{\beta} \}_{\ell}$ is a sequence
such that the number of integers $\beta$ of $1 \le \beta \le \ell$
satisfying $\varepsilon_{\beta} =1$ is given by $j$.
The integers $a(k)$ ($1 \le k \le i)$ and $b(k)$ ($1 \le k \le j)$
satisfying $1 \le a(1) < \cdots < a(i) \le \ell$ and
$1 \le b(1) < \cdots < b(j) \le \ell$, respectively, are related to
the sequences $\{ \varepsilon_{\alpha}' \}_{\ell}$ and
$\{ \varepsilon_{\beta} \}_{\ell}$ by the following relation~\cite{DM5}:
\begin{gather*}
e_1^{\varepsilon_1',   \varepsilon_1} \cdots
e_{\ell}^{\varepsilon_{\ell}',   \varepsilon_{\ell}}
 =
 e_{a(1)}^{1,   0} \cdots e_{a(i)}^{1,   0}
  e^{0,   0}_1 \cdots e^{0,   0}_{\ell}
   e_{b(1)}^{0,   1} \cdots e_{b(j)}^{0,   1}   .  
\end{gather*}

\section{Useful integral formulas}\label{appendixB}

\vspace{-6mm}

\begin{gather}
  \intall\frac{\d x}{\ch x}=\pi, \label{int1}\\
  \intall\frac{\ch 2ax}{\ch^2x}\d x=\frac{2\pi a}{\sin\pi a},
\qquad \text{for} \quad |a|<1  . \label{int2}
\end{gather}

\subsection*{Acknowledgments}

The authors  would like to thank
F.~G{\"o}hmann, C.~Matsui and K.~Motegi for helpful comments.
They are grateful for useful comments to
many participants of the workshop RAQIS'10,  June, 2010, LAPTH,
Annecy, France.
This work is partially supported by
Grant-in-Aid for Scientif\/ic Research (C) No.~20540365.
J.~Sato is supported by Grant-in-Aid for JSPS fellows.

\pdfbookmark[1]{References}{ref}
\LastPageEnding

\end{document}